# The IceCube Neutrino Observatory Part II: Atmospheric and Diffuse UHE Neutrino Searches of All Flavors

The IceCube Collaboration

## Contents

## IceCube Collaboration Member List

M. G. Aartsen[2], R. Abbasi[27], Y. Abdou[22], M. Ackermann[41], J. Adams[15], J. A. Aguilar[21], M. Ahlers[27], D. Altmann[9], J. Auffenberg[27], X. Bai[31,a], M. Baker[27], S. W. Barwick[23], V. Baum[28], R. Bay[7], J. J. Beatty[17,18], S. Bechet[12], J. Becker Tjus[10], K.-H. Becker[40], M. Bell[38], M. L. Benabderrahmane[41], S. BenZvi[27], P. Berghaus[41], D. Berley[16], E. Bernardini[41], A. Bernhard[30], D. Bertrand[12], D. Z. Besson[25], G. Binder[8,7], D. Bindig[40], M. Bissok[1], E. Blaufuss[16], J. Blumenthal[1], D. J. Boersma[39], S. Bohaichuk[20], C. Bohm[34], D. Bose[13], S. Böser[11], O. Botner[39], L. Brayeur[13], H.-P. Bretz[41], A. M. Brown[15], R. Bruijn[24], J. Brunner[41], M. Carson[22], J. Casey[5], M. Casier[13], D. Chirkin[27], A. Christov[21], B. Christy[16], K. Clark[38], F. Clevermann[19], S. Coenders[1], S. Cohen[24], D. F. Cowen[38,37], A. H. Cruz Silva[41], M. Danninger[34], J. Daughhetee[5], J. C. Davis[17], C. De Clercq[13], S. De Ridder[22], P. Desiati[27], K. D. de Vries[13], M. de With[9], T. DeYoung[38], J. C. Díaz-Vélez[27], M. Dunkman[38], R. Eagan[38], B. Eberhardt[28], J. Eisch[27], R. W. Ellsworth[16], S. Euler[1], P. A. Evenson[31], O. Fadiran[27], A. R. Fazely[6], A. Fedynitch[10], J. Feintzeig[27], T. Feusels[22], K. Filimonov[7], C. Finley[34], T. Fischer-Wasels[40], S. Flis[34], A. Franckowiak[11], K. Frantzen[19], T. Fuchs[19], T. K. Gaisser[31], J. Gallagher[26], L. Gerhardt[8,7], L. Gladstone[27], T. Glüsenkamp[41], A. Goldschmidt[8], G. Golup[13], J. G. Gonzalez[31], J. A. Goodman[16], D. Góra[41], D. T. Grandmont[20], D. Grant[20], A. Groß[30], C. Ha[8,7], A. Haj Ismail[22], P. Hallen[1], A. Hallgren[39], F. Halzen[27], K. Hanson[12], D. Heereman[12], D. Heinen[1], K. Helbing[40], R. Hellauer[16], S. Hickford[15], G. C. Hill[2], K. D. Hoffman[16], R. Hoffmann[40], A. Homeier[11], K. Hoshina[27], W. Huelsnitz[16,b], P. O. Hulth[34], K. Hultqvist[34], S. Hussain[31], A. Ishihara[14], E. Jacobi[41], J. Jacobsen[27], K. Jagielski[1], G. S. Japaridze[4], K. Jero[27], O. Jlelati[22], B. Kaminsky[41], A. Kappes[9], T. Karg[41], A. Karle[27], J. L. Kelley[27], J. Kiryluk[35], J. Kläs[40], S. R. Klein[8,7], J.-H. Köhne[19], G. Kohnen[29], H. Kolanoski[9], L. Köpke[28], C. Kopper[27], S. Kopper[40], D. J. Koskinen[38], M. Kowalski[11], M. Krasberg[27], K. Krings[1], G. Kroll[28], J. Kunnen[13], N. Kurahashi[27], T. Kuwabara[31], M. Labare[22], H. Landsman[27], M. J. Larson[36], M. Lesiak-Bzdak[35], M. Leuermann[1], J. Leute[30], J. Lünemann[28], J. Madsen[33], G. Maggi[13], R. Maruyama[27], K. Mase[14], H. S. Matis[8], F. McNally[27], K. Meagher[16], M. Merck[27], P. Mészáros[37,38], T. Meures[12], S. Miarecki[8,7], E. Middell[41], N. Milke[19], J. Miller[13], L. Mohrmann[41], T. Montaruli[21,c], R. Morse[27], R. Nahnhauer[41], U. Naumann[40], H. Niederhausen[35], S. C. Nowicki[20], D. R. Nygren[8], A. Obertacke[40], S. Odrowski[20], A. Olivas[16], M. Olivo[10], A. O'Murchadha[12], L. Paul[1], J. A. Pepper[36], C. Pérez de los Heros[39], C. Pfendner[17], D. Pieloth[19], E. Pinat[12], J. Posselt[40], P. B. Price[7], G. T. Przybylski[8], L. Rädel[1], M. Rameez[21], K. Rawlins[3], P. Redl[16], R. Reimann[1], E. Resconi[30], W. Rhode[19], M. Ribordy[24], M. Richman[16], B. Riedel[27], J. P. Rodrigues[27], C. Rott[17,d], T. Ruhe[19], B. Ruzybayev[31], D. Ryckbosch[22], S. M. Saba[10], T. Salameh[38], H.-G. Sander[28], M. Santander[27], S. Sarkar[32], K. Schatto[28], M. Scheel[1], F. Scheriau[19], T. Schmidt[16], M. Schmitz[19], S. Schoenen[1], S. Schöneberg[10], A. Schönwald[41], A. Schukraft[1], L. Schulte[11], O. Schulz[30], D. Seckel[31], Y. Sestayo[30], S. Seunarine[33], R. Shanidze[41], C. Sheremata[20], M. W. E. Smith[38], D. Soldin[40], G. M. Spiczak[33], C. Spiering[41], M. Stamatikos[17,e], T. Stanev[31], A. Stasik[11], T. Stezelberger[8], R. G. Stokstad[8], A. Stößl[41], E. A. Strahler[13], R. Ström[39], G. W. Sullivan[16], H. Taavola[39], I. Taboada[5], A. Tamburro[31], A. Tepe[40], S. Ter-Antonyan[6], G. Tešić[38], S. Tilav[31], P. A. Toale[36], S. Toscano[27], M. Usner[11], D. van der Drift[8,7], N. van Eijndhoven[13], A. Van Overloop[22], J. van Santen[27], M. Vehring[1], M. Voge[11], M. Vraeghe[22], C. Walck[34], T. Waldenmaier[9], M. Wallraff[1], R. Wasserman[38], Ch. Weaver[27], M. Wellons[27], C. Wendt[27], S. Westerhoff[27], N. Whitehorn[27], K. Wiebe[28], C. H. Wiebusch[1], D. R. Williams[36], H. Wissing[16], M. Wolf[34], T. R. Wood[20], K. Woschnagg[7], D. L. Xu[36], X. W. Xu[6], J. P. Yanez[41], G. Yodh[23], S. Yoshida[14], P. Zarzhitsky[36], J. Ziemann[19], S. Zierke[1], M. Zoll[34]

[1]III. Physikalisches Institut, RWTH Aachen University, D-52056 Aachen, Germany
[2]School of Chemistry & Physics, University of Adelaide, Adelaide SA, 5005 Australia
[3]Dept. of Physics and Astronomy, University of Alaska Anchorage, 3211 Providence Dr., Anchorage, AK 99508, USA
[4]CTSPS, Clark-Atlanta University, Atlanta, GA 30314, USA
[5]School of Physics and Center for Relativistic Astrophysics, Georgia Institute of Technology, Atlanta, GA 30332, USA
[6]Dept. of Physics, Southern University, Baton Rouge, LA 70813, USA
[7]Dept. of Physics, University of California, Berkeley, CA 94720, USA
[8]Lawrence Berkeley National Laboratory, Berkeley, CA 94720, USA
[9]Institut für Physik, Humboldt-Universität zu Berlin, D-12489 Berlin, Germany
[10]Fakultät für Physik & Astronomie, Ruhr-Universität Bochum, D-44780 Bochum, Germany
[11]Physikalisches Institut, Universität Bonn, Nussallee 12, D-53115 Bonn, Germany
[12]Université Libre de Bruxelles, Science Faculty CP230, B-1050 Brussels, Belgium
[13]Vrije Universiteit Brussel, Dienst ELEM, B-1050 Brussels, Belgium
[14]Dept. of Physics, Chiba University, Chiba 263-8522, Japan
[15]Dept. of Physics and Astronomy, University of Canterbury, Private Bag 4800, Christchurch, New Zealand
[16]Dept. of Physics, University of Maryland, College Park, MD 20742, USA
[17]Dept. of Physics and Center for Cosmology and Astro-Particle Physics, Ohio State University, Columbus, OH 43210, USA
[18]Dept. of Astronomy, Ohio State University, Columbus, OH 43210, USA
[19]Dept. of Physics, TU Dortmund University, D-44221 Dortmund, Germany
[20]Dept. of Physics, University of Alberta, Edmonton, Alberta, Canada T6G 2E1
[21]Département de physique nucléaire et corpusculaire, Université de Genève, CH-1211 Genève, Switzerland
[22]Dept. of Physics and Astronomy, University of Gent, B-9000 Gent, Belgium
[23]Dept. of Physics and Astronomy, University of California, Irvine, CA 92697, USA
[24]Laboratory for High Energy Physics, École Polytechnique Fédérale, CH-1015 Lausanne, Switzerland
[25]Dept. of Physics and Astronomy, University of Kansas, Lawrence, KS 66045, USA
[26]Dept. of Astronomy, University of Wisconsin, Madison, WI 53706, USA
[27]Dept. of Physics and Wisconsin IceCube Particle Astrophysics Center, University of Wisconsin, Madison, WI 53706, USA
[28]Institute of Physics, University of Mainz, Staudinger Weg 7, D-55099 Mainz, Germany
[29]Université de Mons, 7000 Mons, Belgium
[30]T.U. Munich, D-85748 Garching, Germany
[31]Bartol Research Institute and Department of Physics and Astronomy, University of Delaware, Newark, DE 19716, USA
[32]Dept. of Physics, University of Oxford, 1 Keble Road, Oxford OX1 3NP, UK
[33]Dept. of Physics, University of Wisconsin, River Falls, WI 54022, USA
[34]Oskar Klein Centre and Dept. of Physics, Stockholm University, SE-10691 Stockholm, Sweden
[35]Department of Physics and Astronomy, Stony Brook University, Stony Brook, NY 11794-3800, USA
[36]Dept. of Physics and Astronomy, University of Alabama, Tuscaloosa, AL 35487, USA
[37]Dept. of Astronomy and Astrophysics, Pennsylvania State University, University Park, PA 16802, USA
[38]Dept. of Physics, Pennsylvania State University, University Park, PA 16802, USA
[39]Dept. of Physics and Astronomy, Uppsala University, Box 516, S-75120 Uppsala, Sweden
[40]Dept. of Physics, University of Wuppertal, D-42119 Wuppertal, Germany
[41]DESY, D-15735 Zeuthen, Germany
[a]Physics Department, South Dakota School of Mines and Technology, Rapid City, SD 57701, USA
[b]Los Alamos National Laboratory, Los Alamos, NM 87545, USA
[c]also Sezione INFN, Dipartimento di Fisica, I-70126, Bari, Italy
[d]Department of Physics, Sungkyunkwan University, Suwon 440-746, Korea
[e]NASA Goddard Space Flight Center, Greenbelt, MD 20771, USA

**Acknowledgements**

We acknowledge the support from the following agencies: U.S. National Science Foundation-Office of Polar Programs, U.S. National Science Foundation-Physics Division, University of Wisconsin Alumni Research Foundation, the Grid Laboratory Of Wisconsin (GLOW) grid infrastructure at the University of Wisconsin - Madison, the Open Science Grid (OSG) grid infrastructure; U.S. Department of Energy, and National Energy Research Scientific Computing Center, the Louisiana Optical Network Initiative (LONI) grid computing resources; Natural Sciences and Engineering Research Council of Canada, WestGrid and Compute/Calcul Canada; Swedish Research Council, Swedish Polar Research Secretariat, Swedish National Infrastructure for Computing (SNIC), and Knut and Alice Wallenberg Foundation, Sweden; German Ministry for Education and Research (BMBF), Deutsche Forschungsgemeinschaft (DFG), Helmholtz Alliance for Astroparticle Physics (HAP), Research Department of Plasmas with Complex Interactions (Bochum), Germany; Fund for Scientific Research (FNRS-FWO), FWO Odysseus programme, Flanders Institute to encourage scientific and technological research in industry (IWT), Belgian Federal Science Policy Office (Belspo); University of Oxford, United Kingdom; Marsden Fund, New Zealand; Australian Research Council; Japan Society for Promotion of Science (JSPS); the Swiss National Science Foundation (SNSF), Switzerland.

# Search for extraterrestrial neutrino-induced cascades using IceCube 79-strings

THE ICECUBE COLLABORATION[1]

[1] *See special section in these proceedings*

*mlesiak-bzdak@icecube.wisc.edu*

**Abstract:** IceCube, a cubic kilometer detector at the South Pole, is the largest neutrino telescope currently taking data. Utilizing the transparent ice of Antarctica as a detection medium, IceCube digital optical sensors observe Cherenkov radiation from secondary particles produced in neutrino interactions inside or near the detector. Charged current $\nu_\mu$ interactions create muon tracks, while charged current $\nu_e$ interactions, and neutral current interactions of all flavors initiate electromagnetic and hadronic showers (cascades). The background coming from atmospheric muons and muon bundles is many orders of magnitude larger than the cascade signal and makes it difficult to observe cascades. However, cascades have better energy resolution and lower atmospheric neutrino background compared to track-like events. The energy spectrum of extraterrestrial neutrinos is expected to be harder than that of atmospheric neutrinos. Thus using cascade events to search for a hardening of the energy spectrum is advantageous compared to using muon tracks. The search for extraterrestrial neutrino-induced cascades with energies in the tens of TeV to a few PeV neutrino energy range using improved reconstruction methods will be presented. The analysis uses 317 days of livetime of the data taken from May 2010 to May 2011 when 79 IceCube strings were operational.

**Corresponding authors:** Mariola Lesiak-Bzdak[2] and Achim Stößl[3]
[2] *Department of Physics and Astronomy, Stony Brook University. Stony Brook, NY 11794-3800*
[3] *DESY, Platanenallee 6, 15738 Zeuthen, Germany*

## 1 Introduction

Extraterrestrial neutrinos, anticipated to be produced together with cosmic rays, might provide information about the mechanism of cosmic ray production and help to unveil cosmic ray sources. Although neutrino fluxes from such sources could be too low to be measured individually, an integrated flux over all sources might be possible to detect with IceCube [1], a cubic kilometer scale neutrino telescope located at the geographic South Pole. Incoming neutrinos interact mostly via deep-inelastic nucleon scattering and produce showers of secondary charged particles. Having relativistic velocities, these particles produce Cherenkov light that is detected by Digital Optical Modules (DOMs).

Neutrino-nucleon reactions are induced by all neutrino flavors via neutral current (NC) or charged current (CC) interactions. In charged current reactions the charged lepton is produced, which carries on average 50% (for $E_\nu \sim$10 GeV) to 80% (at high energies) of the neutrino energy; the remainder of the energy is transferred to the nuclear target. Depending on the charged lepton created in CC reactions, neutrino flavor specific hit-patterns might be observed in the detector which allow the identification of the incoming neutrino flavor. Charged current $\nu_\mu$ interactions create track-like hit patterns while CC $\nu_e$ reactions produce an electromagnetic and hadronic cascade which yields a spherical hit-pattern. The typical cascade analysis searches for $\nu_e$ and $\nu_\tau$ from CC and all neutrino flavors from NC interactions.

A previous cascade analysis searching for an astrophysical neutrino flux in IceCube with 22-strings instrumented [2] set a limit of $3.6\times10^{-7}$ GeV$\cdot$ sr$^{-1}$s$^{-1}$cm$^{-2}$ at 90% C.L. on $E^{-2}$ astrophysical neutrinos (assuming a 1:1:1 flavor ratio) with 90% of events in the energy range between 24 TeV to 6.6 PeV. Another IceCube cascade analyses looking for an extraterrestrial neutrino signal using 40 strings obtained preliminary results [3] and set a limit at 90% confidence level on an astrophysical neutrino flux of $9.5\times10^{-8}$ GeV$\cdot$ sr$^{-1}$s$^{-1}$cm$^{-2}$ with 90% of events in the energy range between 89 TeV to 21 PeV [4]. Preliminary cascade results using the 59-string configuration of IceCube were recently obtained and are presented at this conference [5].

In the recent 79- and 86-string IceCube detector, searches for extremely-high energy (EHE) neutrinos from all flavors from CC and NC interactions, two neutrino-induced cascade events at energies of 1 PeV were observed [6]. As a follow-up analysis, an all-sky search for all flavor neutrino events from CC and NC interactions with energies $E_\nu >$100 TeV and neutrino first interaction well contained in the 79- and 86-string IceCube detector, was performed and the preliminary results are presented in these proceedings [7].

The analysis described here was developed using Monte Carlo simulation and searched for an $E^{-2}$ astrophysical neutrino-induced cascade flux within IceCube with 79 strings instrumented. In these proceedings, we present all flavor sensitivity using high-energy contained cascade events in the IceCube detector. We also discuss adding partially contained events, to increase the effective volume. The neutrino energy range in this analysis is between 44 TeV and 7.7 PeV.

## 2 Data sample

The data used in this analysis were collected from May 2010 to May 2011 with 79 operational strings of IceCube. The analysis was performed as a blind analysis, the selection criteria to reject the background were developed using 10% of the data ("burnsample"). This burnsample consists of data uniformly distributed over the year to avoid biases in muon background rate due to seasonal variations. The burnsample livetime was 33 days. The numbers presented here are based on the remaining 90% of the data, 317 days.

The main background for a search for cascade-like events comes from cosmic ray muons with a faint track and a single catastrophic energy loss from a bremsstrahlung. The background of atmospheric muon events was simulated with the air-shower program CORSIKA [8]. The main goal was to simulate high energy muons that radiate bremsstrahlung secondaries with energies that can mimic cascade events. In this analysis, the CORSIKA background simulation generated for the primary cosmic ray energy higher than 30 TeV per nucleon was used. A sample of 300 days of atmospheric muon events in the energy range above 30 TeV per nucleon was generated.

The signals in this analysis are $\nu_e$ and $\nu_\tau$ from CC and all neutrino flavors cascades from NC interactions. The all flavor neutrino events were simulated with the neutrino generator ANIS [9] for energies from 1 TeV to 1 EeV at the surface of the Earth with $E^{-2}$ energy spectrum. Equal amounts of $\nu$ and $\bar{\nu}$ was produced. IceCube does not distinguish $\nu$ from $\bar{\nu}$ and in this paper $\nu$ denotes the sum of $\nu$ and $\bar{\nu}$. In this analysis we used the flux normalization of signal events of

$$\Phi_{model} = 1.0 \times 10^{-8} (\mathrm{E/GeV})^{-2} \mathrm{GeV}^{-1} \mathrm{s}^{-1} \mathrm{sr}^{-1} \mathrm{cm}^{-2}. \quad (1)$$

The background from atmospheric neutrinos was estimated assuming the conventional [10] and prompt [11] flux contributions.

## 3 Analysis

### 3.1 Cascade reconstruction variables

To isolate the cascade signal from muon background, different selection criteria were applied. Among these were simple quality criteria like the specific topology of cascade-like events, the development of the hit pattern in time, as well as causal and likelihood criteria.

A widely utilized topology criterion for cascade analysis was provided by `TensorOfInertia` [2]. This reconstruction considered the hit-pattern as a rigid body, with the optical modules as mass points with their charge equivalent to their mass. For this rigid body, the mass-eigenstates and corresponding eigenvalues were calculated. The ratio of the highest eigenvalue and the sum of all three eigenvalues is a measure how spherical the hit-pattern is and thus can be used to separate cascade-like from track-like events.

To separate a cascade-like hit pattern, which is a stationary source of light and a track, a moving source of light, the hits in the detector were projected along a track moving through the detector with `LineFitVelocity` [12]. For cascade-like events its value is much smaller then for track-like events and the identification of both hit patterns was possible.

In the analysis chain the following likelihood reconstruction algorithms were used: `ACER` [13], which is a deterministic energy estimator, `CascadeLlh` [2], which uses probability density functions (pdfs) to perform a 4-dimensional fit, and `Credo`, which is more sophisticated algorithm that incorporates a model of light propagation in the ice, the full timing information and reconstructs the energy and direction of the incident neutrino. For `CascadeLlh`, the reduced likelihood `rlogl` defined as a ratio of the logarithm of the likelihood and the number of degrees of freedom was calculated. Smaller values of `rlogl` indicate consistency with the cascade-like hypothesis and this helps to separate signal and background.

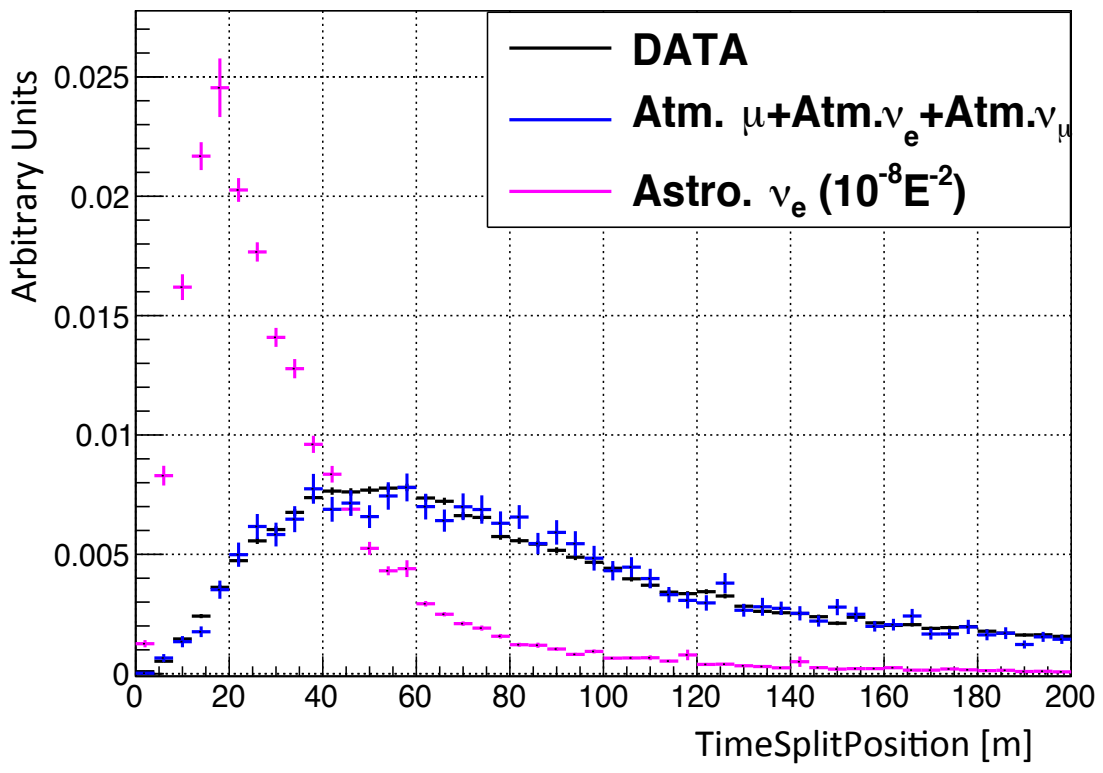

**Figure 1**: Normalized TimeSplitPosition distributions for data (black points), sum of muon and atmospheric neutrino backgrounds (blue) and $E^{-2}$ astrophysical $\nu_e$ signal (magenta).

The `FillRatio` was used to distinguish cascade-like events from muon-like tracks. Firstly, the mean distance between the vertex position and all hit DOMs in an event was calculated. Then, the ratio of number of hit DOMs to the total of all DOMs in the sphere of this mean radius was obtained. For a neutrino signal (cascade-like events) we expect this number to be close to one while for the track-like events this number would be uniformly distributed. This allows the separation of signal and background.

Another topology variable used in this analysis was `TimeSplitPosition`. Each event was split into two halves based on the charge-weighted mean time, and the cascade reconstruction was run on each half separately. Then, the difference `TimeSplitPosition` between reconstructed vertex positions for both halves was calculated. For the events consistent with a signal cascade hit pattern this number has a smaller value than for track-like events and allows the separation of signal and background, as shown in Fig. 1. Figure 1 shows the normalized `TimeSplitPosition` distributions for data, Monte Carlo background and $E^{-2}$ astrophysical $\nu_e$ signal. The shape of the data distribution is nicely reproduced by the sum of muon and atmospheric backgrounds and represents the typical data-Monte Carlo shape agreement at different cut levels in the analysis presented here.

The ratio of maximum total charge on a single DOM in a given event and the total charge in this event `MaxQTotRatio` allowed the identification of the events, where most of the charge was recorded by a single DOM. These events might be created by a low energy muon having a catastrophic energy loss next to a DOM.

The variable `DelayTime`, defined as a minimum of the time difference between the first hit on a DOM and the time of the reconstructed vertex was also used. It allows the separation of a muon-track and cascade-like events as for the former this time difference is bigger than for the latter.

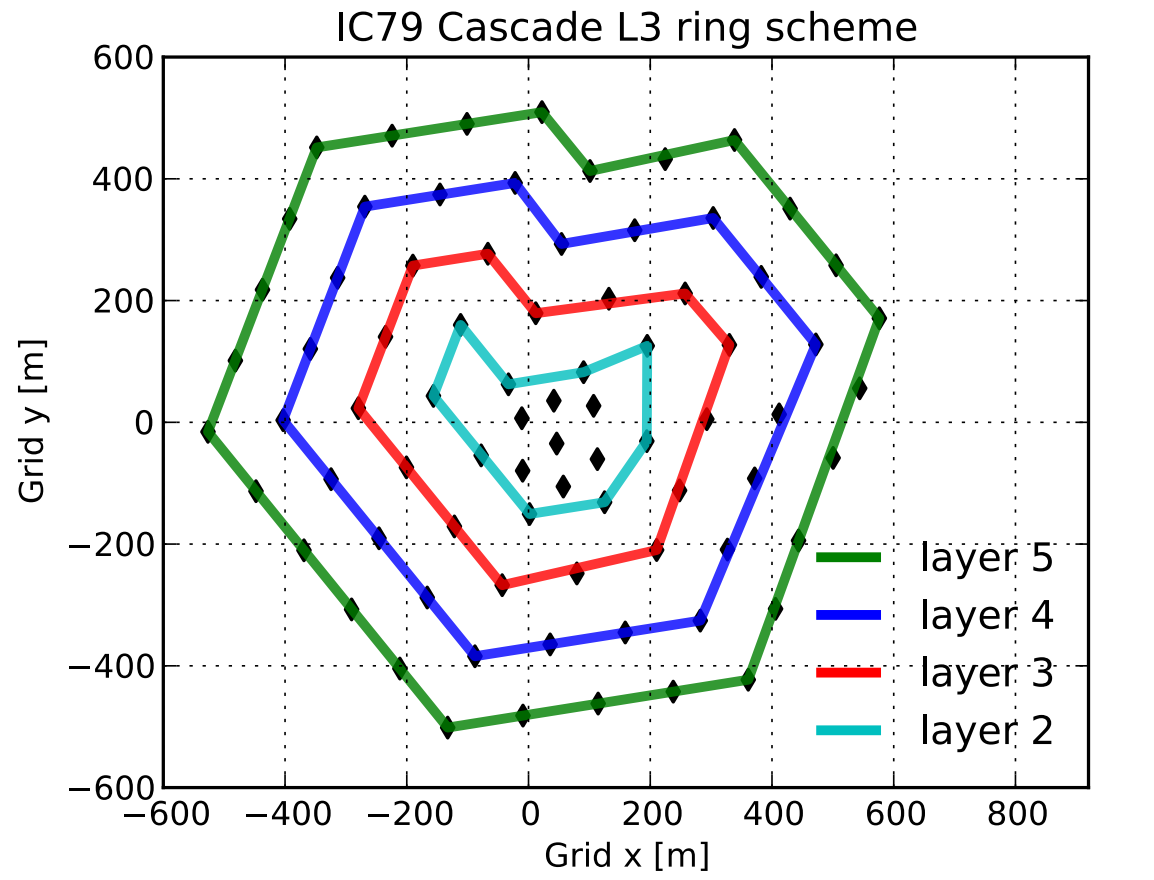

**Figure 2**: Schematic top view of IceCube with 79-strings. The green denotes the most outer layer of strings.

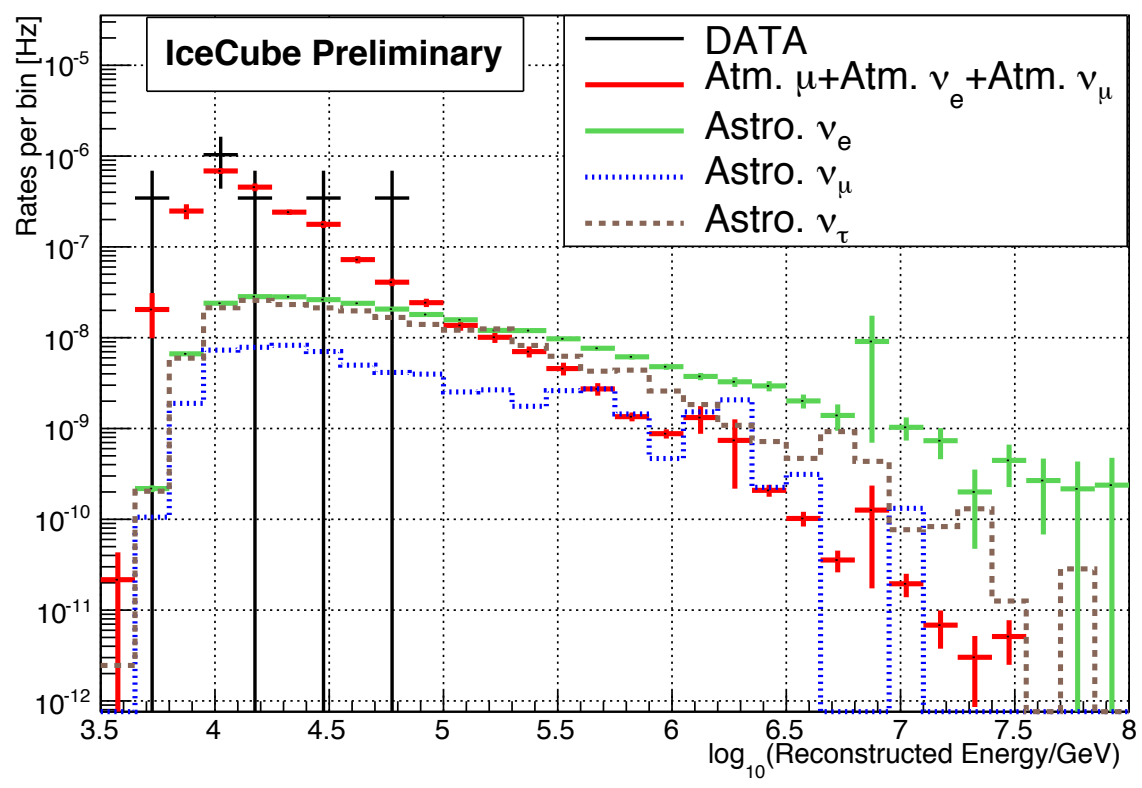

**Figure 3**: Distribution of the reconstructed energy before final energy cut.

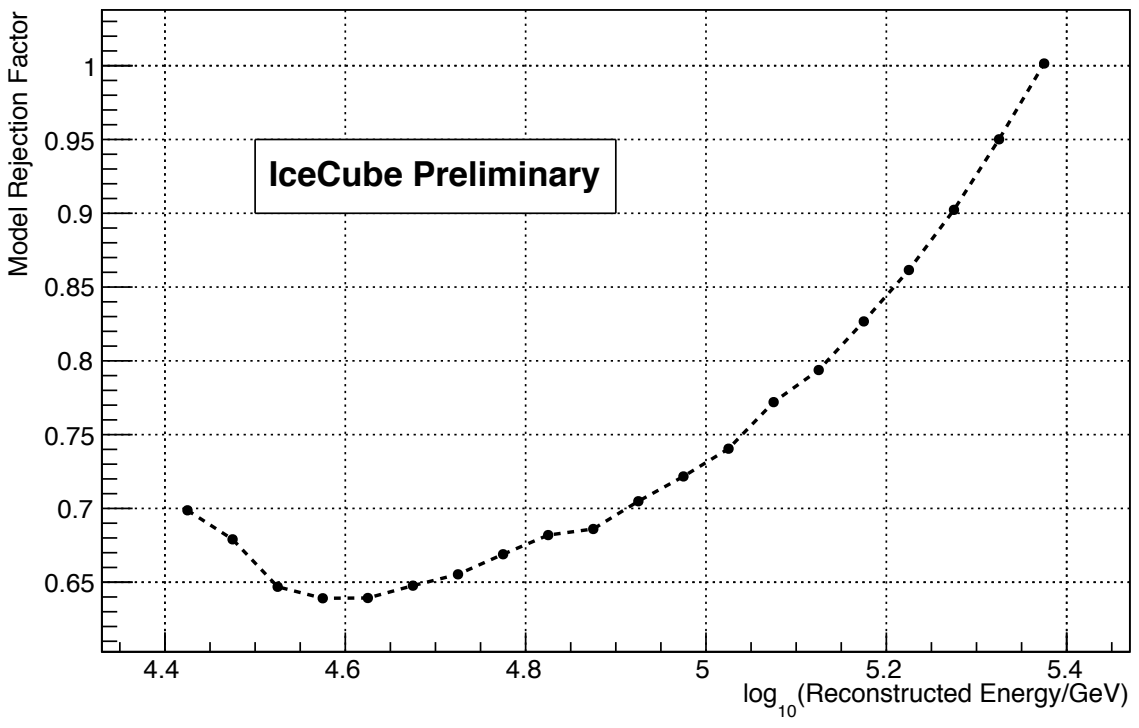

**Figure 4**: Model Rejection Factor (MRF) as a function of reconstructed energy.

### 3.2 Online filters

To reduce the background coming from atmospheric muons and muon bundles several filters were applied to the data. The online filtering process begins at the South Pole with a trigger logic to suppress electronic noise and noise induced by radioactive processes of the detector itself.

The main physics trigger in IceCube is a "Simple Multiplicity Trigger" (SMT) that requires photon signals in at least 8 DOMs. The average trigger rate for the IceCube 79-string configuration was 1970 Hz. In the cascade online filter the cuts on `TensorOfInteria` and `LineFitVelocity` were applied to select cascade-like signal events from track-like background. The online filter reduced the data rate to 21 Hz, about a factor of 100 below the trigger rate. The cascade filter retained 75% of the $\nu_e$ signal. After applying the online filter, the data stream was transferred to the North where more elaborate `CascadeLlh` and `ACER` cascade reconstructions were performed.

### 3.3 Event selection

Selection criteria to reject muon and atmospheric neutrino backgrounds were developed. The Level3 filter retained events that fulfilled either a combined criterion of a cascade and track likelihood ratio `LlhRatio` as well as an energy dependent zenith angle cut or had a reconstructed `ACER` energy larger than 10 TeV.

Then the data stream was split into two branches: fully contained and partially contained events and each branch was analyzed separately. Only the fully contained events selection criteria are described here but the partially contained events were used to enhance the sensitivity of this analysis for neutrino events with energies $E >$100 TeV.

The fully contained events were considered those with both the reconstructed vertex and the first hit inside the most outer string layer of the detector, the green polygon in Fig. 2. In addition, we required that the first hit in the event occurred between $\pm$430 meters in depth and the reconstructed `Credo` vertex position $Z$ was between $\pm$450 meters in the detector. We rejected the event if the earliest hit occurred in the seven topmost DOMs. The `FillRatio` was calculated for this branch and only events with value higher the 0.6 were retained.

At Level4, further cuts were applied to reduce the background from atmospheric muons. Based on the time and position of the pulses in a given event, the events seen by 4 or more strings were selected. In the next step of Level4, we required that the reconstructed `Credo` energy was higher that 10 TeV.

At Level5 we retained events with `TimeSplitPosition` smaller than 40 meters and rejected events with `MaxQTotRatio` bigger than 0.35. In addition, we required that the `DelayTime` was bigger than 100 ns and the `rlogl` was smaller than 7.5.

Finally, using the Feldman-Cousins method [14], a cut on reconstructed energy (see Fig. 3) was optimized and used to suppress remaining muon and atmospheric neutrinos background. The Model Refection Factor (MRF) [15] was calculated as a function of reconstructed energy as shown in Fig. 4. The minimum of the MRF distribution was found at an energy of $E$=40 TeV and the energy cut was placed at this value. The energy resolution for an $E^{-2}$ astrophysical spectrum for fully contained events is $\Delta(\log_{10}E_\nu) \sim 0.04$ and the vertex position resolution is $\sim$ 4 meters.

The analysis aiming at partially contained astrophys-ical neutrino search has a poorer energy resolution of $\Delta(\log_{10}\mathrm{E}_\nu) \sim 0.3$, and the vertex resolution of $\sim$ 10 meters.

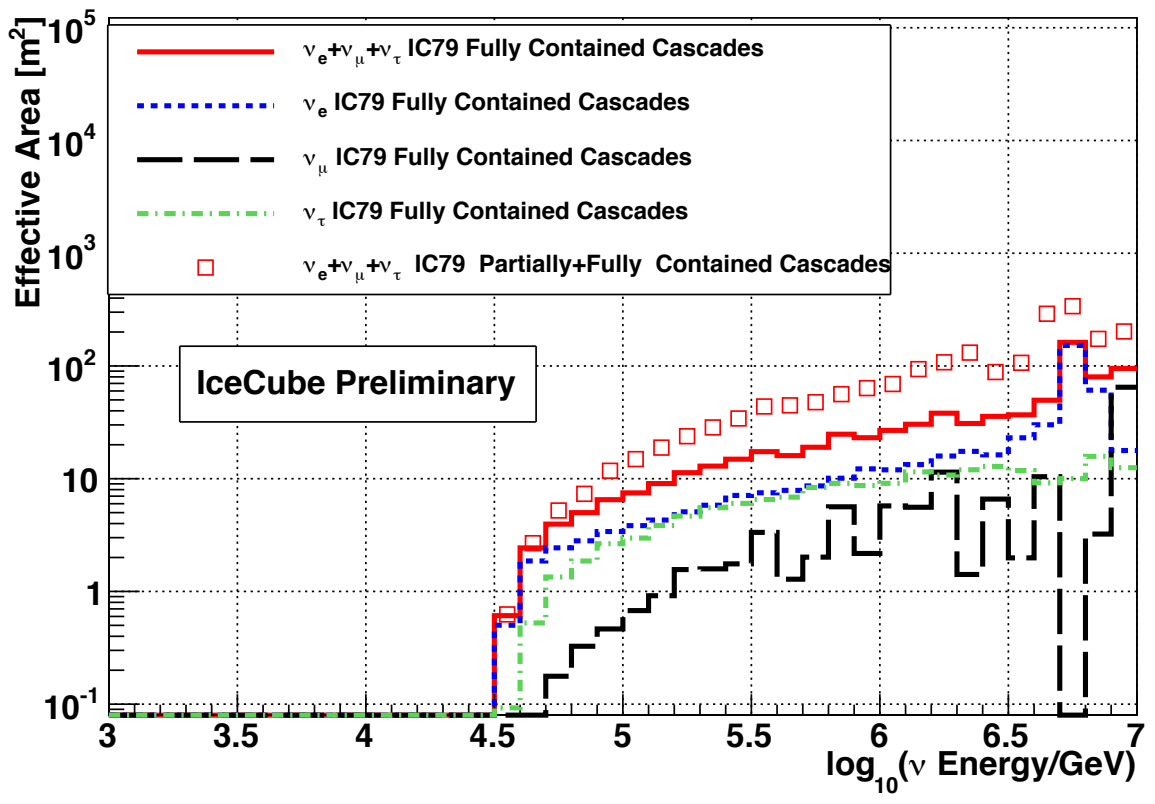

**Figure 5**: Effective area after the final event selection.

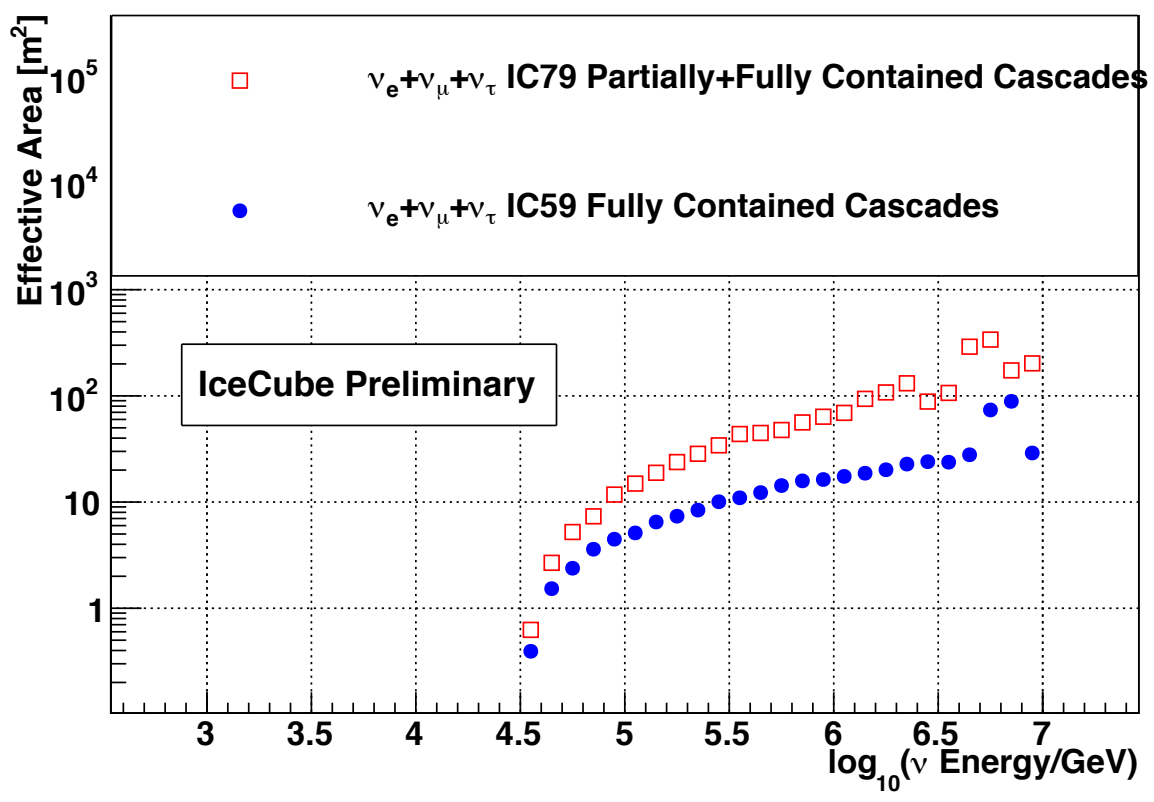

**Figure 6**: Comparison of effective area for sum of all flavor neutrinos (open squares) for the analysis presented here and the cascade neutrino search with IC59 string configuration [5] (filled circles).

## 4 Results

The selection criteria rejected all of the CORSIKA events and the conservative estimate on the number of cosmic-ray muons at the final level was taken as an upper boundry at 90% C.L. interval of 1.6 events. One burn sample data event of 70 TeV reconstructed energy was retained.

From the analysis presented here $4.1 \pm 0.2$ (stat) $\nu_e$, 0.83 $\pm$ 0.06 (stat) $\nu_\mu$ and 2.76 $\pm$ 0.06 (stat) $\nu_\tau$ signal events for an astrophysical flux defined in Eq. (1) are expected in 317 days (90% of the experimental data). Thereby, the predicted number of astrophysical $\nu_\mu$ events from CC interactions is 0.31 $\pm$ 0.04 (stat), while from NC is 0.52 $\pm$ 0.05 (stat).

The expected number of atmospheric neutrino background events from $\nu_e$ is $2.5 \pm 0.2$ (stat) +3.1-2.5 (syst) and from $\nu_\mu$ $1.8 \pm 0.2$ (stat) $\pm$ 0.6 (syst). The statistical uncertainties come from the Monte Carlo statistics. The uncertainties of the theoretical models in the predicted fluxes are dominating sources of systematic uncertainties for estimating atmospheric neutrino background. The uncertainty of 25% for conventional [10] and the factor of two for prompt flux [11] were assumed. These atmospheric background estimates include the neutrino events that would be accompanied by a muon bundle [17] and therefore removed by the analysis selection cuts. The estimated background could hence be lowered by a factor of $\sim$2.

Figure 5 shows the effective area versus neutrino energy after all cuts applied. The Glashow resonance [16] contribution is clearly visible for $\nu_e$. The effective areas for $\nu_e$ and $\nu_\tau$ are higher than for $\nu_\mu$ as this analysis was optimized for cascades and removed muon tracks.

The comparison of the all neutrino flavor effective area for combined fully and partially contained analyses with 79 IceCube strings and the cascade search with the 59-string IceCube configuration [5] is shown in Fig. 6. The effective area for the 79-string configuration is bigger than for a smaller detector, as expected.

The sensitivity for the diffuse all flavor flux of extraterrestrial neutrino signal, defined as the average flux upper limit at 90% C.L. in the absence of signal was calculated and resulted in $2.3 \times 10^{-8}$ GeV s$^{-1}$ sr$^{-1}$ cm$^{-2}$ for the all-flavor neutrino energies between 42 TeV and 6 PeV. No systematic uncertainties were taken into account. Including partially contained events increases the sensitivity to $1.8 \times 10^{-8}$ GeV s$^{-1}$ sr$^{-1}$ cm$^{-2}$ for all-flavor neutrino events with energies between 44 TeV and 7.7 PeV. The obtained result is more stringent than the expected upper limits from previous IceCube cascade analyses with smaller sized detector configurations [2, 4, 5]. The systematic uncertainties are currently being evaluated.

**Acknowledgments**:This work is supported in parts by the National Science Foundation under Grant No. 1205796.

# Measurement of the atmospheric $\nu_\mu$ spectrum with IceCube 59

THE ICECUBE COLLABORATION[1],

[1]*See special section in these proceedings*

*tim.ruhe@udo.edu*

**Abstract:** The energy spectrum of the atmospheric muon neutrino flux was measured with the IceCube detector in the 59-string configuration, using an unfolding procedure. This measurement extended IceCube's reach for atmospheric neutrinos up to 1 PeV in energy. This extension in energy was obtained by using a machine learning algorithm preceeded by a dedicated feature selection for event selection, and by applying the novel unfolding algorithm TRUEE.

**Corresponding authors:**
Tim Ruhe[1],
[1] *Lehrstuhl für Experimentelle Physik V, Technische Universität Dortmund*

## 1 Introduction

IceCube is a state of the art neutrino telescope located at the geograophic South Pole. Its 5160 Digital Optical Modules (DOMs) are mounted on 86 vertical cables called strings, thus forming a three dimensional array of photosensors [1].
Although primarily designed for the detection of high energy neutrinos from astrophysical sources, the detector can be utilised for various other studies, including measurements of the atmospheric neutrino spectrum. Despite the fact that the atmospheric neutrino spectrum was already measured by various experiments, including AMANDA [2] and IceCube in the 40-string configuration [3], the flux at high energies is still subject to rather large uncertainties [4].
The flux of atmospheric muon neutrinos is dominated by neutrinos originating from the decay of pions and kaons, produced by cosmic-ray interactions in the atmosphere, up to engergies of $E_\nu \approx 100$ TeV [2]. Due to their relatively long lifetime pions and kaons lose part of their energy in collisions prior to decaying. The atmospheric neutrino spectrum is therefore expected to follow a power law one power steeper (asymptotically $\frac{d\Phi}{dE} \propto E^{-3.7}$) compared to the spectrum of primary cosmic rays [3].
At energies exceeding 500 TeV neutrinos from the decay of charmed mesons are expected to contribute notably to the spectrum. Due to their short lifetime ($t_{\text{life}} \approx 10^{-12}$ s [5]) these mesons decay before interacting and follow the intial spectrum of cosmic rays more closely, therefore causing a flattening of the overall neutrino flux [2, 3].
As neutrinos cannot be detected directly, neutrino induced muons produced in charged current interactions are used for a measurement of the atmospheric neutrino flux.
Atmospheric muons, produced in cosmic-ray interactions as well, enter the detector from above, thus forming a significant background in the searches for atmospheric neutrinos. As the number of atmospheric muons exceeds the number of neutrino induced muons by several orders of magnitude, a detailed event selection needs to be carried out in order to obtain a high purity neutrino event sample. Within the analysis presented here, a machine learning based event selection was used. Details on this approach are given in the next section.
Although IceCube was finished in December 2010, data was already taken in previous detector configurations. Data for this analysis were taken between May 2009 and May 2010 in the 59-string configuration of the detector.

## 2 Event Selection

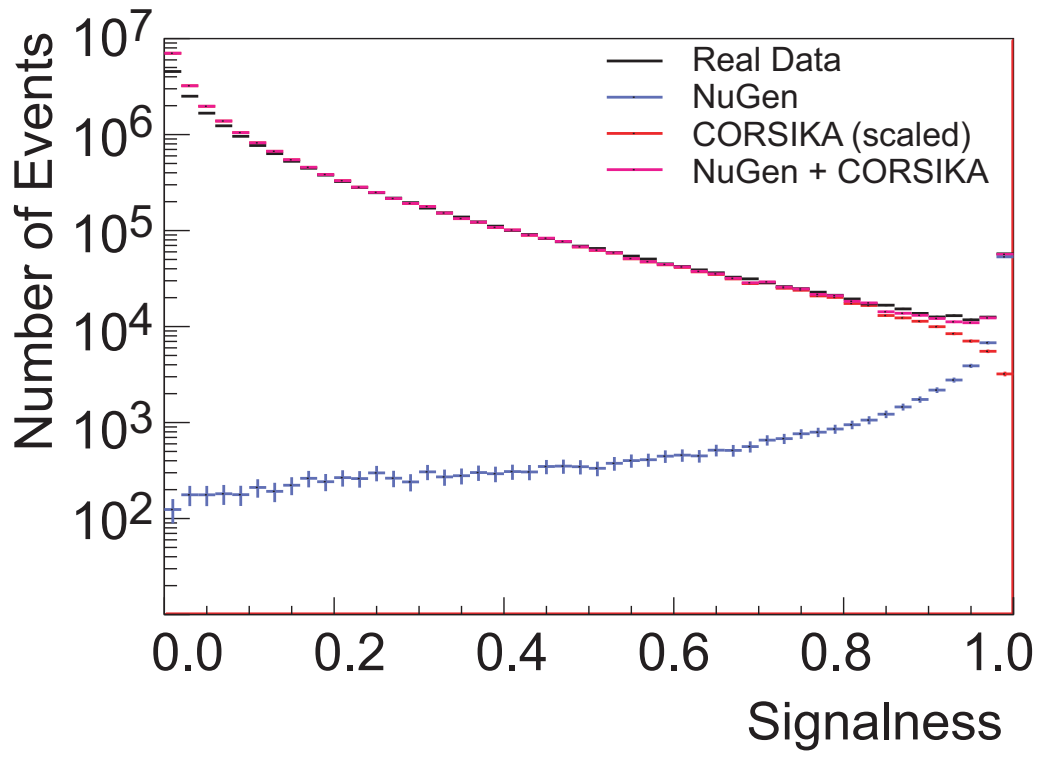

**Fig. 1**: Random Forest output score (signalness) for signal simulation (blue) generated using the IceCube neutrino generator NUGEN and background simulation (red) generated with CORSIKA [10]. Real data is shown in black, whereas the sum of simulated signal and background events is depicted in magenta. The sum of simulated signal and background events is found to agree well with the distribution of real data, indicating a stable performance of the Random Forest.

The event selection used in this analysis consisted of three basic steps, the first one being the application of quality cuts. These quality cuts were followed by a detailed algorithm based feature selection, aiming at the identification of reconstructed track parameters to be used for the training of a Random Forest [8]. The training and testing of the Random Forest was carried out as a third step.

The afore mentioned quality cuts were simultaneously applied to the LineFit velocity ($v_{LineFit} > 0.19$) and the reconstructed zenith angle ($\theta > 88°$). The LineFit velocity is the estimated velocity of the lepton, obtained by fitting a straight line to the spatial and time-distribution of detected light. Cascade like events, originating from charged current $\nu_e$ interactions and neutral current interactions of all neutrino flavors, will produce a spherical light pattern, from which small values of $v_{LineFit}$ are reconstructed. Larger values of $v_{LineFit}$ are obtained for track-like events from $\nu_\mu$ CC interactions. The selection of high quality track-like events is required in order to obtain rather long tracks, from which the energy of the incoming neutrino can be reliably reconstructed.

The zenith-angle cut is mainly aimed at reducing the contamination of atmospheric muons entering the detector at angles $\theta < 90°$. Choosing a cut at $\theta > 88°$ rather than at $\theta > 90°$ aims at slightly extending the field of view in order to detect higher energy neutrinos from above the horizon.

The quality of an automated, machine learning based, event selection largely depends on the utilised set of event parameters. In machine learning, these event parameters are often referred to as features or attributes. As not all attributes are equally well suited for the event selection, a representation in fewer dimensions needs to be found. In general, utilising knowledge about the detector and the classification problem at hand, will result in a good set of features that can be used for the training of a classification algorithm. It will, however, not necessarily result in the *best* set of features.

The Minimum Redundancy Maximum Relevance (MRMR) algorithm [6] was used for the selection of features. Using MRMR is particularly useful when certain quanities (e.g. zenith angle) are obtained from a number of different reconstruction algorithms.

Prior to running MRMR, a number of attributes, which were known to be either useless, redundant, or a source of potential bias, were excluded by hand. This mainly concerned timing information and sky coordinates. All event selection steps regarding machine learning and data mining were carried out using the RAPIDMINER [7] machine learning environment.

Twenty-five attributes were selected in the final event selection as this number represents a reasonable tradeoff between feature selection stability and the anticipated complexity of the learner. Three additional attributes were created and added to the attribute set, according to the findings presented in [3].

A Random Forest [8], which utilises an ensemble of simple decision trees, was chosen as a learning algorithm. Tree based algorithms are well known for their stability and interpretability and were found to perform well in previous IceCube analyses [3]. Moreover, Random Forests were found to outperform other classifiers in [9]. The forest was trained and tested in a five-fold cross validation utilising 70,000 simulated neutrino events and 750,000 simulated background events. The neutrino events were generated by the IceCube neutrino generator NUGEN, according to an $E^{-2}$ spectrum in order to provide a sufficient number of examples at high energies. Background events were simulated according to the poly-gonato model using CORSIKA [10]. In order to avoid overtraining, the number of examples used for the trainig of the forest was limited to 27,000 signal and background events, respectively. The ratio of signal to background events was set at 1:1 in the training process. Although the true distribution of signal and background events differs strongly from 1:1 on real data (93,000 neutrinos in $17.48 \times 10^6$ background events), tests showed that this ratio provided a reasonable tradeoff between signal efficiency and background rejection.

The Random Forest output score (signalness) for simulated signal events (blue) and simulated background events (red) is shown in Fig. 1. Real data is shown in black, whereas the sum of signal and background simulation is depicted in magenta. The sum of simulated signal and background events is found to agree well with the distribution of real data, indicating a stable performance of the Random Forest.

The application of the Random Forest on the full set of IC-59 data was found to yield 27,771 neutrinos in 346 days of IC-59 (80 neutrino events per day). Compared to an expectation of 29,884 neutrino events, derived from Monte Carlo simulations, a slight underfluctuation is observed. This underfluctuation, however, is found to be well within the estimated systematic uncertainties of the event selection. The purity of the final neutrino event sample was estimated to be 99.6%. It should be noted, that no events with a zenith angle $\theta < 90°$ were observed in the sample, after the application of the Random Forest.

## 3 Spectrum Unfolding

As the neutrino energy spectrum cannot be accessed directly, it needs to be inferred from the reconstructed energy of the muons. This task is generally referred to as an inverse- or ill posed problem and described by the Fredholm integral equation of first kind [11]:

$$g(y) = \int_b^a A(y,E) f(E) dE. \quad (1)$$

For the discrete case this is transformed into:

$$\vec{g}(y) = \underline{A}(E,y)\vec{f}(E), \quad (2)$$

where $\vec{f}(E)$ represents the sought energy distribution, whereas the measured energy dependent distribution is given as $\vec{g}(y)$. $\underline{A}(x,E)$ represents the response matrix of the detector, which also accounts for the physics of neutrino interactions in or near the detector as well as for the propagation of the muon.

Several approaches to the solution of inverse problems exist. The unfolding program TRUEE [11], which is an extension of the well known $\mathscr{RUN}$ [12] algorithm, was used for unfolding in this analysis.

Five unfolding settings (three different sets of input variables and two different settings for the regularisation parameter) were found to produce stable results, when tested on Monte Carlo simulation. Compatible results were obtained for all of these five settings, when applied to real data. The setting least sensitive to the ice model was chosen for the final unfolding (see section 4), in order to keep the systematic uncertainties as small as possible.

Three variables (track length, number of channels, number of direct photons) were used as input for the unfolding as TRUEE allows for the use of up to three input parameters.

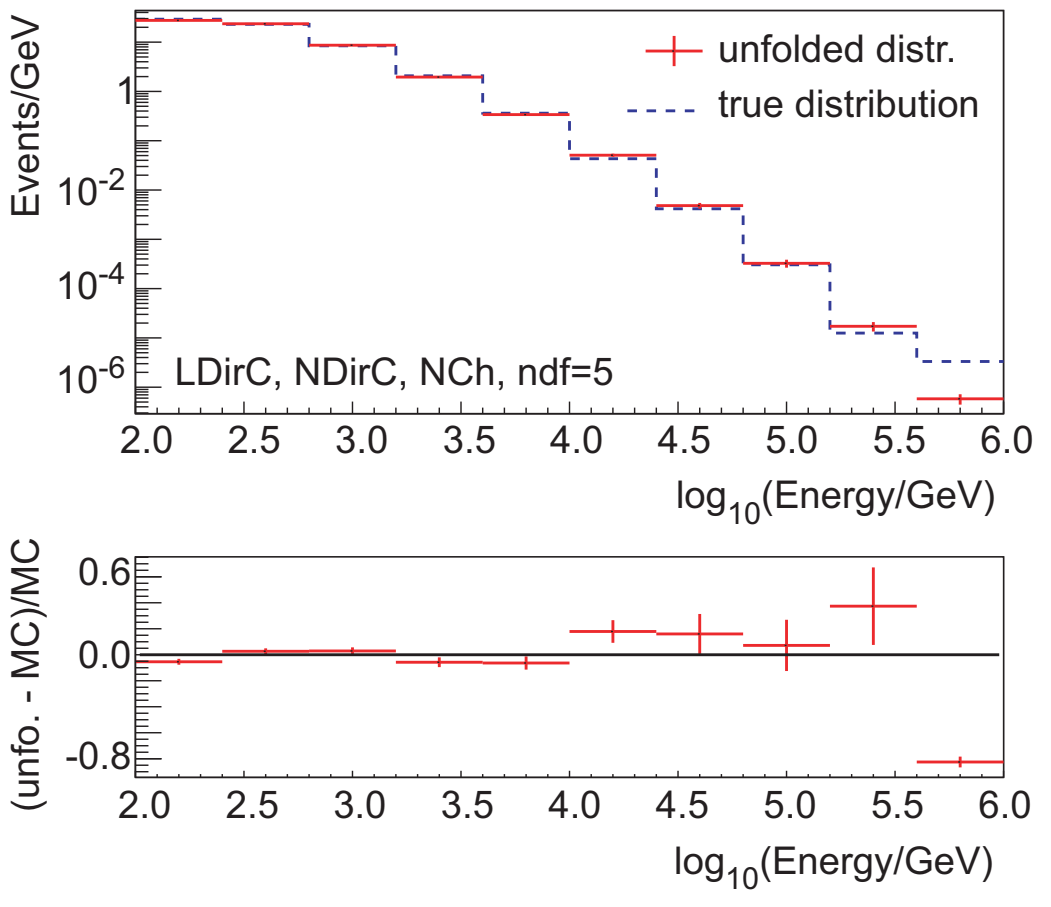

**Fig. 2**: Selected test mode result. The true distribution is represented by the blue dashed line, whereas the unfolding result is depicted by the red line. A good agreement between both distributions is observed.

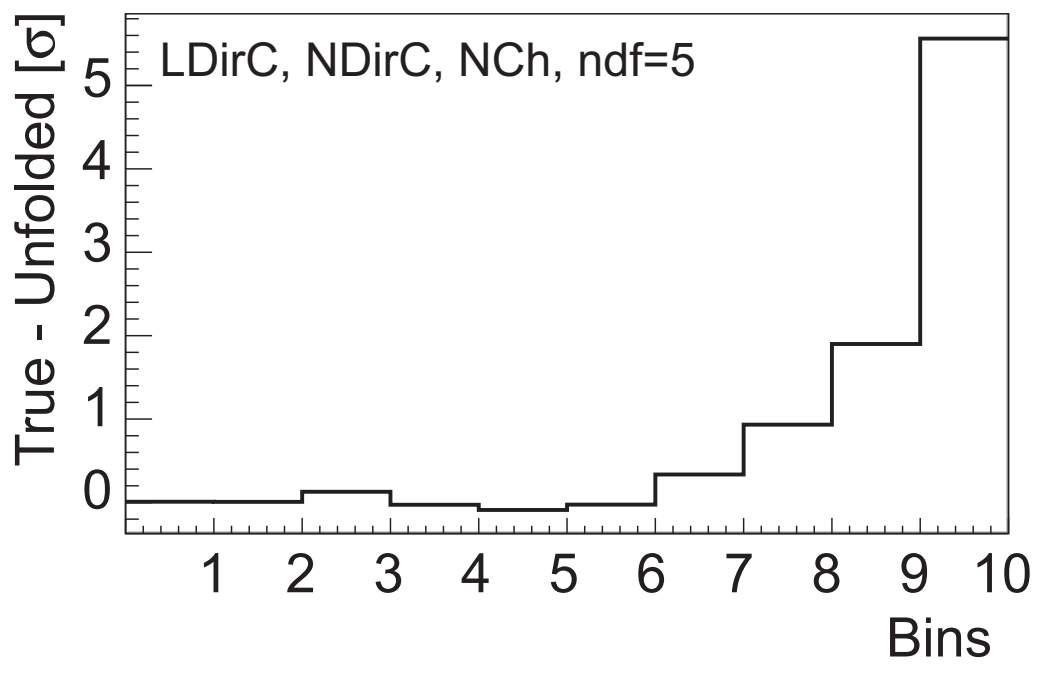

**Fig. 3**: Average deviation of the unfolding result from the true distribution in units of the statistical uncertainty $\sigma$. Only small deviations are observed for the first eight bins. The discrepancies are found to increase towards higher energy, due to the steeply falling spectrum of atmospheric neutrinos.

In IceCube, photons are considered direct when they are detected within a certain time window, computed with respect to the reconstructed track. An estimate of the track length inside the detector is obtained by projecting all DOMs that recorded direct photons onto the reconstructed track. The number of channels corresponds to the number of DOMs hit during an event.

Good data to Monte Carlo agreement, as well as good correlation with energy were observed for all variables. The stability of the unfolding as well as the results obtained on real data are addressed in the following.

A selected test mode result comparing the unfolding result to the true distribution of events is shown in Fig, 2. The true distribution is represented by the blue dashed line, whereas the unfolding result is shown in red. In general both distributions were found to agree well. Discrepancies were observed for the last bin. Whether this poses a potential problem to the stability of the result cannot be determined from the outcome of a single unfolding.

The stability of the unfolding was validated in a bootstrapping procedure implemented in the pull mode of TRUEE. Within this pull validation 500 test unfoldings were carried out, each treating 30,000 events as pseudodata. For all of these unfoldings, the deviation between the unfolding result and the true distribution is computed binwise and in units of the statistical uncertainty $\sigma$ [11]. The average deviation of the individual bins is shown in Fig. 3. Only small deviations (well below the $1\sigma$ limit) were observed for the first seven bins, indicating a stable behaviour of the unfolding.

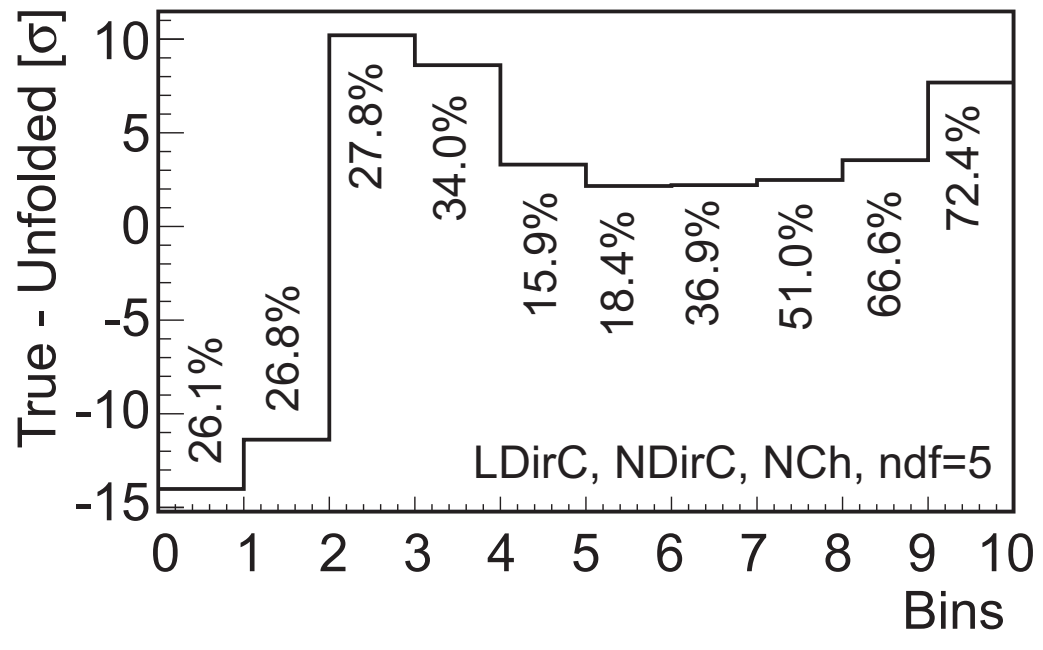

**Fig. 4**: Estimated systematics due to uncertainties in the ice model obtained by applying the pull mode on a different set of Monte Carlo simulations. The deviation between the unfolding result and the true distribution in units of the statistical uncertainty is depicted on the y-axis. Systematic uncertainties of the order of 30% are observed for the first couple of bins. Larger errors are observed towards higher energies.

The rather large deviation obtained for the highest energy bins is a result of the steeply falling spectrum of atmospheric neutrinos and the bootstrapping procedure applied in the pull mode. Due to the small number of events in the last bin, either 0 or 1 events are drawn randomly from the true distribution. Two or more events are only drawn in rather rare cases. Based on the response matrix, which accounts for the limited statistics in the highest energy bins by using ten times more events compared to real data, only a fraction of an event is reconstructed for the highest energy bin. As the statistical uncertainties derived in TRUEE fail to cover the distance of the predicted bin content to the true bin content, large deviations are observed. This further implies that an overestimation is obtained in case no events are present in the last bin on real data. An underestimation is observed in case one event is present in this bin in real data. As there is no way to determine the number of events in the last bin on real data prior to unblinding, the statistical uncertainty for the last bin should cover both of the cases discussed above. The pull mode in TRUEE, therefore, simultaneously serves as a cross check for the size of the statistical uncertainties derived as part of the algorithm.

Thus, taking into account the pull mode results for the last and next to last bin, one finds that the uncertainties derived in TRUEE are estimated too small for these two bins, as possible statistical fluctuations in real data are not covered. The statistical uncertainties of these bins were, therefore, scaled up by 1.9 and 5.56, respectively.

The normalisation of the atmospheric flux, as well as the spectral index were found to be retained during the unfolding.

## 4 Estimation of Systematic Uncertainties

Since the pull mode in TRUEE can be used on two different sets of Monte Carlo simulation, it offers the possibility to study systematic effects in a statistically reliable manner. Within this study the Monte Carlo set used for the determination of the response matrix is kept constant with respect to the unfolding of real data. Monte Carlo sets differing in certain systematic parameters were then treated as pseudodata. Similar to the pull mode discussed above, these pseudodata were unfolded 500 times and the deviation between the unfolding result and the true distribution was calculated in units of the statistical uncertainty. The obtained deviation can be easily converted into a relative uncertainty, as the statistical uncertainty returned by the unfolding algorithm was found to vary by less than 2% level between different unfoldings.

One of the main sources of systematic uncertainties is the modelling of the ice used in the Monte Carlo production. The outcome of using the pull mode on simulation generated with different ice models is shown in Fig. 4. Error bars on the order of 30% or below are observed for the first seven bins. The uncertainties were found to increase in the highest energy bins. One should note that large uncertainties in units of the statistical uncertainty correspond to rather small relative errors in the first bins. This behaviour is due to the large statistics obtained in the first couple of bins, which in turn leads to rather small statistical errors.

An increase and decrease in the pair production cross section, respectively, was used to investigate the effect of uncertainties on the amount of light detected in IceCube. As the observation of more or less light, respectively, can in principle be caused by various effects that cannot be disentangled on real data, a double counting of the same systematic uncertainty needs to be avoided.

Cross checks on the size of the systematic uncertainty were performed by dividing the detector into two distinct subdetectors according to the z-coordinate of the center of gravity of the charge distribution of the event (COGZ). COGZ is calculated with respect to the center of the detector. In these checks the detector is split up into an inner and an outer layer, which aims at maximizing the difference in the ice for both detectors. The inner layer, which contains a large layer of dust, contains all events for which COGZ $> -225$ m and COGZ $\leq 225$ m. The outer layer of the detector contains all events for which COGZ $\geq 275$ m or COGZ $\leq -275$ m. Buffer zones of 50 m were introduced between the subdetectors in order to avoid a random counting of events into one of the subdetectors due to small uncertainties in the ice.

This cross check yielded very positive results, as the observed spectrum obtained using the full IceCube detector was found to agree with the two subdetector spectra within the estimated systematic uncertainties. This result was confirmed by an additional cross check, which divided the detector into an upper- and a lower layer. It can therefore be concluded that the systematic uncertainties have been correctly and reliably estimated.

## 5 Final Result

Figure 5 shows the zenith-averaged and acceptance corrected flux of atmospheric neutrinos obtained with IceCube in the 59-string configuration. Two theoretical model calculations are shown for comparison. The model using Honda et al. [13] (conventional) and Enberg et al. [14] (prompt) is depicted in red, wheras the conventional model by Barr et al. [15] is shown in black.

Good agreement between the measured flux of atmospheric neutrinos and the model calculations is observed. The systematic uncertainties have been reduced, compared to previous measurements of the atmospheric neutrino flux with IceCube, especially in the intermediate energy region. Furthermore, a measurement of the atmospheric neutrino flux up to an energy of 1 PeV was obtained. Thus, the energy range accessible using the IceCube detector has been extended from 400 TeV to 1 PeV compared to a measurement obtained using IceCube in the 40-string configuration [3].

No statement on a possible contribution of neutrinos from the decay of charmed mesons can be made, due to the limited statistics and large systematic uncertainties in the high energy region.

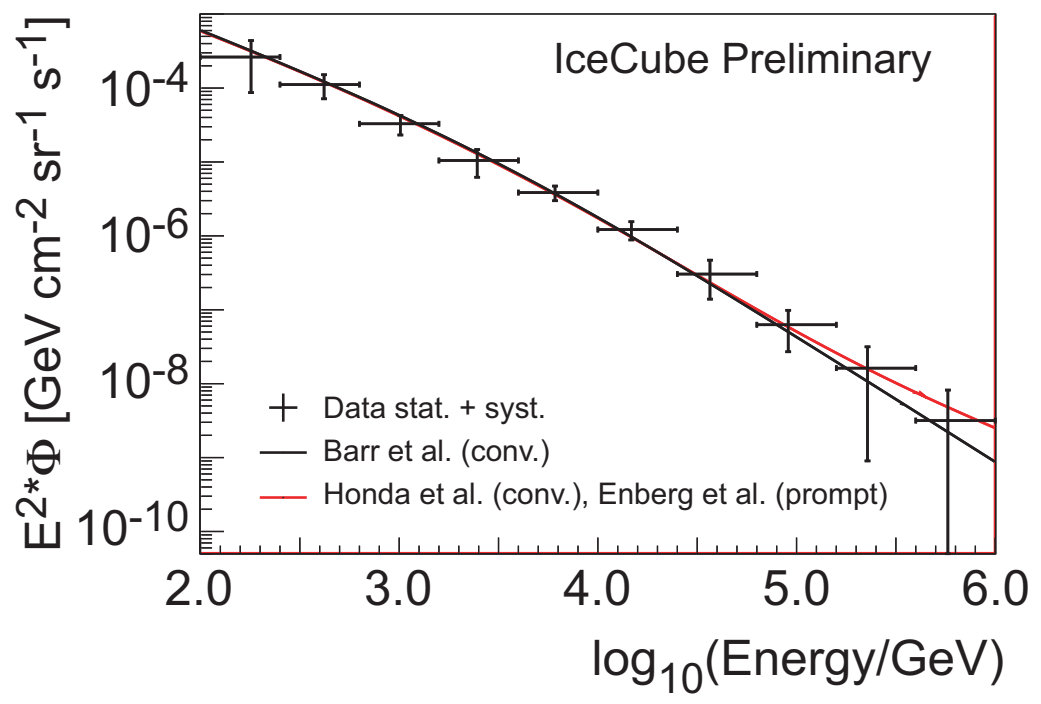

**Fig. 5**: Acceptance corrected and zenith-averaged atmospheric neutrino spectrum obtained with IceCube, compared to theoretical predictions. The model using Honda al. [13] (conv.) and Enberg et al. [14] (prompt) is depicted in red. The conventional model by Barr et al. [15] is shown in black. Good agreement between the unfolded flux and the theoretical models is observed. No statements on a contribution of neutrinos from the decay of charmed mesons can be made due to the rather large systematic uncertainties in the highest energy bins.

# Seasonal variation of atmospheric neutrinos in IceCube

THE ICECUBE COLLABORATION[1]

[1]*See special section in these proceedings*

*desiati@wipac.wisc.edu*

**Abstract:** The IceCube Observatory is a cubic-kilometer neutrino telescope at the South Pole, which currently collects about 170 well-reconstructed neutrinos per day with energies above 100 GeV. These neutrinos are generated by cosmic ray interactions in the atmosphere, and their rate is expected to correlate with the atmospheric density, which depends on the temperature. A large portion of upward moving neutrinos reconstructed within 30° of the horizon is produced above the Antarctic continent. This component of the upward neutrino flux is therefore expected to correlate with the stratospheric temperature in a similar way as downward-going muons produced in the atmosphere above IceCube. We report the first observation of an annual modulation of the atmospheric neutrino flux in correlation with the upper atmospheric temperature. Its amplitude of about ±5% is inconsistent with a constant rate at a confidence level of 3.4 sigma.

**Corresponding authors:** P. Desiati[1], K. Jagielski[2], A. Schukraft[2], G.C. Hill[3], T. Kuwabara[4], T. Gaisser[4]
[1]*Wisconsin IceCube Particle Astrophysics Center, University of Wisconsin, Madison, WI 53706, U.S.A.*
[2]*III. Physikalisches Institut, RWTH Aachen University, 52056 Aachen, Germany*
[3]*School of Chemistry and Physics, University of Adelaide, Australia*
[4]*Bartol Research Institute and Dept. of Physics and Astronomy, University of Delaware, Newark, DE 19716, U.S.A.*

## 1 Introduction

Understanding the lepton flux produced by the interaction of cosmic rays with the Earth's atmosphere is important for neutrino observatories. These events are the main sources of background in the search for astrophysical neutrinos, and they are also important for calibrating the detector. In addition, the fluxes of atmospheric muons and neutrinos provide an indirect probe of the particle physics of hadronic showers in the atmosphere.

The correlation of the lepton fluxes with temperature in the upper atmosphere is an interesting detail to study, in part because the general features of the normalization (such as the primary spectrum) cancel to a large extent. Pions and kaons produced by interactions of cosmic rays either interact again or decay into muons and neutrinos. The competition between the two processes depends on the local density of the atmosphere in the production region, which changes with temperature [1]. The correlation of the intense muon flux with the upper atmospheric temperature has been extensively studied by various experiments at different energy thresholds [2, 3, 4, 5]. The IceCube Observatory provides observations with unprecedented statistics that show correlations on short time scales with variations in the stratosphere over Antarctica [6] as well as the yearly modulation of the muon flux [7]. Study of seasonal variations can provide a constraint on the kaon to pion production ratio in the extensive air showers [8, 7], and is also a tool to probe charm production [9].

With the kilometer-scale IceCube Observatory it is possible for the first time to obtain enough events to observe the correlation of the neutrino flux with the upper atmospheric temperature. A similar study done with the smaller AMANDA detector [10] lacked the statistics to demonstrate the expected correlation. In this paper we present a correlation study of the atmospheric neutrino flux with the stratospheric temperature, along with the theoretical expectations.

## 2 Neutrino and Temperature Data

IceCube consists of 5160 optical sensors viewing a cubic kilometer of ice at a depth of 1450 to 2450 meters in the Antarctic glacier (see [11] for an overview of IceCube). This study uses about 90,000 neutrino-induced, upward muon events collected by IceCube in 1040 days of operation, from April 2008 to May 2011. During this period IceCube was under construction, and the instrumented volume increased from 40 deployed strings in 2008 (IC40) to 59 strings in 2009 (IC59) and 79 in 2010 (IC79). The neutrino event samples summarized in Table 1 were selected by independent data analyses to determine their energy spectrum [12] or to search for neutrinos of extra-terrestrial origin [13, 14]. In these data samples, muon neutrino induced events were separated from the large background of atmospheric muons from above by selecting well reconstructed upward-going tracks. Most of the neutrino events are of atmospheric origin. Their flux is therefore correlated with the temperature variations in the atmosphere where they were produced. The contamination of mis-reconstructed atmospheric muon events is less than 1%.

| configuration | time period | events | livetime |
|---|---|---|---|
| IC40 (40 strings) | 4/2008 - 5/2009 | 12877 | 375.5 d |
| IC59 (59 strings) | 5/2009 - 5/2010 | 21943 | 348.1 d |
| IC79 (79 strings) | 6/2010 - 5/2011 | 54999 | 315.5 d |

**Table 1**: Neutrino event samples selected from the three detector configurations considered in this analysis, with data collection time period, number of selected events and the corresponding livetime (in days) [12, 13, 14].

For the purpose of studying correlation with temperature in the upper atmosphere where the neutrinos are produced, it is convenient to divide the hemisphere below the detector into three zones. Zone 1 contains events with zenith angles in the range $90^\circ < \theta < 120^\circ$, which covers a solid angle of $\pi$ sr corresponding to latitudes between $-30^\circ$ and $-90^\circ$. Zone 2 ($120^\circ < \theta < 150^\circ$), with a solid angle of $0.73\pi$ corresponds to the equatorial region with latitudes in the range $\pm 30^\circ$. Zone 3 ($150^\circ < \theta < 90^\circ$) covers the Northern temperate latitudes and the Arctic region. The total solid angle of Zone 3 as seen from the South Pole is small ($0.27\pi$ sr), and the Arctic region is only 15% of this. The seasonal temperature variation in the equatorial zone is small, so it is not suited for measuring correlation with variations in temperature. Zone 1 has half the total solid angle and contains more than half the neutrino-induced muons because the high-energy atmospheric neutrino flux is largest near the horizon. The temperature variation in Zone 1 has the same phase as that at the South Pole, and about 35% of this region is over the Antarctic continent. In this paper only neutrino events in Zone 1 are considered.

## 3 Temperature Correlation

The relevant temperatures and densities are those where the neutrinos are produced. It is therefore necessary to convolve the temperature profile with the muon production spectrum along each direction considered. A simple analytic approximation for neutrino production is used to obtain a single effective temperature for each direction at each time.

In this approach [9], the differential flux of $\nu_\mu + \bar{\nu}_\mu$ is approximated as

$$\phi_\nu(E_\nu,\theta) = \phi_N(E_\nu) \times \left\{ \frac{A_{\pi\nu}}{1 + B_{\pi\nu}\cos\theta^\star E_\nu/\varepsilon_\pi} + \frac{A_{K\nu}}{1 + B_{K\nu}\cos\theta^\star E_\nu/\varepsilon_K} \right\}, \quad (1)$$

where $\phi_N(E_\nu)$ is the primary spectrum of nucleons ($N$) evaluated at the energy of the neutrino. The first term in Eq. 1 corresponds to neutrino production from leptonic and semi-leptonic decays of pions, while the second term is related to kaons. The constants $A_{\pi\nu}$ and $A_{K\nu}$ depend on the branching ratio for meson decay into neutrinos, the spectrum weighted moments of the cross section for a nucleon to produce secondary mesons, and those of the meson decay distribution. The denominators in Eq. 1 reflect the competition between decay and interaction of secondary mesons in the atmosphere. At energies below $\varepsilon_{\pi,K}/\cos\theta^\star$ (with the neutrino zenith angle evaluated at its point of production) meson decay is the dominant process, and neutrinos are produced with the same spectral index as the parent cosmic rays. At high energies meson interaction dominates and the corresponding neutrino spectrum becomes asymptotically one power steeper than the primary spectrum.

The characteristic critical energies $\varepsilon_{\pi,K}$ at a given atmospheric depth are inversely proportional to the atmospheric density at that point, and therefore are affected by temperature variations. In an isothermal approximation of the atmosphere, the density profile is described by an exponential with a scale height of $h_0 \sim 6.19$ km (over Antarctica). This numerical value corresponds to the lower stratosphere, where most of the neutrinos are generated. In the ideal gas law approximation, $\varepsilon_{\pi,K}$ are proportional to the atmospheric temperature in the isothermal approximation. At a mean atmospheric temperature of $T_0 = 224$ K (average over Zone 1) the critical energies are $\varepsilon_\pi = 117$ GeV and $\varepsilon_K = 871$ GeV. At energies far above $\varepsilon_{\pi,K}$, the terms in Eq. 1 reach the asymptotic regime where the flux is proportional to the mesons critical energy and, therefore, to the atmospheric temperature. This dependency is the source of the seasonal modulation of the neutrino flux.

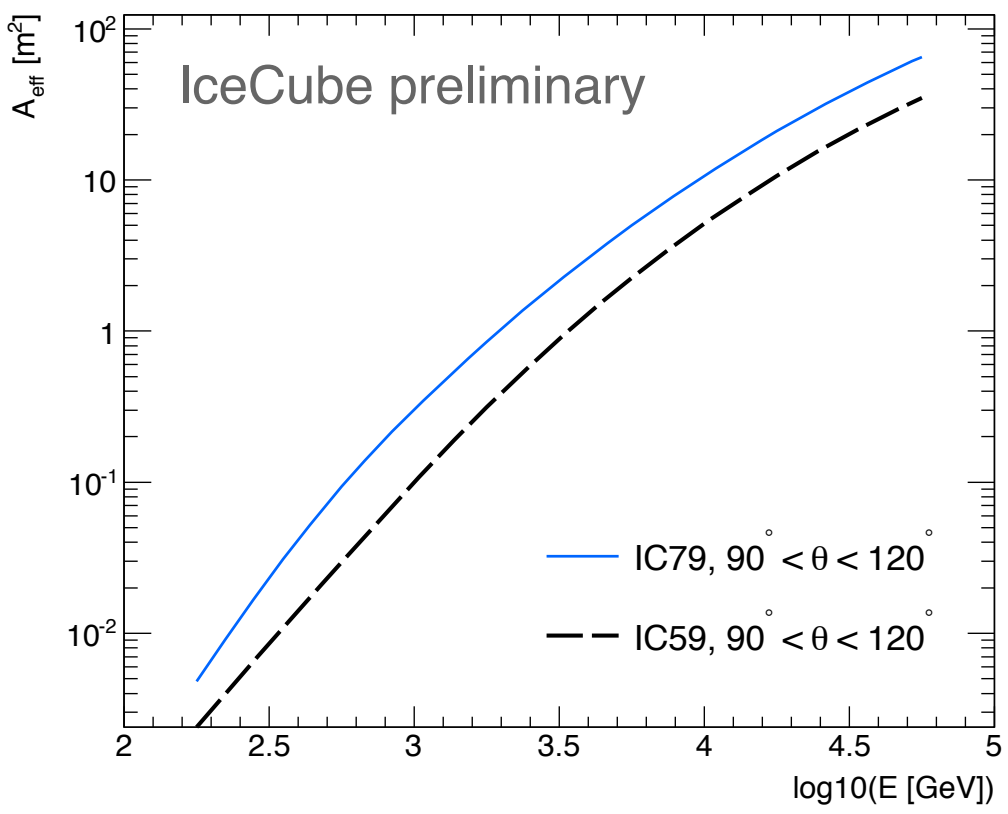

**Fig. 1**: Neutrino effective area, averaged over $\nu_\mu$ and $\bar{\nu}_\mu$ as a function of neutrino energy (at the interaction point) for the sample selected with the 79-string detector configuration [14] (blue continuous line) and for the sample selected with the 59-string detector configuration [13] (black dashed line). The effective areas are averaged over the zenith angle ranges $90^\circ < \theta < 120^\circ$.

The effective temperature is the convolution of the actual atmospheric temperature profile over the atmospheric slant depth $X$ (in g/cm$^2$) with the neutrino production spectrum profile $P_\nu(E_\nu,\theta,X)$, where critical energies are evaluated at the actual temperature at atmospheric depth $X$ [7, 9]. Taking into account the detector response, the effective temperature is given by

$$T_{\mathrm{eff}}(\theta) = \frac{\int dE_\nu \int dX\, P_\nu(E_\nu,\theta,X) A_{\mathrm{eff}}(E_\nu,\theta)\, T(X)}{\int dE_\nu \int dX\, P_\nu(E_\nu,\theta,X) A_{\mathrm{eff}}(E_\nu,\theta)}, \quad (2)$$

where $A_{\mathrm{eff}}(E_\nu,\theta)$ is the neutrino effective area obtained from simulation which contains the detector acceptance, event selection, and the neutrino interaction cross section. The denominator in Eq. 2 is the total measured neutrino intensity. The total effective temperature $T_{\mathrm{eff}}$ is the weighted average of Eq. 2 over the actual zenith distribution of the neutrino-induced events.

Figure 1 shows the neutrino effective area for the IC59 and IC79 detector configurations for Zone 1. The selected neutrino events have a mean energy of the order of 1 TeV, therefore the high energy behavior of the effective area is not relevant for this analysis. Figure 1 shows that the events in the IC79 sample have a lower energy threshold than those in the IC59 sample. Because $\alpha_T^{th}$ increases with increasing

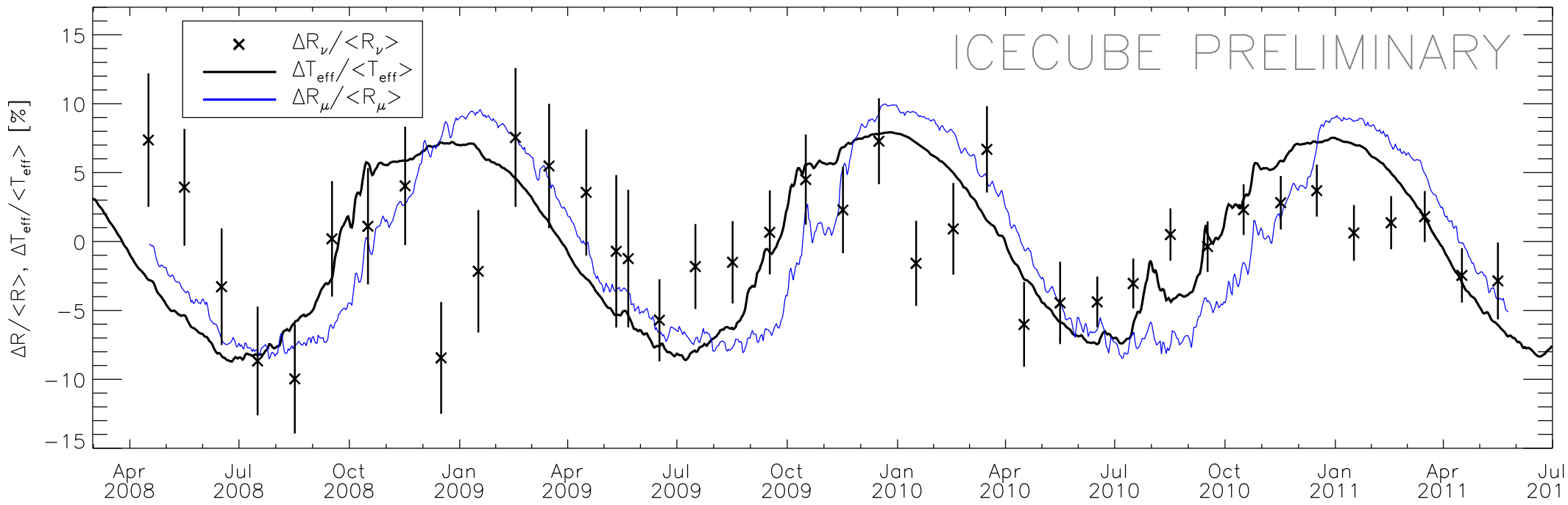

**Fig. 2**: The relative modulation of the effective temperature calculated for neutrinos collected within the zenith angle range $90^\circ < \theta < 120^\circ$ between April 2008 and July 2011 (black line), compared with the corresponding relative variation in the monthly neutrino rate (points with statistical errors). The blue line shows the downward muon event rate collected in the same time period. The statistical errors in the muon rates are small and not visible. The modulation of the nearly horizontal upward neutrinos is somewhat ahead of that for muons (see text).

energy threshold, the correlation analysis was performed separately for each event sample.

The atmospheric temperature profile data used in this analysis were collected by the NASA Atmospheric Infrared Sounder (AIRS) on board the Aqua satellite. Daily atmospheric temperatures at 24 different pressure levels from 1 to 1000 hPa at geographic locations around the globe were obtained from the AIRS Level 3 Daily Gridded Product available on NASA Goddard Earth Sciences, Data and Information Services Center (GES DISC) [15]. Using these data the daily effective temperature $T_{\mathrm{eff}}$ was calculated based on the zenith-weighted average of Eq. 2.

As with the muon case [7, 9], the relation between the variation of temperature and the variation of neutrino intensity at a given energy and zenith angle can be expressed in terms of a theoretical correlation coefficient calculated from Eq. 1 as [7]

$$\alpha_\nu(E_\nu,\theta) = \frac{T}{\phi_\nu(E_\nu,\theta)} \frac{\partial \phi_\nu(E_\nu,\theta)}{\partial T}, \tag{3}$$

which depends explicitly on the characteristic critical energies $\varepsilon_{\pi,K}$. With increasing energy, the temperature correlation coefficient increases until it reaches a constant value at sufficiently high energy.

To compare the prediction with measurements, it is necessary to convolve the neutrino differential spectrum with the detector response. The corresponding weighted correlation coefficient is

$$\alpha_T^{th}(\theta) = \frac{T \cdot \frac{\partial}{\partial T} \int dE_\nu \, \phi_\nu(E_\nu,\theta) A_{\mathrm{eff}}(E_\nu,\theta)}{\int dE_\nu \, \phi_\nu(E_\nu,\theta) A_{\mathrm{eff}}(E_\nu,\theta)}. \tag{4}$$

This equation defines the correlation coefficient for a particular zenith angle $\theta$. The total correlation coefficient is then obtained by averaging $\alpha_T^{th}(\theta)$ over $\theta$ with a weight given by the observed event angular distribution. With this definition the relative variation in neutrino intensity $I_\nu$ is given by

$$\frac{\Delta I_\nu}{I_\nu} = \alpha_T^{th} \frac{\Delta T_{\mathrm{eff}}}{T_{\mathrm{eff}}}. \tag{5}$$

Since the rate $R_\nu$ of observed neutrinos is proportional to the incident neutrino intensity $I_\nu$, it is correlated with the effective temperature as well

$$\frac{\Delta R_\nu}{\langle R_\nu \rangle} = \alpha_T^{exp} \frac{\Delta T_{\mathrm{eff}}}{\langle T_{\mathrm{eff}} \rangle}, \tag{6}$$

where $\alpha_T^{exp}$ is the experimentally determined correlation coefficient.

## 4 Results

Figure 2 shows the monthly rates of neutrino events with $90^\circ < \theta < 120^\circ$ relative to the mean annual rate, along with the corresponding effective temperatures relative to the mean. The monthly rate is calculated as the number of events divided by livetime in the corresponding month. The effective temperature is calculated with Eq. 2 using the neutrino effective area corresponding to each detector configuration. A yearly modulation of the neutrino rate is clearly observed, and a $\chi^2$ analysis with the three years of IceCube data rejects a constant rate of neutrinos at the $3.4\sigma$ level. The apparently reduced rates during the months of January and February (when Antarctic summer operations occurred during construction) are under investigation. Figure 2 shows that the modulation in neutrino rate is correlated with the variation of the effective temperature. To quantify this correlation, a linear fit is performed, as shown in Fig. 3. The results are shown in Table 2. The decrease of the uncertainty in the correlation coefficient from IC40 to IC79 reflects the larger event samples collected with the bigger instrumented volume. As mentioned, the temperature correlation coefficient was not determined by stacking the three data samples because of their different energy thresholds.

| configuration | $\alpha_T^{exp}$ | $\chi^2$/ndf | $\alpha_T^{th}$ |
|---|---|---|---|
| IC40 | 0.27±0.21 | 22.85/12 | $0.557^{+0.008}_{-0.007}$ |
| IC59 | 0.50±0.15 | 12.30/11 | $0.518^{+0.008}_{-0.007}$ |
| IC79 | 0.45±0.11 | 4.48/10 | $0.489^{+0.007}_{-0.005}$ |

**Table 2**: Experimental and theoretical neutrino temperature correlation coefficients corresponding to the three detector configurations and the $\chi^2$/ndf of for the experimental coefficient. Errors on $\alpha_T^{exp}$ are statistical and those on $\alpha_T^{th}$ are from the seasonal change of critical energies $\varepsilon_{\pi,K}$.

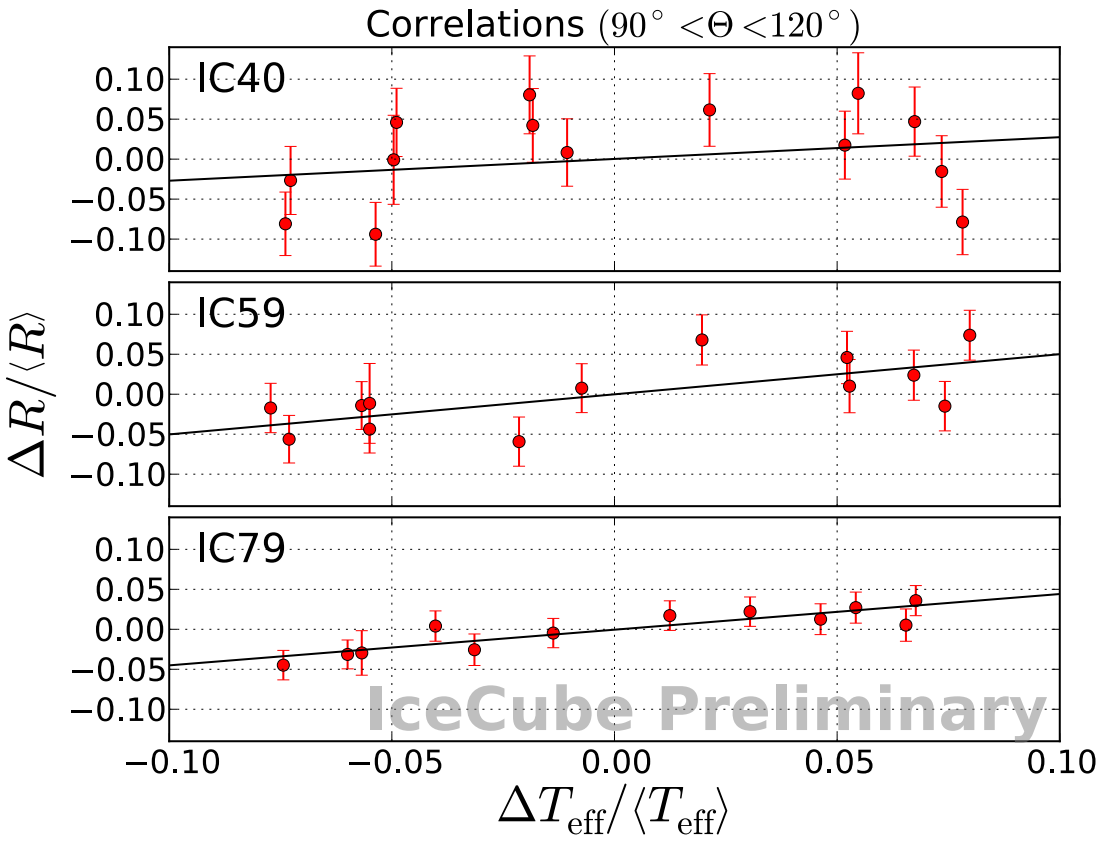

**Fig. 3**: Correlation between the the measured monthly relative rate variation and that of the corresponding effective temperature for the three detector configurations datasets for $90° < \theta < 120°$. The linear fits are also shown and the results shown in Table 2.

Systematic uncertainties of the analysis arise from understanding of the detector (light yield, ice properties, efficiency of the optical sensors) and from uncertainties in the theoretical parameters such as the spectral index and the $K/\pi$ ratio. The latter is particularly important because the charged kaon channel is the main source of muon neutrinos above 100 GeV. For both sources of experimental error, however, the effects are relatively small because the same uncertainties occur in the numerator and the denominator of key quantities such as $T_{\text{eff}}$ (Eq. 2) and the correlation coefficient, Eq. 4.

One of the main experimental uncertainties is the optical sensitivity of the detector, which includes e.g. the photon light yield of propagating particles in the ice, efficiency of IceCube's optical sensors and the global transparency of the ice. A 20% uncertainty on this parameter results in a variation of $\alpha_T^{exp}$ of less than $\pm 1\%$ and a variation of $\alpha_T^{th}$ of $\pm 4\%$.

Events in Zone 1 ($90° < \theta < 120°$) are produced in the southern atmosphere so their correlation with temperature is expected to be similar to that of the downward cosmic ray induced muons [7] produced locally in the atmosphere above the detector. The apparent difference in phase between the muons and neutrinos in Fig. 2 is likely due to the fact that at more horizontal directions the pions and kaons decay higher in the atmosphere, making the peak of neutrino production spectrum shift to smaller atmospheric depths. Since the atmospheric temperature increases sooner in the upper atmosphere than in the lower layers when the Sun rises in the austral spring, the modulation of the horizontal neutrino flux would be expected to precede that of the more vertical downward muons. At the same time, temperature modulations are larger at higher altitudes. This makes the variation in $T_{\text{eff}}$ for neutrinos comparable to that for muons even though neutrinos are produced in more temperate latitudes within Zone 1.

The differential flux of muons is represented by an expression similar to Eq. 1 at sufficiently high energy (>100 GeV). However, the fractional contributions of the main hadronic channels to the production of leptons in the atmosphere are different for muons and neutrinos. The kinematics of $\pi^{\pm} \to \mu^{\pm} + \nu$ decay in flight favors the transfer of most of the pion energy to the muon, since its mass is comparable to that of the pion. On the contrary, in the corresponding kaon decay the energy is equally distributed between the muons and the neutrinos. This means that kaons become the dominant source of neutrinos above ~100 GeV. On the other hand, muons are always dominated by pion decay, although the relative kaon contribution increases with energy. As mentioned earlier, due to the higher critical energy of kaons than pions, the kaon term in Eq. 1 reaches the asymptotic regime at higher energy than the pion term, making it relatively less sensitive to temperature variations [9]. Therefore the kaon-dominated neutrinos have a smaller correlation with temperature than muons [7].

## 5 Conclusions

Using neutrino-induced muon events reconstructed and selected with three years of IceCube data from April 2008 and May 2011, a seasonal variation in the neutrino event rate is observed for the first time. The neutrino rate for events in the horizontal region of $90° < \theta < 120°$ is observed to correlate with the effective temperature in the Earth's atmosphere in a manner that is consistent with studies of downward atmospheric muons performed with a much larger statistical data sample. Because of the importance of the kaon channel for production of muon neutrinos, further studies can contribute to understanding the kaon/pion ratio in the atmospheric cascade.

## 6 Acknowledgements

The temperature data used in this study were acquired as part of the NASA's Earth-Sun System Division and archived and distributed by the Goddard Earth Sciences (GES) Data and Information Services Center (DISC) Distributed Active Archive Center (DAAC).

# Cascade Reconstruction at the Glashow Resonance in IceCube

THE ICECUBE COLLABORATION[1]

[1] *See special section in these proceedings*

*joanna.kiryluk@stonybrook.edu*

**Abstract:** IceCube is a cubic kilometer neutrino detector located at the South Pole. Observation of the spectrum near the characteristic energy $E_\nu \simeq 6.3\,\mathrm{PeV}$ of the Glashow resonance, the interaction of anti-neutrinos with atomic electrons via $\bar{\nu}_e + e^- \to W^-$, is of particular interest, since it offers the unique possibility to determine the contribution from electron anti-neutrinos to the diffuse flux of astrophysical neutrinos. The flux of electron anti-neutrinos, if observed, provides new constraints on the possible production mechanisms for high-energy neutrinos in astrophysical sources. The dominant signatures of neutrino interactions at the Glashow resonance are particle showers (cascades), originating from hadronic $W^-$ decay. The corresponding signal is anticipated to exceed the continuum of deep-inelastic-scattering induced cascades by up to a factor of 10. Assuming an extraterrestrial electron neutrino flux $E^2\Phi = 1 \times 10^{-8}\,\mathrm{GeV} \cdot \mathrm{cm}^{-2} \cdot \mathrm{s}^{-1} \cdot \mathrm{sr}^{-1}$ and a neutrino to anti-neutrino ratio of $(\nu_e : \bar{\nu}_e) = (1:1)$, which is a generic prediction for pure proton-proton (pp) sources, we expect to detect 0.9 events per year that are contained within the instrumented volume of IceCube.

**Corresponding authors:** J. Kiryluk[2] and H. Niederhausen[2]
[2]*Department of Physics and Astronomy, Stony Brook University, Stony Brook NY 11794-3800*

## 1 Introduction

IceCube is a cubic-kilometer neutrino detector at the South Pole [1]. IceCube's primary science goal is the discovery and study of high energy extraterrestrial neutrinos of all flavors. IceCube detects neutrinos indirectly, typically through deep inelastic scattering (DIS) off the nucleons in the ice. Electron anti-neutrinos of $E_\nu \sim 6.3\,\mathrm{PeV}$ have an enhanced probability to scatter off atomic electrons in the ice by forming an on-shell $W^-$-boson, the so-called Glashow resonance (GR) [2]. The GR events can be used to quantify the contribution of $\bar{\nu}_e$ to the electron neutrino flux $\Phi_{\nu_e+\bar{\nu}_e}$ at earth, which mainly relates to the ratio between the abundances of charged pions ($\pi^-$,$\pi^+$) at source region. For purely hadronic pp sources, $(\nu_e : \bar{\nu}_e) \approx (1:1)$ while $(\nu_e : \bar{\nu}_e) \approx (1:0)$ for purely photo hadronic p$\gamma$ sources [3]. Assuming tribimaximal neutrino mixing[1], this relates to the following electron type neutrino flux composition at earth: $(\nu_e : \bar{\nu}_e) \approx (1:1)$ for pp sources and $(\nu_e : \bar{\nu}_e) \approx (0.78:0.22)$ for p$\gamma$ sources [3].

## 2 Glashow Resonance: Cascade Channel

The $W^-$ produced in $\bar{\nu}_e + e^-$ interaction decays predominantly into a hadronic particle shower with a combined branching ratio of $\Gamma_{\mathrm{hadr}}/\Gamma_{tot} = 0.68$. The $W^-$ decay modes into a charged lepton and the corresponding anti-neutrino have a branching ratio of $\Gamma_{l\bar{\nu}_l}/\Gamma_{tot} = 0.11$ [6]. Thus, the fraction of $W^-$ decays that produces final state muons and the associated track-like signature is small and a search for resonance events is most promising in the cascade detection channel. Cascades from particle showers produce a nearly spherical light pattern in the IceCube detector, as illustrated in Fig. 1 which shows a simulated GR hadronic cascade event. Cascades offer superior accuracy in energy measurement compared to muon tracks, which deposit only a fraction of their energy in the detector. With the recent observation of two $\sim$ 1 PeV cascades [7], IceCube has already begun to explore the interesting high energy range that will eventually extend to the energy of the Glashow resonance peak.

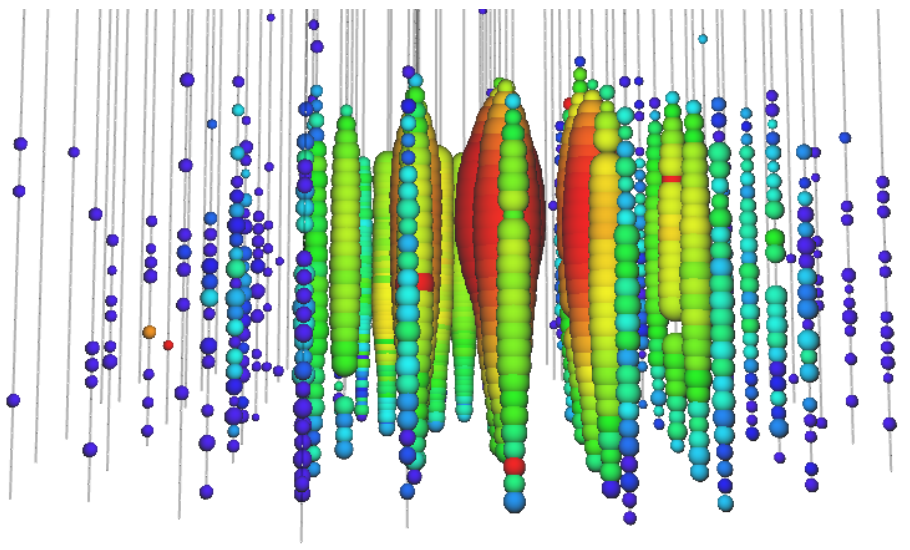

**Fig. 1**: Event display for a simulated Glashow resonance hadronic cascade, $\bar{\nu}_e + e^- \to W^- \to$ hadrons for the IceCube 79-string detector configuration.

Figure 2 shows the expected neutrino primary spectra for pp sources (top) and p$\gamma$ sources (bottom) for an assumed isotropic electron (anti-) neutrino flux of $E^2\Phi = 1 \times 10^{-8}\,\mathrm{GeV} \cdot \mathrm{cm}^{-2} \cdot \mathrm{s}^{-1} \cdot \mathrm{sr}^{-1}$ at the Earth as obtained from Monte Carlo simulations for the IceCube 79-string detector configuration (IC79). The GR events (red: hadr., pink: $e^-\bar{\nu}_e$) exceed the continuum of DIS induced cascades (blue: Charged Current CC events, green: Neutral Current NC events) around $E_\nu = 6.3\,\mathrm{PeV}$. In this figure we require the events to pass the online cascade filter [8], which is designed to select events with almost spherical light deposition. Furthermore, we require that all events develop their associated shower maxima, where most of the Cherenkov photons are emitted, within the fiducial volume of IceCube.

1. Now ruled out because $\theta_{13}$ is large [4, 5].

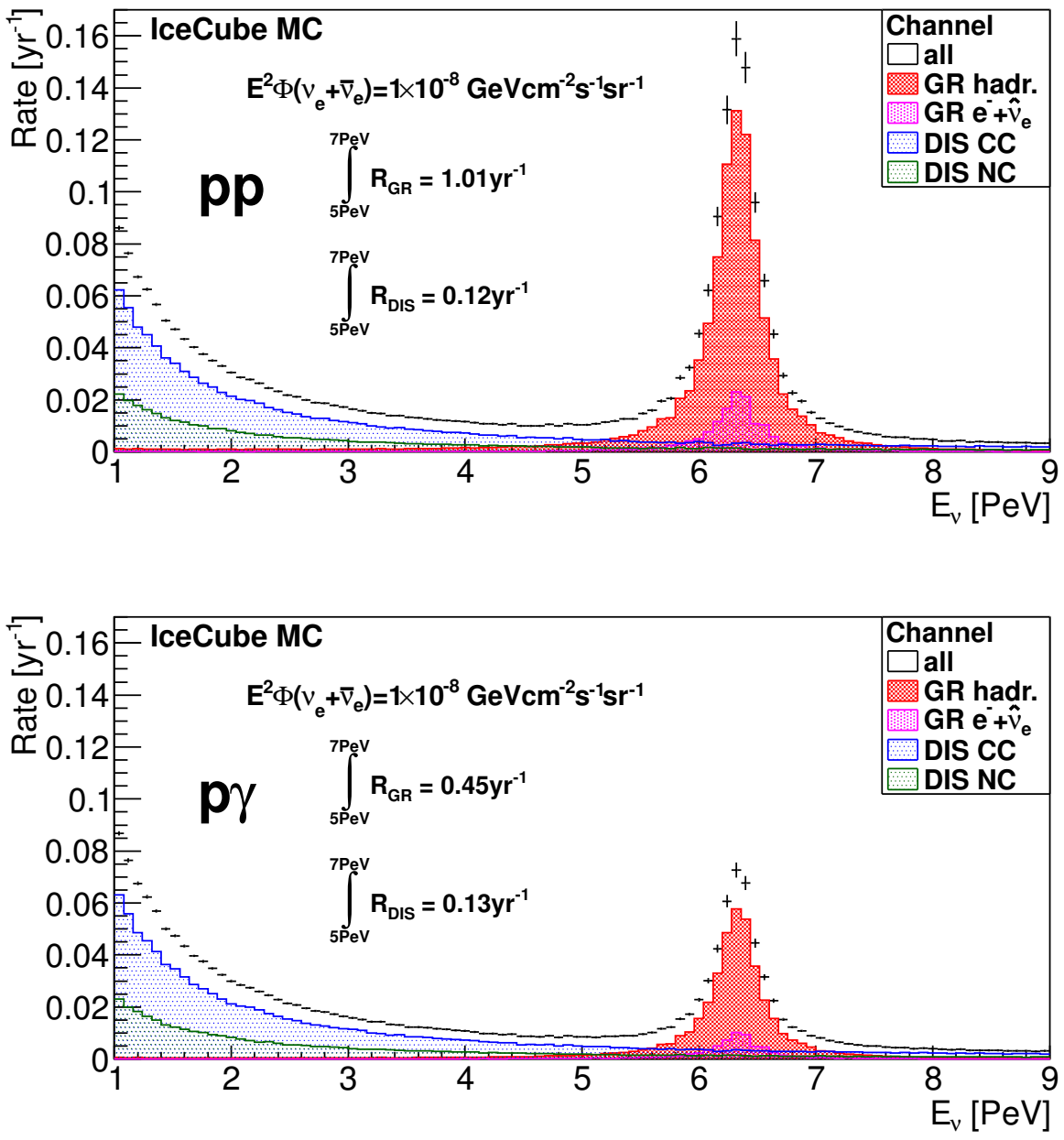

**Fig. 2**: Anticipated electron (anti)neutrino interaction rates in the IceCube 79-string detector configuration versus true Monte Carlo electron anti-neutrino energy $E_\nu$ for pure pp (top) and pure p$\gamma$ (bottom) sources. All events are required to pass the IceCube online cascade selection filtering and shower containment criteria.

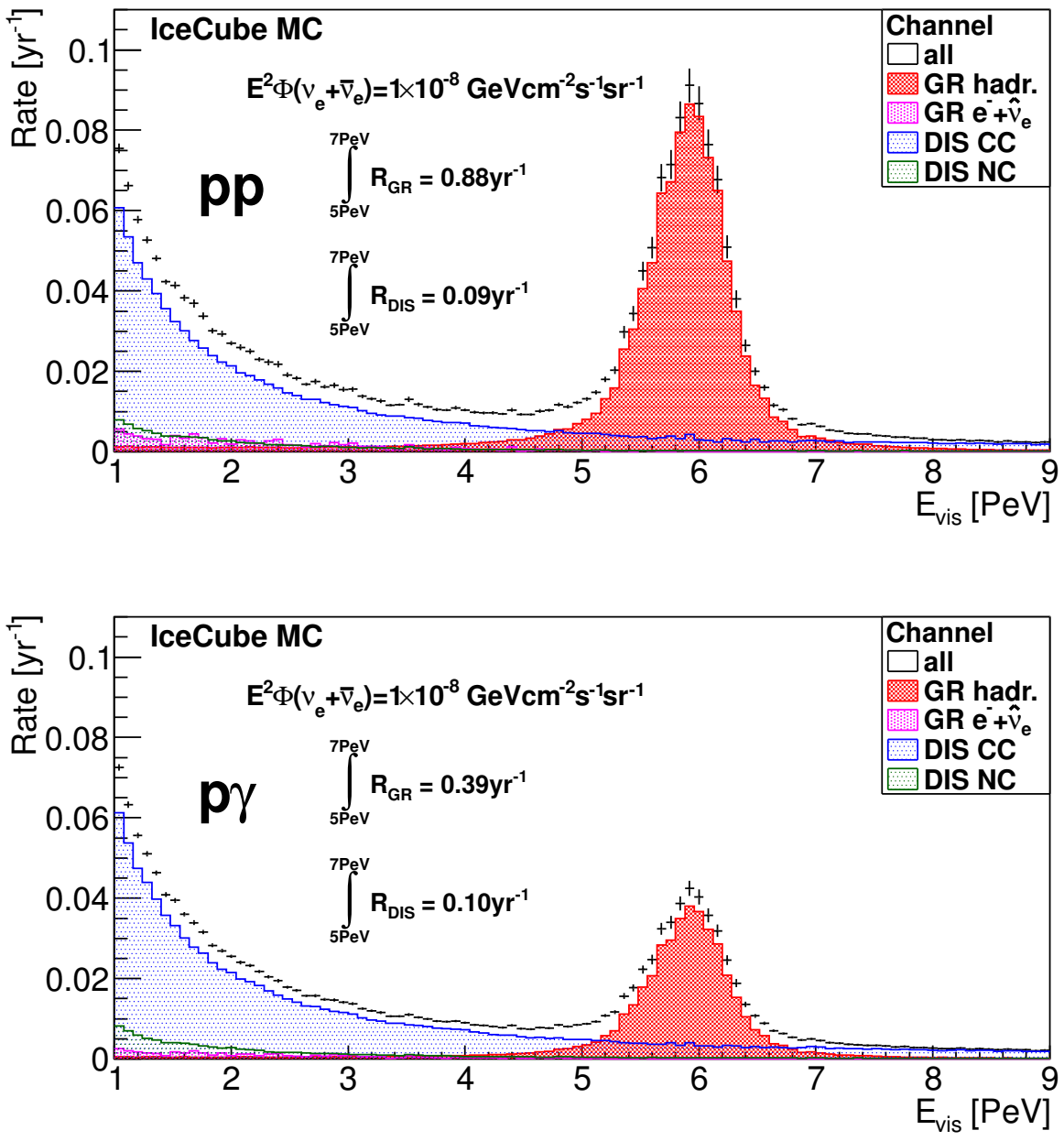

**Fig. 3**: Electron (anti)neutrino interaction rates in IceCube 79-string detector configuration versus true Monte Carlo visible cascade energy deposit, $E_{vis}$, expected from pure pp (top) and pure p$\gamma$ (bottom) sources. All events are required to pass IceCube online cascade selection filtering criteria, as well as the shower containment criteria.

For pure pp sources we expect to observe 1.0 GR cascades per year induced by electron anti-neutrinos with energies between 5 PeV and 7 PeV. Due to the suppressed electron anti-neutrino contribution in the case of pure p$\gamma$ sources this expectation is lowered to 0.5 events pear year while the DIS expectation remains stable at 0.1 events in the same neutrino energy range. On the other hand, processes like multiple-pion production can increase the expected anti-neutrino contribution at the source region by up to 20% [9] while muon cooling could almost completely suppress any electron anti-neutrino flux [10]. Corrections due to the recently established finite value of $\theta_{13}$ [4, 5] remain to be evaluated.

IceCube determines the neutrino energy from the "visible" energy $E_{vis}$, defined as the energy deposit of a purely electromagnetic cascade that produces the observed Cherenkov light yield. Figure 3 shows the distribution of the visible energy $E_{vis}$ for the cases of pure pp and pure p$\gamma$ sources. Due to hadronic physics in the $W^- \rightarrow q\bar{q}$ channel we expect to observe the resonant peak at slightly lower energies ($\sim 7\%$) compared to the neutrino spectrum shown in Fig. 2) and with a marginally larger width due to fluctuations in the Cherenkov light yield of hadronic cascades ($\sim 4\%$). Because of final state kinematics in the $W^- \rightarrow e^- + \bar{\nu}_e$ mode involving a neutrino with "invisible" energy, this channel does not contribute to the peak region of the GR, $5\,\text{PeV} < E_{vis} < 7\,\text{PeV}$, thus lowering the signal expectation by $\sim 10\%$ to 0.9 events per year. Measurement of the flux in the $5\,\text{PeV} < E_{vis} < 7\,\text{PeV}$ energy range is expected to provide insight in and constraints on the physics of cosmic ray accelerators.

## 3 Cascade Reconstruction

Events with a cascade topology are described by a set of seven parameters $C_0 = \{E_{vis}, \vec{x}_{vis}, t_0, \theta_0, \phi_0\}$ assuming point-like Cherenkov emission. Here, $\vec{x}_{vis}$, $t_0$ denote the position of the shower maximum and the corresponding time, and $\theta_0$ and $\phi_0$ the orientation of the shower axis, the neutrino arrival direction. We obtain the values for these parameters by matching the photon arrival time distributions measured in each photomultiplier tube to the expectation for a hypothetical cascade characterized by known parameter values $C_h$. The expected probability density functions (pdfs) corresponding to $C_h$ are obtained from Photonics simulation [11]. The matching is done by using a Poisson likelihood technique [12].

## 4 Performance on simulated GR Cascades

In order to estimate the accuracy of the reconstruction method we rely on simulations of neutrino interactions, Cherenkov light emission and propagation, and detector response.

The same processing and filtering was applied to simulated events as to the data recorded with the IceCube 79-string configuration at the South Pole.

In this analysis we use only hadronic GR cascades that satisfy the following selection criteria: $1\,\text{PeV} < E_{vis} < 10\,\text{PeV}$ and ($-350\,\text{m} < Z_{vis} < -200\,\text{m}$ or $100\,\text{m} < Z_{vis} < 350\,\text{m}$). The former ensures that the events cover the energy range of interest while the latter ensures that the showers do not lose significant fractions of their energy in the known dust layer within the detection volume or in regions below and above the detector. The level of the shower containment for the analyzed samples is controlled in the xy-plane by using the detector scaling variable (XYScale). The XYScale value depends on the position of the shower maximum in the xy-plane, and XYScale = 1 specifies the set of points defined by the polygon of the outermost layer of IceCube strings. This polygon then scales linearly in that variable; XYScale < 1 implies that the event develops its shower maximum within the instrumented region of IceCube, as illustrated in Fig. 4 (bottom).

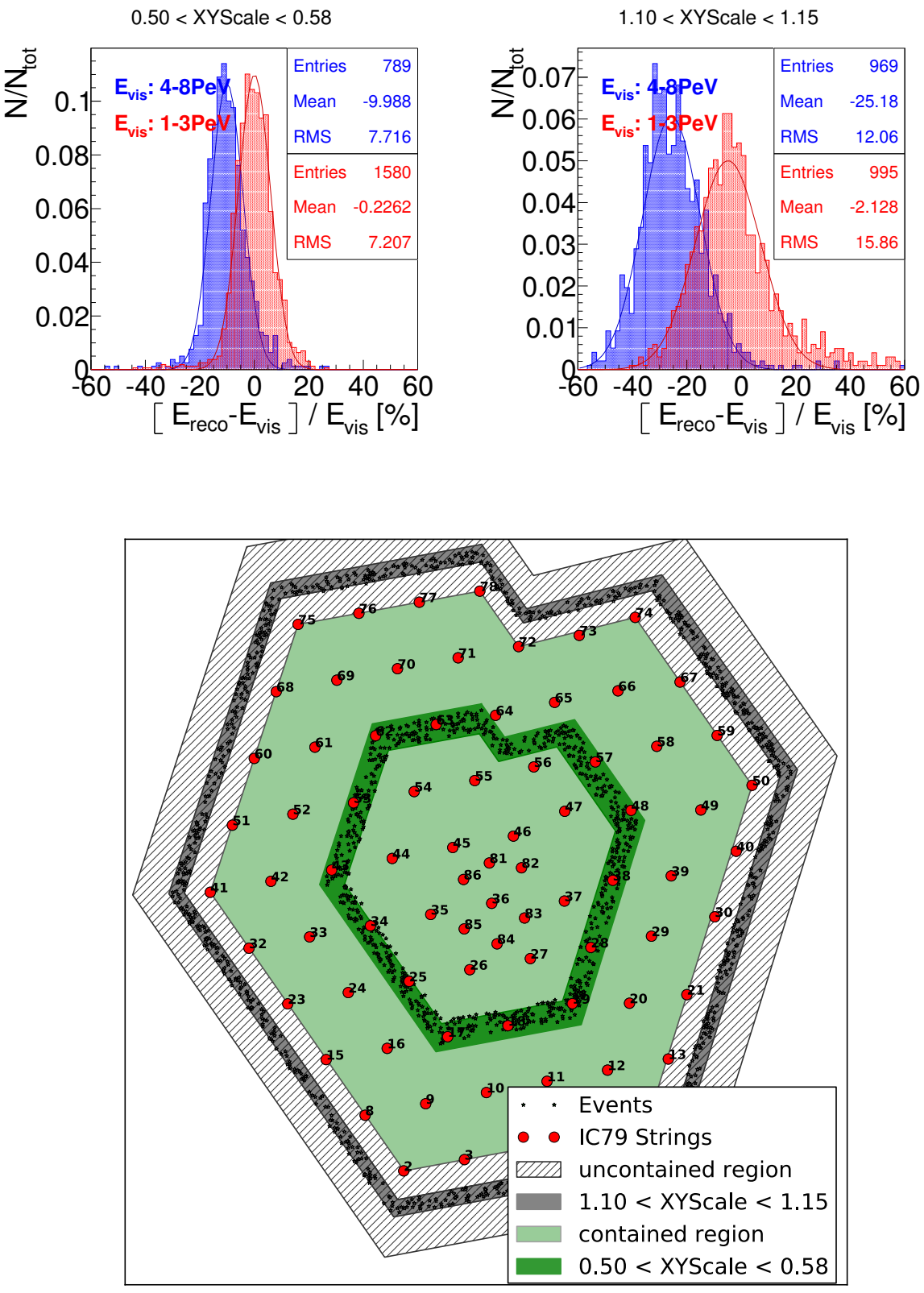

**Fig. 4**: Top: the distribution of $(E_{redo}-E_{vis})/E_{vis}[\%]$ in the energy range 1 PeV $< E_{vis} <$ 3 PeV (red histogram) and 4 PeV $< E_{vis} <$ 8 PeV (blue histogram) for contained events (left) and partially contained events (right) from electron-neutrino Monte Carlo simulation. Bottom: the corresponding location of the shower maximum in $x$ and $y$.

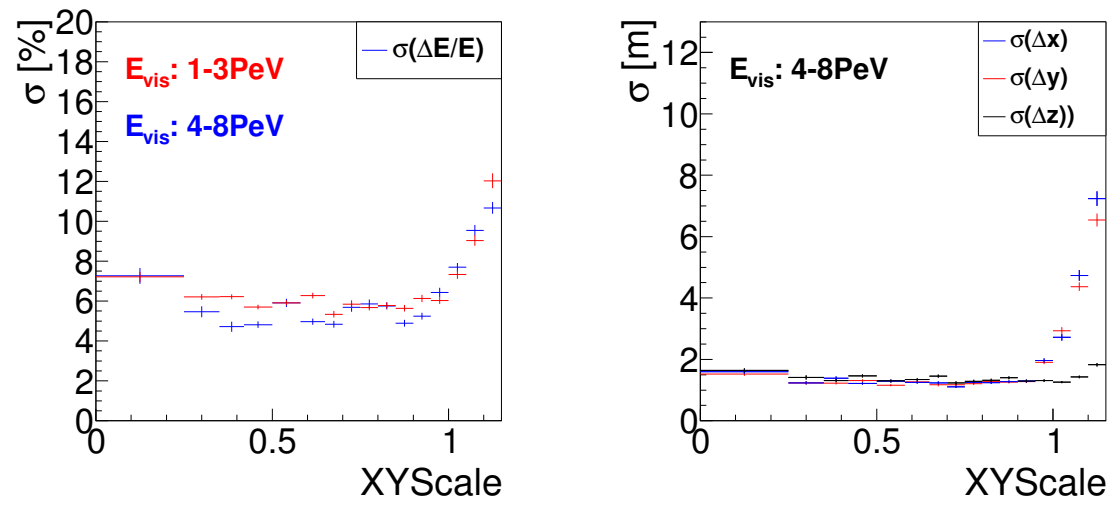

**Fig. 5**: Electron neutrino Monte Carlo. Resolution of the reconstructed energy (left) and of the reconstructed vertex (right) as a function of the detector containment variable XYScale.

Figure 4 (top) shows the distributions of the difference between the reconstructed visible energy and the MC truth relative to the MC true visible energy for contained showers with $0.50 <$ XYScale $< 0.58$ (left) and partially contained showers with $1.10 <$ XYScale $< 1.15$ (right) in two energy bins: 1 PeV $< E_{vis} <$ 3 PeV (red histogram) and 4 PeV $< E_{vis} <$ 8 PeV (blue histogram). Figure 5 shows the resolution ($\sigma_{\Delta E}$) of the energy reconstruction as determined from the parameters of a gaussian fit to the relative differences in Fig. 4 (top) for different XYScale values. Within the detection volume we achieve an energy resolution of $5-8\%$ for GR cascades which is sufficient to observe the (anticipated) GR signal above the DIS continuum at PeV energies. For partially contained cascades ($1.0 <$ XYScale $< 1.15$) we find better than 15% energy resolution. The MC simulations show that the reconstructed energies in the 4 PeV $< E_{vis} <$ 8 PeV bin are underestimated by $\sim 10\%$ (25%) for fully (partially) contained cascades. This effect is not observed in the lower energy bin 1 PeV $< E_{vis} <$ 3 PeV. Further studies are needed. The inclusion of partially contained cascades in the analyzed sample can enhance our expected GR signal rate by up to $\sim 50\%$.

Figure 5 quantifies how well the position of the shower maximum of the GR signal cascades is resolved. The results were obtained from the parameters of a gaussian fit to the difference between true and reconstructed x,y and z positions. No bias is found in the vertex reconstruction. The vertex resolutions along each of the three coordinates within the detection volume is about 1.5 m. While the $z$-position of the shower is still well constrained beyond the detection volume with an accuracy of better than 2 m up to XYScale = 1.15, the resolution in $x$ and $y$ quickly worsens by a factor of up to five. This is due to the difference in spacing between adjacent detector elements in these coordinates.

The zenith angle $\theta$ is of interest in the reconstruction of the cascade direction, as is the opening angle $\Psi$ between the true and reconstructed cascade direction. For contained events (XYScale$< 0.95$) we find a median angular (zenith) resolution $\theta_{0.5} = \text{Median}\{|\theta_{reco}-\theta_0|\}$ of $\sim 5°$ and a median directional resolution $\Psi_{0.5} = \text{Median}\{|\Psi|\}$ of $\sim 8°$. The median zenith resolution is $\theta_{0.5} \simeq 9°$ for uncontained cascades with $0.95 <$XYScale$< 1.15$, and the median directional resolution is $\Psi_{0.5}$ between $8°$ and $27°$.

## 5 Performance on laser data

In order to verify the cascade reconstruction performance independent from Monte Carlo simulation, we analyzed calibration laser data recorded with the 79-string detector configuration in 2011. This nitrogen laser, deployed in clear ice at a depth of 2153 m, emits photons at the near-UV wavelength of 337 nm and its light deposition mimics the point like light pattern of highly energetic cascades. A reflective cone ensures that photons are emitted at the Cherenkov angle of 41°. Different attenuation filters make it possible to adjust the laser brightness. In this study we reconstructed 1000 laser flashes for each of the three brightness levels (1.6%, 3.2% and 8.9% of the maximum brightness) that cover the PeV energy region of interest for this analysis. Figure 6 (top) shows the distribution of the fractional deviation of the reconstructed energy $E_{reco}$ from the mean reconstructed energy $E_{reco}^{mean}$ for the lowest (left) and highest (right) laser brightness used in this analysis. We find excellent energy resolutions of better than 2% for these events. This result is comparable to the per-pulse fluctuations of the laser brightness, which were found to be less than 3% in laboratory measurements before its deployment at the South Pole. The best-fit energies range from 1.1 PeV to 6.3 PeV. In Fig. 6 (bottom) the dependence of the best-fit values and their statistical uncertainties is shown as a function of the filter transmittance (laser brightness). The three data points are well described by a one-parameter linear fit, as expected in view of the linearity of the device. In addition, we found that the vertex resolutions are better than 0.35 m in $x$ and $y$, independent of the laser brightness for filter transmittance settings between 1.6% and 9%. We observe a systematic shift of the best-fit value from the known laser position of up to 0.80 m. This may be due to systematic uncertainties associated with the description in the analysis of the in-ice light propagation at a wavelength of 337 nm, which is obtained by extrapolation from the ice properties (scattering and absorption lengths) determined using the LED calibration system at 400 nm [13].

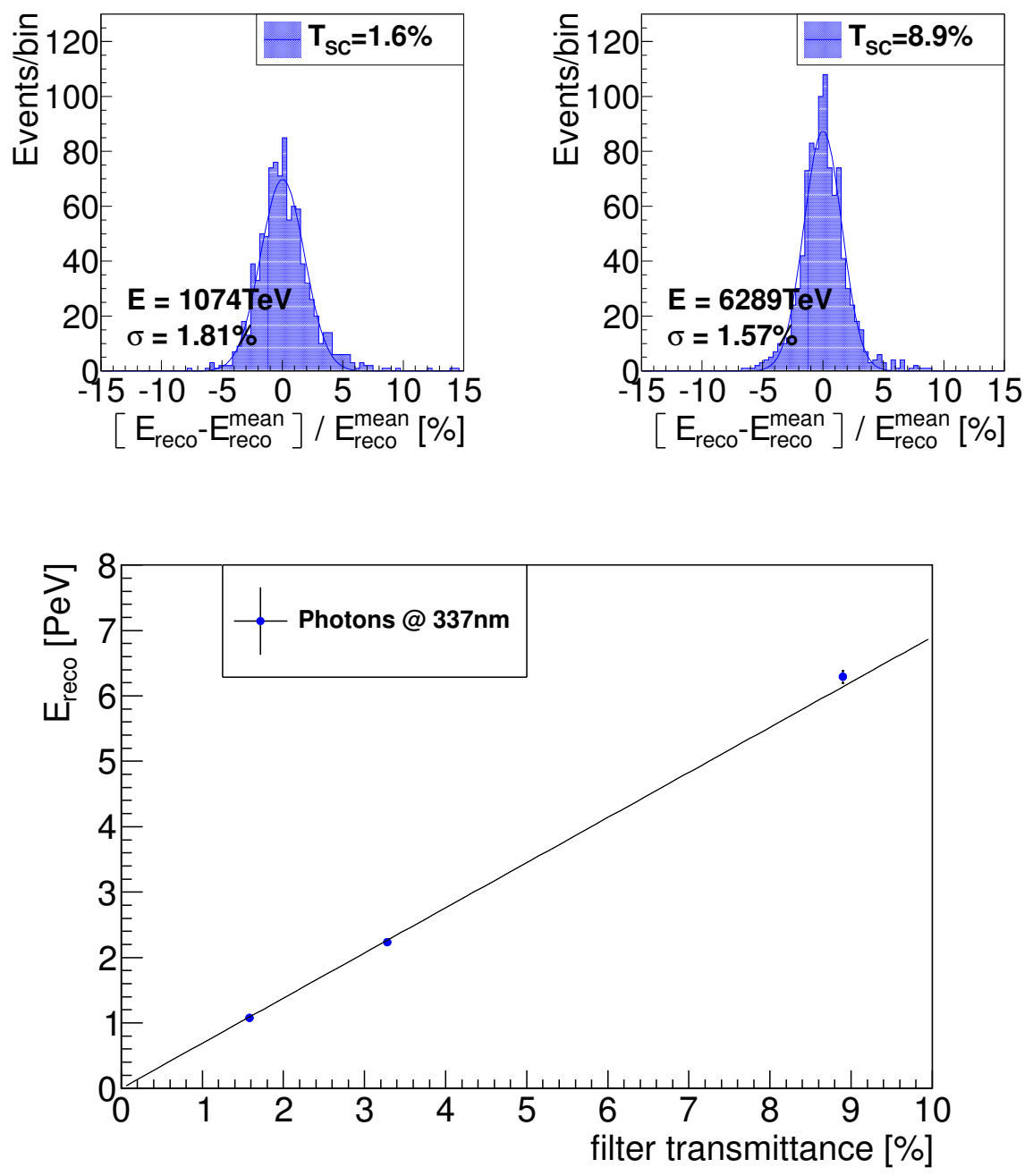

**Fig. 6**: Top: the fractional deviation of the reconstructed energy $E_{reco}$ from the mean reconstructed energy $E_{reco}^{mean}$ for the 1.6% laser transmittance (left) and the 9% filter transmittance (right). Bottom: best fit reconstructed energy values [PeV] versus filter transmittance [%] .

## 6 Summary

One of the primary goals of IceCube is to observe the flux of high-energy cosmic neutrinos and anti-neutrinos from astrophysical sources. The (anti-)neutrino spectrum near the characteristic energy $E \sim 6.3$ PeV of the Glashow resonance offers the unique possibility to determine the contribution from electron anti-neutrinos and provides new constraints on the possible production mechanisms for high-energy (anti-)neutrinos in astrophysical sources. The dominant signatures of neutrino interactions at the Glashow resonance in IceCube are particle showers, originating from hadronic $W^-$ decay with a combined branching ratio of 70%. Assuming a neutrino to anti-neutrino ratio of $(\nu_e : \bar{\nu}_e) = (1:1)$, which is a generic prediction for pure pp sources, we expect the integrated GR signal rate to be an order of magnitude larger than its DIS counterpart in the 5 PeV to 7 PeV energy range. The GR signal is less pronounced in the case of pure p$\gamma$ sources, but is still anticipated to be four times larger than the DIS contribution. Assuming an extraterrestrial electron type neutrino flux of $E^2\Phi = 1\times10^{-8}\,\mathrm{GeV}\cdot\mathrm{cm}^{-2}\cdot\mathrm{s}^{-1}\cdot\mathrm{sr}^{-1}$ that is based purely on pp (p$\gamma$) collisions we expect to detect 0.9 (0.4) GR signal cascade events per year that are contained within the instrumented volume of IceCube.

For fully contained GR cascades we expect to reconstruct the energy $E_{vis}$ with a resolution of 5% to 8%, not including systematic uncertainties. Even in the region beyond the outermost IceCube strings with XYScale $< 1.15$ we obtain useable energy resolutions of better than 15%. Including this region into a future GR analysis of data is simulated to increase our expected signal rates by up to $\sim 50\%$.

We have verified our simulation methods using experimental data obtained by pulsing an in-situ calibration laser at three different brightness settings, corresponding to the energy range of 1.1 PeV $< E_{reco} <$ 6.3 PeV. We find energy resolutions of better than 2% for these data. Within statistical uncertainties, the best-fit energies increase linearly with the laser intensity.

We conclude that IceCube has the capability to resolve a GR signal above the DIS continuum. The recent observation of two $\sim$ 1 PeV cascades in IceCube indicates that a flux may exist.

**Acknowledgments:**This work is supported by the National Science Foundation under Grant No. 1205796.

# Detecting Tau Neutrinos in IceCube with Double Pulses

THE ICECUBE COLLABORATION[1]
[1] *See special section in these proceedings*

*dxu@crimson.ua.edu*

**Abstract:** Neutrinos produced in astrophysical sources such as active galactic nuclei and gamma ray bursts are thoroughly mixed en route to earth. Therefore we expect a 1:1:1 neutrino flavor ratio from astrophysical sources at the earth's surface. IceCube is designed to detect all flavors of astrophysical neutrinos, each of which produce distinctive light patterns in the detector. At sufficiently high energies the charged current interaction of the tau neutrino and subsequent decay of the tau lepton within the detector, may appear as one or more double pulses in the IceCube module waveforms. We will review the sensitivity of the double pulse signature to astrophysical tau neutrinos.

**Corresponding authors:** Donglian Xu[2], Dawn Williams[2] and Pavel Zarzhitsky[2]
[2] *Department of Physics and Astronomy, University of Alabama, Tuscaloosa, AL 35487, USA*

## 1 Introduction

If extremely energetic cosmic rays are accelerated in astrophysical sources such as active galactic nuclei (AGN) [1] and gamma ray bursts (GRBs) [2], then there should be a corresponding flux of extremely energetic neutrinos from these sources. While charged cosmic ray particles are deflected en route to earth by magnetic fields, neutrinos are neutral and can travel astronomical distances without change of trajectory. Thus neutrinos can point back to their astrophysical origin. The three flavors of neutrinos $(\nu + \bar{\nu})$[1] from the Standard Model, $\nu_e$, $\nu_\mu$ and $\nu_\tau$, with an expected flux ratio of 1:2:0 at source of production, would mix thoroughly over astronomical distances. Therefore we expect a 1:1:1 neutrino flavor ratio from astrophysical sources in terrestrial detectors [3].

The IceCube Neutrino Observatory, located at the geographic South Pole [4] is designed to detect astrophysical neutrinos and is sensitive to all three neutrino flavors. Each flavor makes distinctive light patterns in the detector. There are two major types of neutrino interaction light patterns in IceCube. One is track-like, produced by a muon resulting from the charged current (CC) interaction of a muon neutrino. The other is cascade-like, made by electron neutrino and low energy tau neutrino CC interactions and neutral current (NC) interactions of all three flavors. The primary background to astrophysical neutrino searches are muons and neutrinos from the atmosphere which are produced in air showers induced by cosmic rays. At sufficiently high energies, the tau lepton produced by the CC interaction of a tau neutrino may travel substantially long distances ($O(10)$ m at 100 TeV) before decaying. The hadronic shower from the interaction vertex and the electromagnetic/hadronic shower from the tau decay are well separated in time, and hence photons from these two showers registered at the nearby IceCube photon sensors may appear as double pulses in these sensors, see Fig. 1. The tau neutrino channel is of particular interest for astrophysical neutrino searches because the flux of tau neutrinos from the atmosphere due to oscillation and prompt charm production is very low [5]. Therefore, a high energy tau neutrino detected in IceCube is likely to be of astrophysical origin. A dedicated analysis searching for astrophysical tau neutrinos was carried out during IceCube's construction phase when only 22 strings were instrumented [6]. Neutrino flux upper limit set by this analysis was superseded by later IceCube analyses searching for extremely-high-energy (EHE) neutrinos [7, 8].

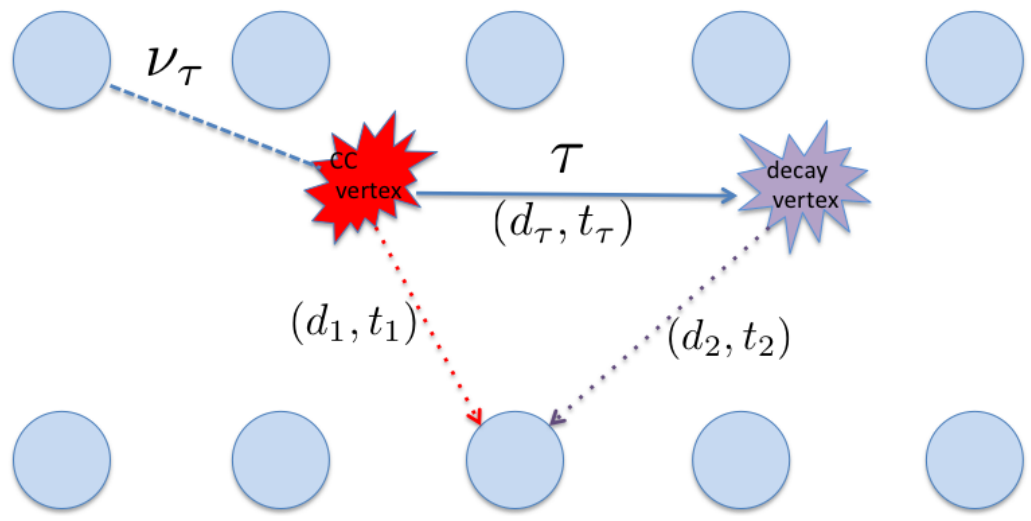

**Figure 1**: Sketch of a tau neutrino undergoing CC interaction in the IceCube PMT array.

## 2 The IceCube Waveforms

IceCube consists of 86 strings, each of which is instrumented with 60 Digital Optical Modules (DOMs). Most strings are deployed on a hexagonal grid with 120 m horizontal string-to-string spacing and 17 m vertical DOM-to-DOM spacing. The inner DeepCore detector includes 8 strings with smaller string-to-string and DOM-to-DOM spacing. The DOM consists of a glass pressure vessel enclosing a 10-inch photomultiplier tube (PMT) and a main board which contains the digitizing electronics [9]. The PMT signal is split into three amplifier channels, each with a different gain. All three channels are digitized by a separate Analog Transient Waveform Recorder (ATWD) which digitizes at a rate of 3.3 ns per sample for 128 samples. A longer signal is recorded by the Fast Analog to Digital Con-

1. IceCube doesn't discriminate $\nu$ and $\bar{\nu}$

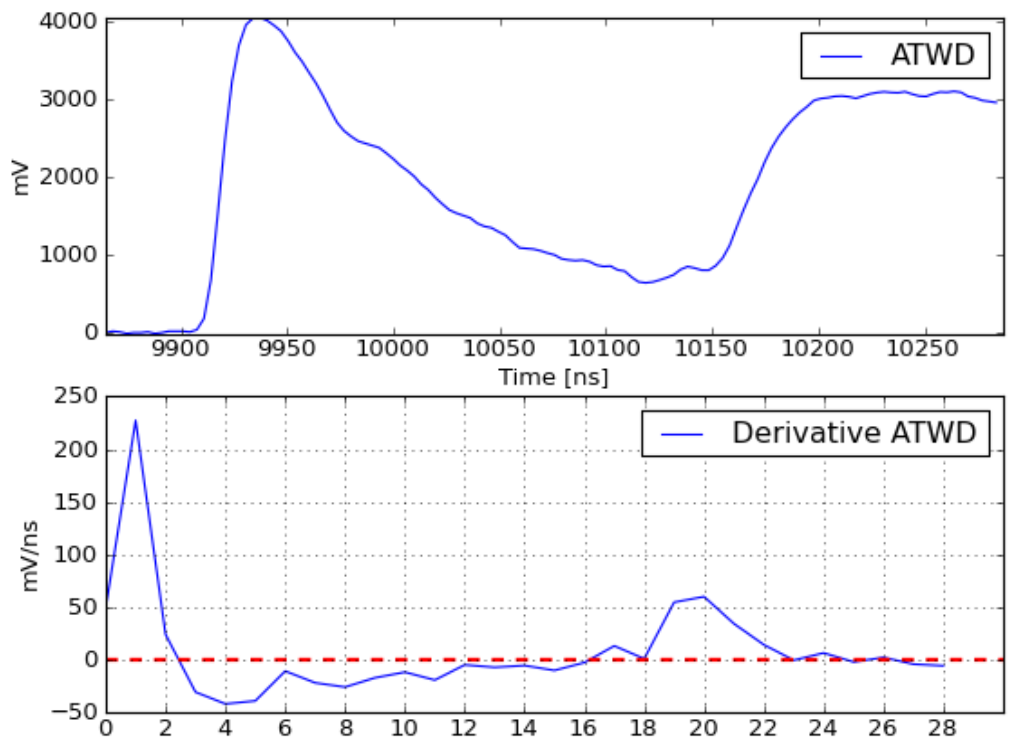

**Figure 2**: Top: a simulated ATWD double pulse waveform from a $\nu_\tau$ CC interaction. Bottom: derivative of the above double pulse waveform, the horizontal axis of the waveform derivative is plotted in terms of bin numbers, each of which consists of four ATWD sampling bins ($4 \times 3.3$ ns/sample = 13.2 ns).

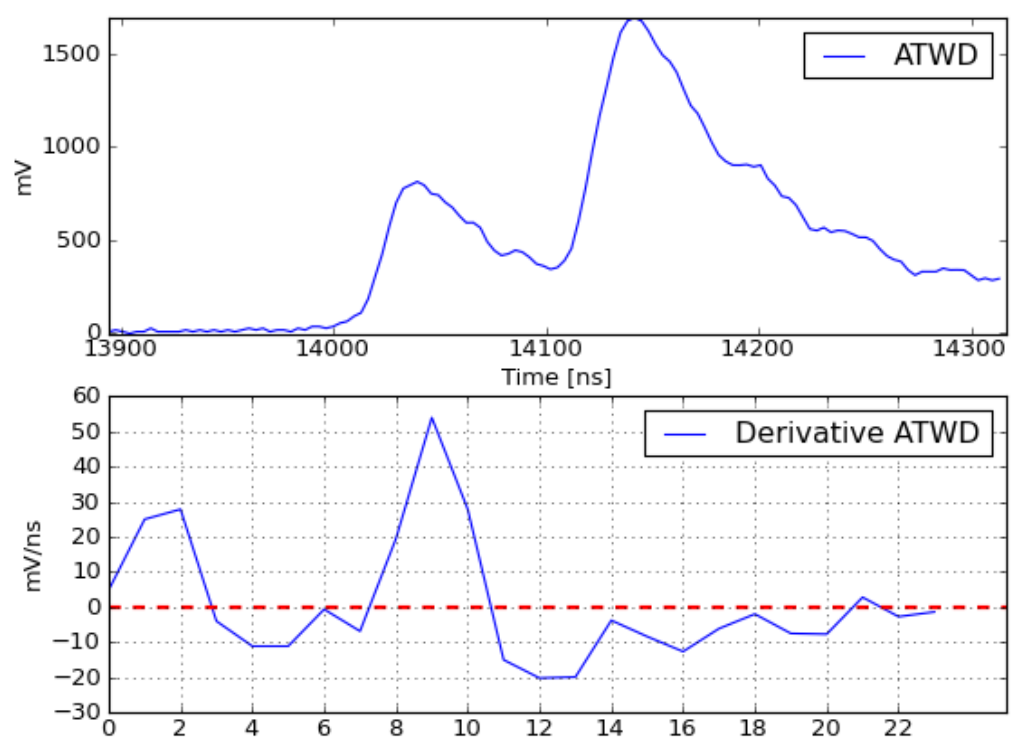

**Figure 3**: Top: a simulated ATWD double pulse waveform of a muon bundle event. Bottom: derivative of the above double pulse waveform, the horizontal axis of the waveform derivative is plotted in terms of bin numbers, each of which consists of four ATWD sampling bins ($4 \times 3.3$ ns/sample = 13.2 ns).

verter (FADC) at 25 ns per sample for 256 samples. Typically, a DOM launch is a single photoelectron (SPE) which only uses the highest-gain channel of the ATWD. However, high-charge waveforms such as those which are used in this analysis are clipped in the high-gain channel, so they launch the low-gain channels as well. Since this analysis depends on the shape of individual waveforms, it should be noted that although IceCube employs a lossy compression algorithm for transmitting most waveforms to the Northern hemisphere, all waveforms which use the lower-gain channels are transmitted uncompressed, so there is no loss of information.

The basic unit of IceCube data is a hit, or a signal recorded on a single PMT. The DOMs communicate with each other via a local coincidence (LC) signal and send the full digitized waveforms to the surface if a DOM is launched in coincidence with its nearest or next-to-nearest neighbor. On the surface, trigger algorithms are applied; the most common being a simple majority trigger of 8 or more hits within a time window. A set of hits within a trigger window is denoted an "event", which is then sent to a filtering algorithm to determine whether the event is sent to the Northern hemisphere for analysis. This analysis uses the Extremely High Energy (EHE) filter which selects events where more than 1000 photoelectrons (PE) are deposited in the detector.

## 3 Potential $\nu_\tau$ Double-Pulse Events

Whether photons from the two showers from a $\nu_\tau$ CC event will appear in the waveform as double-pulse feature or not, is a combination of three effects - time separation between the two showers, distances of the two vertices from the DOM and orientation of the two vertices with respect to the DOM. We denote distances from both cascades to a nearby DOM as $d_1$ and $d_2$, time for photons to travel from both vertices to that DOM as $t_1$ and $t_2$, tau length and lifetime as $d_\tau$ and $t_\tau$. For unscattered photons, the arrival time difference between photons from the two cascades arriving at a nearby DOM is $\Delta t_{arr} = |t_2 - t_1| = |\frac{d_2 - d_1}{v_{ice}}|$, where $v_{ice} = \frac{c}{n_{ice}}$ is the speed of light in ice, $n_{ice} = 1.36$ is the refractive index of Antarctic glacial ice. Photons from the two cascades may appear as double pulses in the digitized waveforms if $\Delta t_{arr}$ is within the range such that the double-peak feature can be resolved, which is at least 100 ns given the typical time profile of the light distribution from a single cascade. Photons traveling in the glacial ice will undergo complex scattering and absorption [10], therefore the double pulse feature is harder to resolve in distant DOMs due to the smearing effect of scattering. Therefore, the most likely DOMs with double pulse waveforms are the ones close to the interaction and decay vertices. In order to study double pulse events a sample of $\nu_\tau$ Monte Carlo simulation with energies between 1 TeV and 10 PeV was generated with an $E^{-1}$ neutrino spectrum to maximize the statistics of high energy events. This sample can be subsequently re-weighted to a more realistic $E^{-2}$ spectrum to calculate event rates. To select double pulse candidates, we performed a geometrical selection of $d_1 < 200$ m and $d_2 < 200$ m and $\Delta t_{arr} > 100$ ns on a sample. Any $\nu_\tau$ CC event that has DOMs satisfying this criteria is considered as a potential double pulse event. All events passed these criteria are found out to have neutrino energies greater than 100 TeV. In order to develop an automate algorithm to search for double pulse waveforms, we began with a sample of visually selected double pulse waveforms from these events. An example of a $\nu_\tau$ double pulse waveform is shown in top panel of Fig. 2. Based on features of these waveforms, a computational double pulse algorithm was developed and optimized to identify the $\nu_\tau$ double pulse signature. The algorithm is described in the following section.

## 4 $\nu_\tau$ Event Selection

### 4.1 Filter

The first step in data processing is the EHE filter which retains 100% of the simulated $\nu_\tau$ double pulse events. As shown in Fig. 4, the total event charge of $\nu_\tau$ double pulse events is typically well over 1000 PE, which is the threshold of the EHE filter.

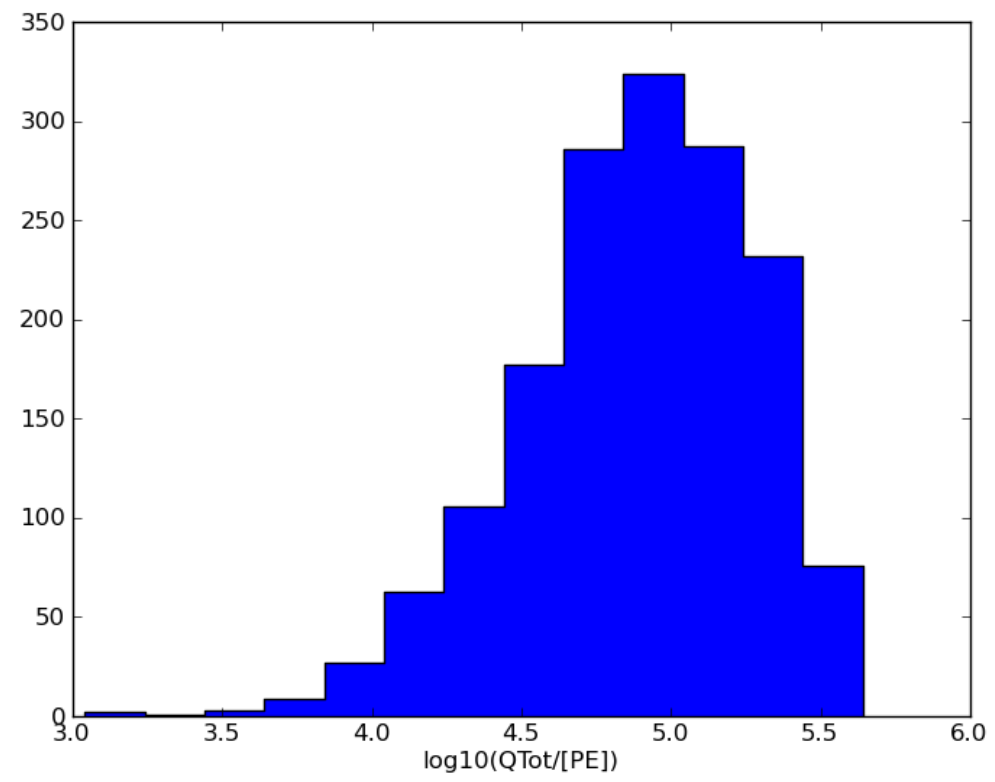

**Figure 4**: Distribution of total charge of $\nu_\tau$ CC double pulse events (no filter). Y-axis is event numbers per bin.

## 4.2 Double Pulse Algorithm

The goal of the algorithm is to identify waveforms with double pulse features which are consistent with a $\nu_\tau$ double pulse waveform, while rejecting features such as late-arriving scattered photons from a single cascade event.

- We use waveforms from the ATWD digitizer in the lowest gain channel available, since higher gain channels are generally clipped for high-amplitude waveforms. Waveforms with integrated amplitude less than 10000 mV·ns are rejected. With base impedance of 47 ohms and nomial gain of $10^7$ [9], this translates to $\sim$ 131 PE. FADC waveforms are not used since they do not have multiple gain channels available and since their coarser timing causes double pulse features to be blended together or clipped.

- The beginning of the waveform is detected by a sliding time window which searches for a monotonic increase in the waveform amplitude within the window.

- Once the beginning of the waveform is found, the waveform is divided into 13.2 ns segments and the derivative is calculated for each segment. The bottom panel of Fig. 2 shows an example of an ATWD waveform derivative vector.

- If the derivative is positive in two consecutive segments, this is considered the rising edge of the first pulse. When the subsequent derivative is negative for two consecutive segments, this is considered the trailing edge of the first pulse. The rising edge of the first pulse is required to have an integrated charge of at least 3.5 PE. The integrated charge sums up all the charge corresponding to the entire rising edge, which usually lasts longer than two segments (26.4 ns) for a big pulse.

- The second pulse rising edge is defined when the derivative after the trailing edge of the first pulse is positive again for three consecutive segments. This requirement is due to the fact that the second pulse is often more scattered and therefore has a less steep rising edge than the first pulse. The second pulse trailing edge is often outside the ATWD time window, and is therefore not calculated. The rising edge of the second pulse is required to have an integrated charge of at least 5.3 PE.

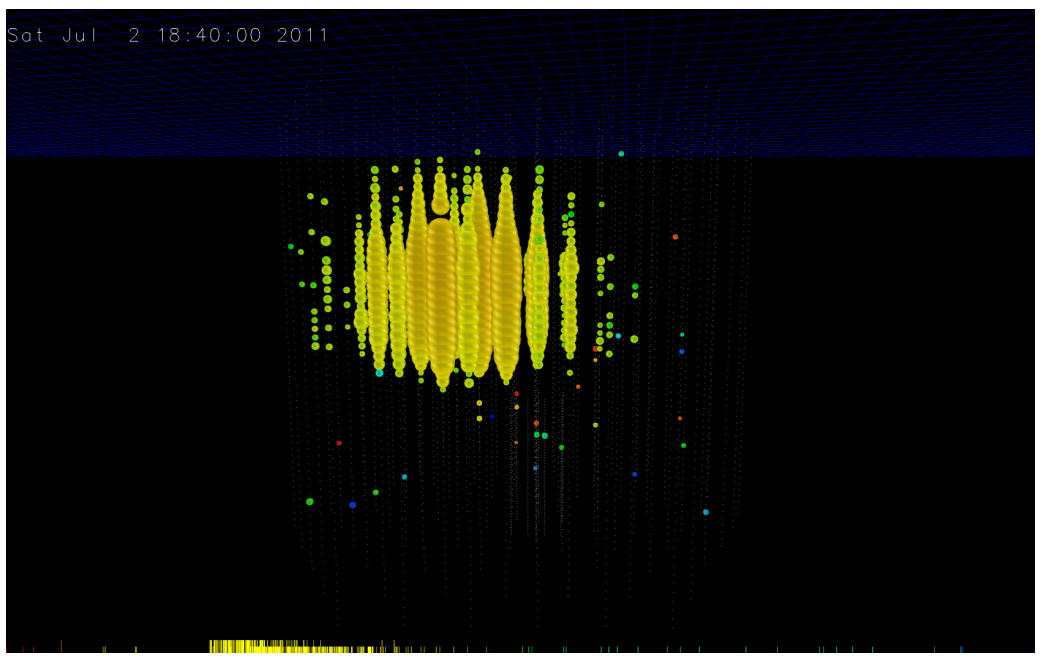

**Figure 5**: Simulated $\nu_\tau$ CC double pulse event with $E_\nu$=2.67 PeV and 109740 PE deposition in IceCube. Color denotes photon deposition timing, red: early, blue: late.

A waveform which has a first rising edge, and first trailing edge and a second rising edge according to the above criteria is considered a double pulse waveform.

## 4.3 Efficiency

With the above criteria, the double pulse algorithm is able to pick up $\sim$ 90% of the double pulse waveforms from the 270 waveforms selected from the aforementioned $\nu_\tau$ MC sample. The algorithm also picks up about 15% of weak double pulse waveforms. Cuts to exclude the weaker double pulse waveforms are under study. The second pulse in these waveforms is very weak and may be eliminated with cuts on the width of the second pulse or the charge in the second pulse relative to the first.

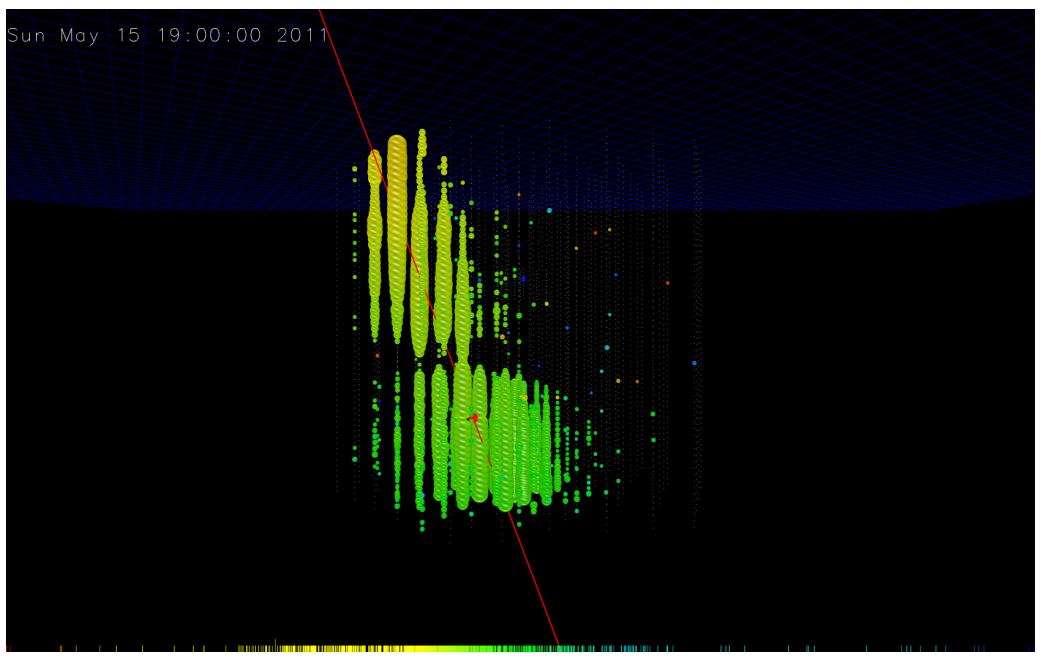

**Figure 6**: Simulated muon bundle double pulse event with primary nucleus energy of 280 PeV and 53384 PE deposition in IceCube. Color denotes photon deposition timing, red: early, blue: late.

## 4.4 Veto

The major background for this analysis is throughgoing muon bundles from cosmic ray induced air showers. A veto is applied in order to reject background events. This veto is the same that was used for the high energy starting events analysis in IceCube which is discussed elsewhere in these proceedings [11]. The main goal of this veto is to reject throughgoing muons. The veto region includes the top 90 m (equivalent to 5 layers of DOMs) of the detector, the outermost layer of strings, the bottom layer of DOMs and the DOMs in the 80 m thick dust layer which is located about 2100 m from the surface of the glacial ice cap [10]. The veto also determines the event start time as the time at which 250 PE have been deposited in the detector within a time window of 3000 ns. An event is rejected if more than 3 PE are deposited in the veto region before the event starts.

## 5 Preliminary Event Rates

The processing chain is applied to all Monte Carlo samples (both signal and background) and 10% of the total data collected between May 13 2011 and May 15 2012 in order to calculate signal and background event rates.

### 5.1 Signal

From the simulated $\nu_\tau$ sample, the rate of CC events above 100 TeV is estimated to be about 16 per year based on a flux of $E^2\Phi_{\nu_e+\nu_\mu+\nu_\tau} = 3.6\times10^{-8}$ GeV sr$^{-1}$ s$^{-1}$ cm$^{-2}$. This flux corresponds to the upper limit from a previously published IceCube extremely-high energy cosmic neutrinos search [7]. Applying the double pulse requirement yields 0.72±0.06 (stat) signal events per year. With additional veto applied, 0.32±0.04 (stat) events per year survive.

### 5.2 Background

The following simulated backgrounds are considered:

- Atmospheric muons and muon bundles from cosmic ray air showers with primary energy from $10^5$ to $10^{11}$ GeV
- Muon and electron neutrinos with an $E^{-2}$ spectrum with energies between 10 and $10^9$ GeV.
- Muon and electron neutrinos with energies between 10 and $10^9$ GeV weighted to atmospheric neutrino models [5, 12].

Before the veto is applied, 8800±1900 (stat) simulated muon events per year have at least one double pulse waveform. Typically these events are muon bundles rather than single muons. The cause of double pulses in muon bundle events is stochastic TeV-scale energy loss within $O(10)$ m of a DOM. Fig. 3 shows a simulated ATWD double pulse waveform from a muon bundle event. Although the shape of the waveform is similar to that of a $\nu_\tau$ CC double pulse waveform, Fig. 2, the overall event topology of the event is very distinctive. While light deposition from a $\nu_\tau$ CC double pulse event is well contained within the detector, see Fig. 5, a muon bundle double pulse event clearly comes from outside the detector and has a track-like structure, see Fig. 6. The veto algorithm rejects 100% of these double pulse muon and muon bundle events, while retaining $\sim$ 44% of the $\nu_\tau$ CC double pulse events.

The remaining sources of background are NC events from all three neutrino flavors and CC events from $\nu_e$ and $\nu_\mu$, both of astrophysical and atmospheric origin. Atmospheric neutrino flux consists of a conventional [12] and a prompt components [5]. Atmopheric neutrino rates are estimated by weighting the aforementioned neutrino Monte Carlo samples to these atmospheric neutrino flux models. The flux of high energy (>100 TeV) atmospheric $\nu_\tau$ which is mainly from prompt charm mesons decay, is $\sim$ 10 times lower than that of $\nu_\mu$ and $\nu_e$ [5], and hence is not considered in this analysis. A total of 1.6±0.7 (stat) background neutrino events per year survive the veto and the double pulse cut. A $\nu_\mu$ CC event which produces an outgoing muon track may have double pulse waveforms due to stochastic energy loss of the muon. Such events should have a track-like appearance which may be rejected by a cut on the event topology. Cascade-like events from NC events of all three flavors or $\nu_e$ CC events tend to have very weak double pulses, which may be rejected by including cuts on the width and amplitude of the second pulse.

## 6 Conclusion

Preliminary studies of a simple double pulse algorithm search for $\nu_\tau$ double pulse signature indicate that we should be able to see $O(1)$ $\nu_\tau$ double pulse events in 3 years of IceCube data, based on a flux of $E^2\Phi_{\nu_e+\nu_\mu+\nu_\tau} = 3.6\times10^{-8}$ GeV sr$^{-1}$ s$^{-1}$ cm$^{-2}$ [7], with no background from atmospheric muons surviving the initial cuts. Future refinements will include cuts on the width of the pulses, as well as veto optimization to reject fewer signal events and cuts on event topology to reject $\nu_\mu$ CC double pulse events.

This work is supported in part by the National Science Foundation under Grant #1205600.

# Evidence for High-Energy Extraterrestrial Neutrinos at the IceCube Detector

THE ICECUBE COLLABORATION[1],

[1] *See special section in these proceedings*

*claudio.kopper@icecube.wisc.edu*

**Abstract:** We present a follow-up search on the previous detection of two PeV neutrino events with the IceCube detector during an observation period from May 2010 to May 2012. Selecting for high energy neutrino events with vertices well contained in the detector volume, our analysis has improved sensitivity and extended energy coverage down to approximately 50 TeV. We observed 26 new events in addition to the two events already observed earlier. The combined preliminary significance from both searches represents a $4.1\sigma$ excess above expected backgrounds from atmospheric muons and neutrinos. The entire sample of 28 events includes the highest energy neutrinos ever observed, and has properties consistent in flavor, arrival direction, and energy with generic expectations for neutrinos of extraterrestrial origin.

**Corresponding authors:** Claudio Kopper[1], Nathan Whitehorn[1], Naoko Kurahashi Neilson[1],
[1] *Wisconsin IceCube Particle Astrophysics Center, University of Wisconsin, Madison, WI 53706, U.S.A.*

## 1 Introduction

Observation of high-energy neutrinos provides insight into the problem of the origin and acceleration mechanism of high-energy cosmic rays. Cosmic-ray protons and nuclei produce neutrinos in interactions with gas and photons present in the environment of sources and in the interstellar space through decay of charged pions and kaons. These neutrinos have energies proportional to the cosmic rays that produced them and point back to their sources since they are neither affected by magnetic fields nor absorbed by matter opaque for radiation. Large-volume Cherenkov detectors like IceCube [1] can detect these neutrinos through production of secondary leptons and hadronic showers when they interact with the detector material.

Here we present a follow-up analysis to a recent IceCube search for neutrinos of EeV energies [2], which found two neutrino events with energies around 1 PeV, just above its total charge threshold. Their topologies were consistent with either a neutral-current interaction or a charged-current interaction of $\nu_e$ or $\nu_\tau$. By selecting for events with vertices well contained in the detector volume, our analysis has a lower energy threshold (starting at about 50 TeV), a higher sensitivity at energies up to 10 PeV and is sensitive to all neutrino flavors from all directions. The goal of this analysis is to characterize the flux responsible for these events.

Both analyses share the same data-taking period, starting in May 2010 using 79 strings and continuing with the completed detector (86 strings) from May 2011 to May 2012 for a total livetime of 662 days.

## 2 Event Selection

Backgrounds for cosmic neutrino searches arise entirely from interactions of cosmic rays in the Earth's atmosphere. These produce secondary muons that penetrate into underground neutrino detectors from above as well as atmospheric neutrinos that reach the detector from all directions due to the low neutrino cross-section which allows them to penetrate the Earth from the opposite hemisphere.

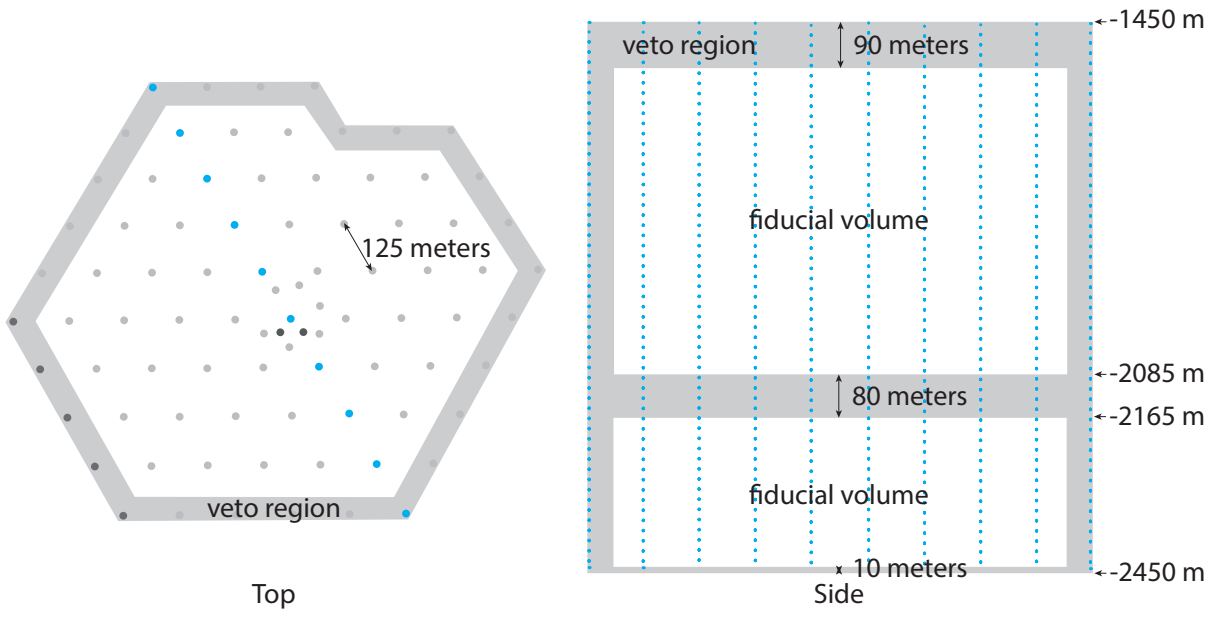

**Figure 1**: Illustration of the anticoincidence veto region used for this analysis. Events producing first light in the veto region are discarded as entering tracks. Most background events are nearly vertical, requiring a thick veto cap at the top of the detector. The excluded region in the middle contains ice of high dust concentration [3]. Due to the high degree of light absorption in this region, near horizontal events could have entered here without being tagged at the sides of the detector without a dedicated veto region.

Neutrino candidates were selected by finding events that originated within the detector interior. Included were those events that produced their first light within the fiducial volume (Fig. 1) and were of sufficiently high energy such that an entering muon track would have been reliably identified if present. In particular, we required that each event have fewer than three of its first 250 observed photoelectrons (p.e.) detected at the IceCube boundary. In addition, we required that the event produce at least 6000 p.e. overall to ensure that statistical fluctuations in the light yield were low enough to reliably veto entering muons. This event selection rejects 99.999% of the muon background above 6000 p.e. (Fig. 2) while retaining approximately 98% of all neutrino events interacting within the fiducial volume at energies above a few hundred TeV. This selection is largely independent of neutrino flavor, event topology, or arrival direction. It also removes 70% of atmospheric neutrinos [4] in the Southern Hemisphere, where atmospheric neutrinos are usually accompanied into the detector by muons produced in the same parent air shower. To prevent confirmation bias, we conducted a blind analysis designed on a subsample of 10% of the full dataset.

## 3 Event Reconstruction

Neutrino interactions in IceCube have two primary topologies: showers and muon tracks. Secondary muon tracks are created primarily in $\nu_\mu$ charged-current interactions and have a typical range that is on the order of kilometers, larger than the dimensions of the detector. Showers are created by the secondary leptons produced in $\nu_e$ and $\nu_\tau$ charged-current interactions and in the neutral current interactions of neutrinos of all flavors. At the relevant energies, showers have a length of roughly 10 meters in ice and are, to a good approximation, point sources of light.

Using the timing patterns of photon arrival times in individual PMTs allows for reconstruction of shower and track directions and deposited energies. The main systematic uncertainties arise from uncertainties in modeling of photon propagation in the natural ice [3] and from uncertainties on the absolute energy scale. Overall, we estimate an uncertainty of better than 15% on the reconstruction of deposited energy. The typical median angular resolution for showers is 10°-15°, whereas it is much better for tracks due to their extension (around 1° or better, depending on their energy and length).

## 4 Atmospheric Muon and Neutrino Background

Remaining atmospheric muon background in the analysis comes from tracks that produce too little light at the edge of the detector to be vetoed and instead emit their first detected photons in the interior volume, mimicking a starting neutrino. These events usually produce an observable muon track in the detector like that from a $\nu_\mu$ charged-current event.

The veto passing rate for throughgoing muons was evaluated from data by tagging entering events using the outer layer of IceCube. The rate of these known background events that pass the veto one layer of PMTs deeper can be used to estimate the veto efficiency of the original veto (without an outer tagging region) by correcting for the differences in fiducial volume. The resulting predicted veto passing rate agrees well with data at low energies where we expect the event rate to be background dominated (Fig. 2). In our signal region above 6000 p.e., we observed three tagged events passing the inner veto and so predict $6 \pm 3.4$ veto-penetrating muon events in the two-year data set.

Atmospheric neutrino backgrounds were estimated based on IceCube measurements of the Northern Hemisphere muon neutrino flux [5]. We have also included a suppression of the atmospheric neutrino background from the Southern Hemisphere resulting from the fact that accompanying high-energy muons produced in the same air shower can trigger our muon veto if they penetrate to the depth of the detector. This suppression factor has been determined from simulations using the CORSIKA [6] air-shower simulation.

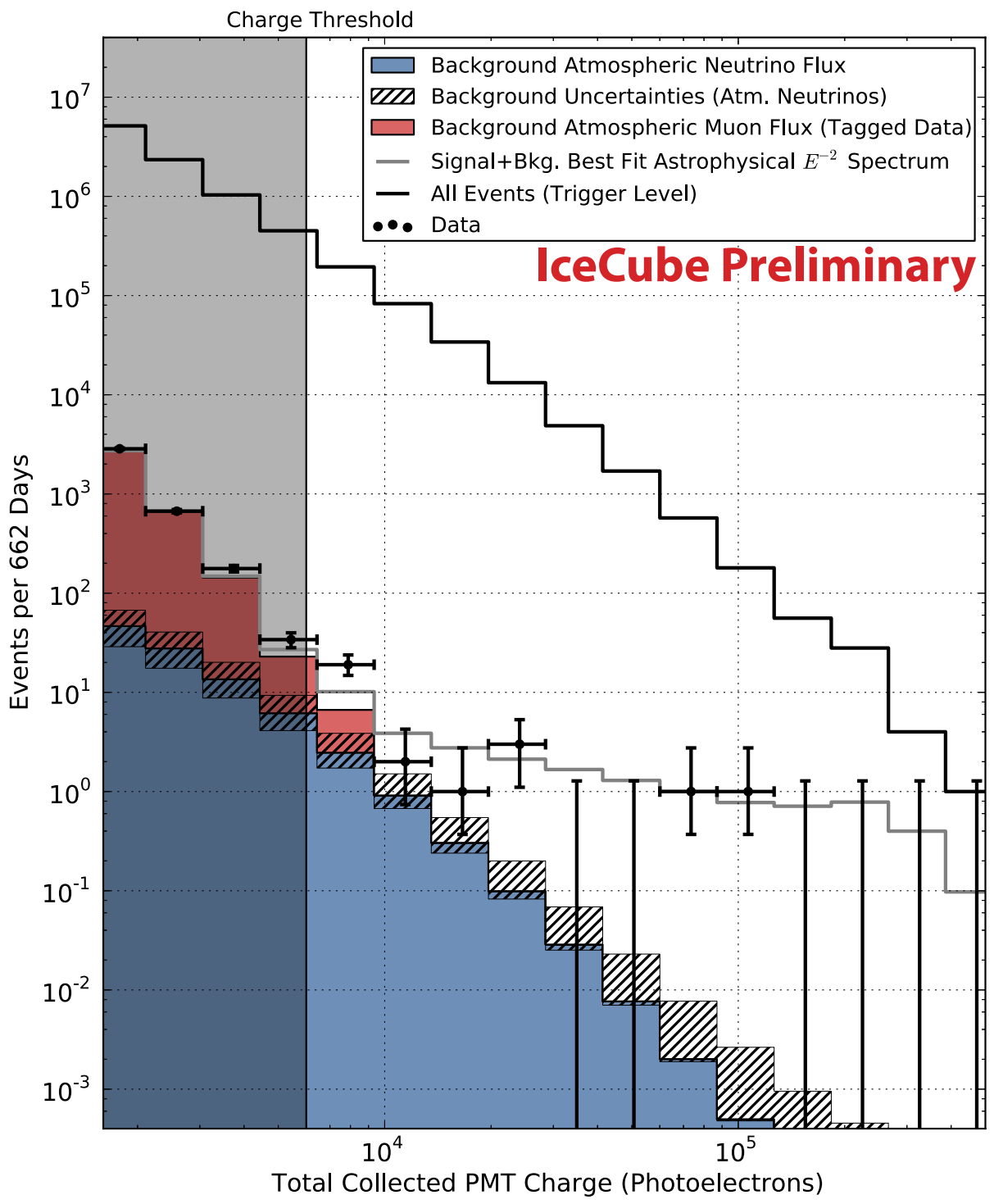

**Figure 2**: Distribution of deposited PMT charge ($Q_{tot}$) for events in IceCube. Muons at higher total charges are less likely to pass the veto layer undetected, causing the muon background (red, estimated from data) to fall faster than the overall cosmic ray spectrum (uppermost line). The data events in the unshaded region, at $Q_{tot} > 6000$, are the events reported in this work. The atmospheric neutrino flux (blue) and best-fit astrophysical spectrum have been determined using Monte Carlo simulations. The hatched region shows uncertainties on the atmospheric neutrino background including a potential component from charmed meson decays (the best-fit charm component included in the blue region is zero) and uncertainties on the normalization of the conventional atmospheric neutrino spectrum.

We estimate an atmospheric neutrino background of $4.6^{+3.3}_{-1.2}$ events in our livetime of 662 days. These events would be concentrated near the energy threshold of the analysis due to the steeply falling atmospheric neutrino spectrum. Uncertainties are dominated by a potential component from charmed meson decays, chemical composition of cosmic rays, hadronic interaction models and the detector energy scale.

## 5 Results

In the two-year dataset, 28 events with deposited energies between 30 and 1200 TeV were observed (Fig. 3) on an expected background of $10.6^{+4.7}_{-3.6}$ events from atmospheric muons and neutrinos. The two highest-energy of these are the previously reported PeV events [2]. Seven events contained clearly identifiable muon tracks, while the remaining twenty-one were shower-like.

The significance of the excess over atmospheric backgrounds was evaluated based on both the total rate and properties of the observed events. From each event, the total deposited PMT charge, arrival angle, and reconstructed energy were used to compute the probability that it belonged to either the atmospheric muon or neutrino backgrounds. Overall significance was computed using the product of these per-event probabilities as a test statistic.

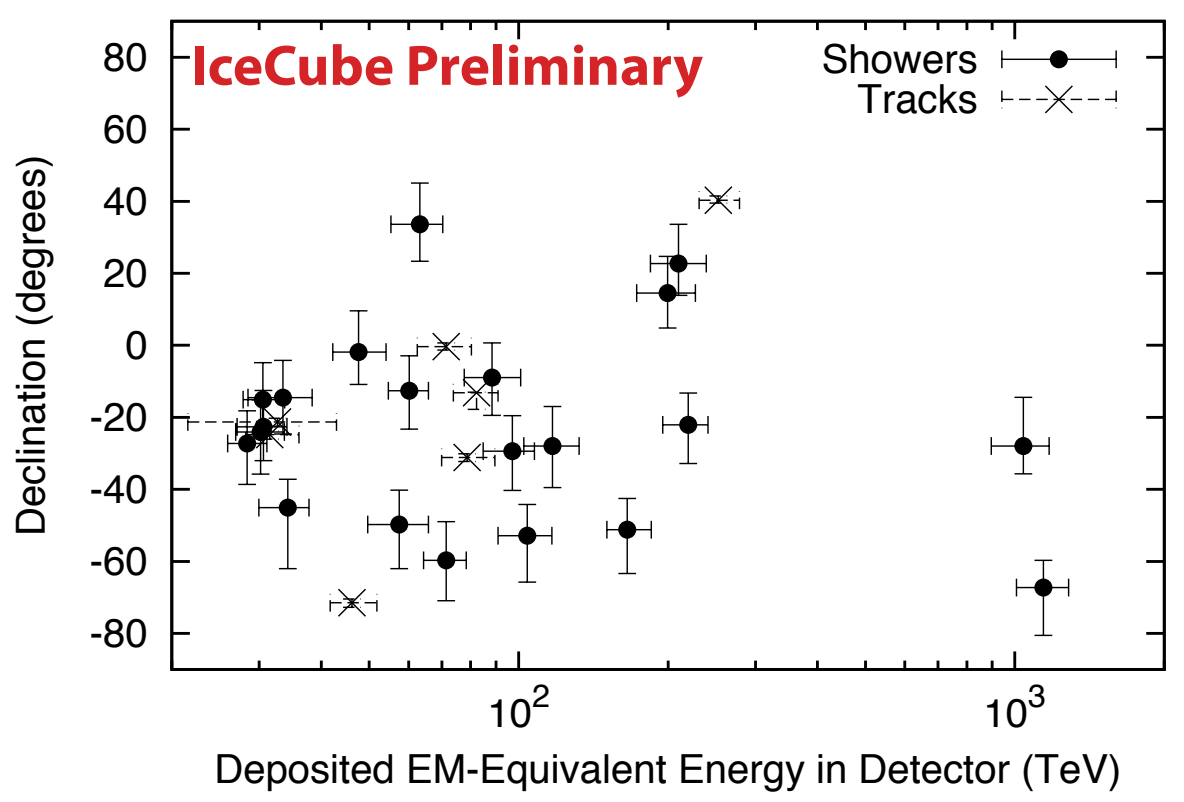

**Figure 3**: Distribution of best-fit deposited energies and declinations. Seven of the events are muons (crosses) with an angular resolution of about 1°, while the remainder are either electromagnetic or hadronic showers (filled circles) with an energy-dependent resolution of about 15°. Error bars are 68% confidence intervals including both statistical and systematic uncertainties. Energies shown are the energy deposited in the detector assuming all light emission is from electromagnetic showers. For $\nu_e$ charged-current events this equals the neutrino energy; otherwise it is a lower limit on the neutrino energy.

Because our two highest energy events were previously reported, they were removed from the sample for purposes of calculating significance. The benchmark atmospheric neutrino model used there [7, 2], which is compatible with experimental bounds but contains a non-zero prompt flux, adds an additional 1.5 prompt neutrinos to our background estimate and predicts $12.1 \pm 3.4$ background events total. Note that this is for a fixed atmospheric flux model, including only uncertainties on the atmospheric muon background. Relative to this model, we obtain a preliminary $3.3\sigma$ (one-sided) significance for those events first reported here. Combined with the $2.8\sigma$ observation of the earlier analysis [2] by Fisher's method, we thereby obtain a preliminary significance for the entire data set of $4.1\sigma$.

## 6 Discussion

Although there is some uncertainty in the expected atmospheric background rates, in particular for the potential contribution from charmed meson decays, the energy spectrum, zenith distribution, and shower to muon track ratio of the observed events strongly constrain the possibility that our events are entirely of atmospheric origin. Almost all of the observed excess is in showers instead of muon tracks, ruling out an increase in penetrating muon background. Atmospheric neutrinos are a poor fit to the data for a variety of reasons. The observed events are much higher in energy, with a harder spectrum (Fig. 4), than expected from an extrapolation of the well-measured conventional atmospheric background at lower energies [8, 5]: nine had reconstructed deposited energies above 100 TeV, with two events above 1 PeV, relative to an expected background from atmospheric neutrinos of 1-2 events above 100 TeV. Raising the normalization of this flux both violates previous limits and, due to $\nu_\mu$ bias in $\pi$ and $K$ decay, predicts too many muon tracks in our data (2/3 tracks vs. 1/4 observed, including muon backgrounds).

Another possibility is that the high energy events result from charmed meson production in air showers [7, 9]. These produce higher energy events with no bias toward muon neutrinos, matching our observed muon track fraction reasonably well. However, our event rates are substantially higher than even optimistic models [9] and the energy spectrum even from charm production is too soft to explain the data. More importantly, increasing charm production to the level required to explain our observations violates existing experimental bounds [5]. The additional events added by an increased flux from charm would originate predominantly from the northern rather than the southern sky, whereas the majority of our events are contained in the southern sky where atmospheric neutrinos produced by any mechanism are suppressed by detection of their accompanying air showers.

By comparison, a neutrino flux produced in extraterrestrial sources would, like our data, be heavily biased toward showers because neutrino oscillations over astronomical baselines tend to equalize neutrino flavors [10, 11]. The observed zenith distribution is also typical of such a flux: as a result of absorption in the Earth above tens of TeV energy, most events (approximately 60%, depending on the energy spectrum) from even an isotropic high-energy extraterrestrial population would be expected to appear in the Southern Hemisphere. Although the zenith distribution is well explained (Fig. 4) by an isotropic flux, a slight southern excess remains, which could be explained either as a statistical fluctuation or by a source population that is either relatively small or unevenly distributed through the sky.

This discussion can be quantified by an a posteriori fit to the energy and zenith distributions of the events above 60 TeV as a combination of the conventional atmospheric background, a possible prompt component, and an isotropic equal-flavor extraterrestrial power-law flux. The best-fit energy spectrum of the extraterrestrial component is similar ($E^{-2.2}$) to the $E^{-2}$ spectrum generically expected from a primary cosmic ray accelerator. For such a generic $E^{-2}$ spectrum we obtain a best-fit normalization of $E^2\Phi(E) = (3.6 \pm 1.2) \cdot 10^{-8}\,\mathrm{GeV\,cm^{-2}\,s^{-1}\,sr^{-1}}$ (shown in Fig. 4) and a cutoff energy of $1.6^{+1.5}_{-0.4}$ PeV. The cutoff fitted here is a hard energy cutoff, but due to low statistics its shape cannot be constrained by our fit. A flux at this level is compatible with previous IceCube observations [2, 5].

In order to test for spatial clustering of the events, a significance against the hypothesis that all events in this sample are uniformly distributed in right ascension was calculated. This test (performed once on the full sample and again on the subset of shower-like events) did not yield a significant result. Several other tests (among them a galactic plane correlation study and multiple time clustering tests) did not yield significant results, either. A future larger dataset containing more well-resolved events than in our sample may allow identification of potential galactic or extragalactic sources.

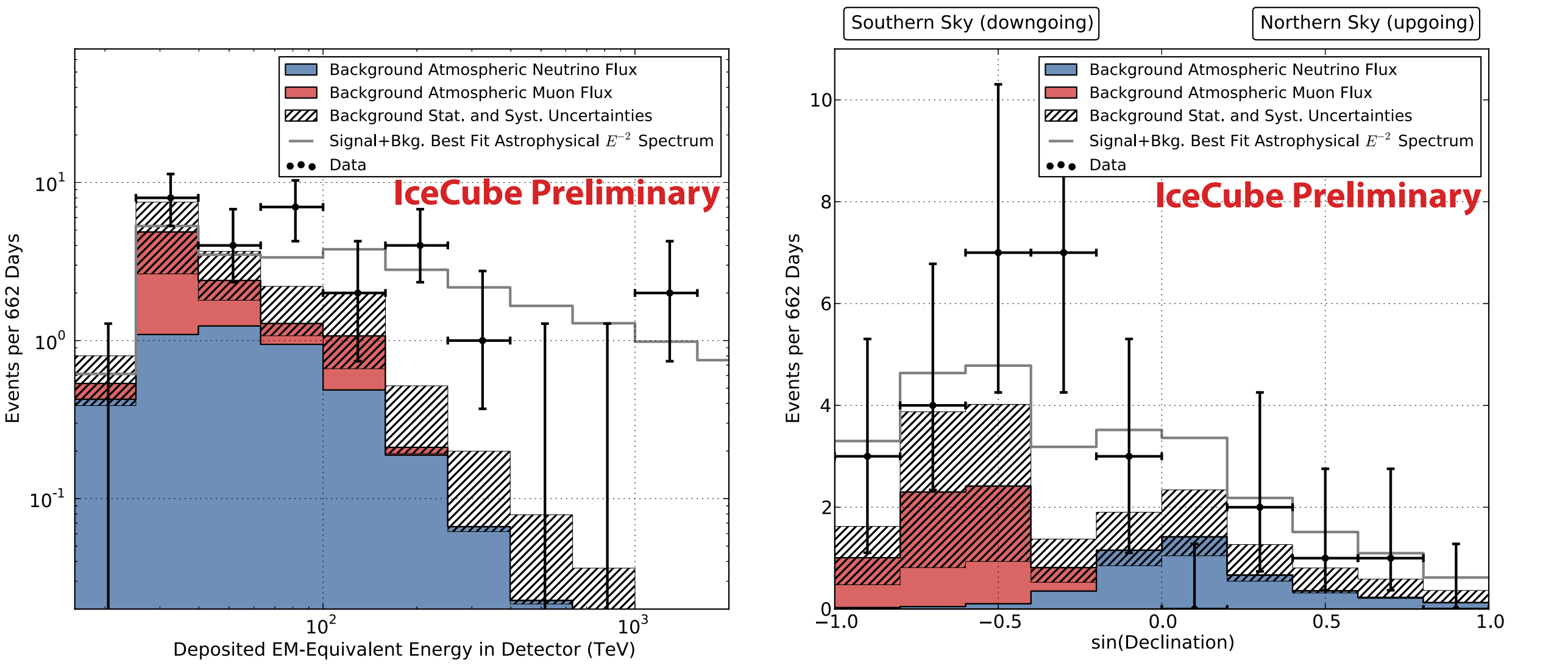

**Figure 4**: Distribution of the deposited energies (left) and declination angles (right) of the observed events compared to model predictions. Zenith angle entries for data are the best-fit zenith position for each of the 28 events. For some of them the angular uncertainty can lead to zenith widths wider than the shown bin width. Energies plotted are in-detector visible energies, which are lower limits on the neutrino energy. The estimated distribution of the background from atmospheric muons is shown in red. Due to lack of statistics from data far above our cut threshold, the shape of the distributions from muons in this figure has been determined using Monte Carlo simulations with total rate normalized to the estimate obtained from our in-data control sample. Combined statistical and systematic uncertainties on the sum of backgrounds are indicated with a hatched area. The gray line shows the best-fit $E^{-2}$ astrophysical spectrum with all-flavor normalization (1:1:1) of $E^2\Phi_\nu(E) = 3.6 \cdot 10^{-8}\,\mathrm{GeV\,cm^{-2}\,s^{-1}\,sr^{-1}}$ and a cutoff energy of 2 PeV.

## 7 Conclusion

An analysis of two years of IceCube data from 2010 to 2012 has revealed 28 events with deposited energies between 30 and 1200 TeV, including the most energetic neutrinos ever observed. The set of events does not seem to be compatible with expectations for terrestrial processes, deviating at a preliminary significance of $4.1\sigma$ from standard assumptions. It contains a mixture of neutrino flavors with events originating primarily from the Southern Hemisphere where high energy neutrinos are not absorbed by the Earth. The events are compatible with a flux proportional to $E^{-2}$, a spectrum expected for neutrinos associated with primary cosmic ray acceleration. The sample is thus consistent with generic expectations for a neutrino population with origins outside the solar system. We did not observe significant spatial clustering of the events, although this study is currently limited by low statistics and poor angular resolution for the majority of the observed events. Future observations with IceCube will provide improved measurements of the energy spectrum and origins of this flux, providing insight into the underlying processes responsible for these events.

# Search for diffuse astrophysical neutrinos with cascade events in the IceCube-59 detector

THE ICECUBE COLLABORATION[1]

[1] *See special section in these proceedings*

*arne.schoenwald@desy.de*

**Abstract:** We report the results of a search for a diffuse flux of high-energy (> 30 TeV) astrophysical neutrinos using data from the IceCube neutrino observatory in its 59 string configuration. The data was taken between May 2009 and April 2010, yielding a total live time of 335 days. Two complementary analyses were performed on this dataset following different strategies for the selection of shower-type events that arise from all neutrino flavors. One of these analysis finds 8 neutrino candidates (for an expected background of $4.0 \pm 0.3_{\mathrm{stat}}$ events) above the energy threshold of 38 TeV (the results of the second analysis will be reported later somewhere else). A likelihood analysis of the energy spectrum of these events shows that their spectrum is compatible with expectations from the background of atmospheric muons and neutrinos. Therefore an upper limit on a diffuse isotropic astrophysical neutrino flux with a power-law spectrum $d\Phi/dE = \Phi_0 \cdot E^{-2}$ is derived from the likelihood fit. Assuming a neutrino flavor ratio of $\nu_e : \nu_\mu : \nu_\tau = 1:1:1$, we preliminary constrain the astrophysical neutrino flux normalisation ($\Phi_0$) to be smaller than $1.7 \cdot 10^{-8}\,\mathrm{GeV\,s^{-1}\,sr^{-1}\,cm^{-2}}$ in the energy range between 43 TeV and 6310 TeV.

**Corresponding authors:** Arne Schönwald[1], Anthony M. Brown[2], Lars Mohrmann[1]
[1] *DESY (Zeuthen), Platanenallee 6, 15738 Zeuthen, Germany*
[2] *Department of Physics and Astronomy, University of Canterbury, Christchurch, 8140, New Zealand*

## 1 Introduction

Astrophysical neutrinos are diagnostic probes of possible acceleration processes in the universe. In contrast to gamma rays, they are only produced in hadronic interaction processes, mainly via p-p or p-$\gamma$ interactions. Therefore a detection of sources of astrophysical neutrinos would probe the galactic and extragalactic acceleration sites of cosmic rays. No such sources have been found yet [1]. Even if individual sources are too weak to be resolved, the cumulative emission from all the sources in the universe might be discovered. We expect a harder energy spectrum from such astrophysical neutrino sources than from the atmospheric neutrino and muon backgrounds that are often present in such a search when performed with a ground-based telescope. The Fermi shock acceleration process [2] that is usually assumed to be responsible for the acceleration of cosmic rays produces a power-law spectral index of $\gamma \approx -2$. Neutrinos in astrophysical environments are dominantly produced by the decay of pions and therefore follow a flavor ratio of $\nu_e : \nu_\mu : \nu_\tau = 1:2:0$. Flavor oscillations during their propagation to Earth then change the flavor ratio to $\nu_e : \nu_\mu : \nu_\tau = 1:1:1$ [3, 4].

Shower-type neutrino events, i.e. neutrino interactions that produce a localized hadronic or electromagnetic shower within a neutrino detector, are particularly well suited to a search for diffuse astrophysical neutrinos as they allow a calorimetric measurement of the full energy deposited in the neutrino interaction. They arise from the neutral-current (NC) interactions of all neutrino flavors as well as from charged-current (CC) interactions of $\nu_e$ and $\nu_\tau$.

IceCube is a neutrino observatory located at South Pole and consists of 5160 Digital Optical Modules (DOMs) instrumenting a volume of 1 km$^3$ between depths of 1450 m and 2450 m. The DOMs are designed to detect the Cherenkov light emitted from muons and showers produced in neutrino interactions. See [5, 6] for detailed information on the IceCube detector and DOMs.

In this proceeding, two complementary searches for shower-type events are presented. Both use data from the IceCube Neutrino Observatory in its 59-string configuration between May 2009 and April 2010 (335 days of live time). They follow different strategies to remove the abundant background of atmospheric muons and neutrinos produced in cosmic-ray interactions in the atmosphere. Both astrophysical neutrino event selections were developed on a 10% sub-sample of the full dataset to avoid the introduction of statistical biases in the event selection. One of the searches (denoted as "Analysis A" below) has recently been applied to the full dataset and we report the results from this search. The results of the second analysis ("Analysis B") will be reported elsewhere.

## 2 Event selection

Most of the events recorded by the IceCube detector are muons from cosmic-ray air showers. Their rate was 1.6 kHz in the 59-string IceCube configuration. These events usually produce track-like patterns and the vast majority of them can be distinguished well from shower-like events. Both analyses presented here use a common filter designed to remove such track-like events and to select candidate shower-like events. It is based on simple reconstruction algorithms, e.g. to exploit their different geometrical patterns in the detector, described here [7]. This common filter reduces the data rate to about 1 Hz. 78% of simulated $E^{-2}$ neutrino showers which triggered the detector survive this filtering.

Events remaining after the application of the common filter are still dominated by muons from cosmic-ray air showers. However, these muons mimic shower-type events, either because they feature a large catastrophic energy loss via emitting a bremsstrahlung photon along their track, or because the muons only graze the corners and edges of the instrumented volume. Different strategies have been employed to remove these muons and to obtain a sample of neutrino induced showers. Both strategies have been developed based on the event patterns observed in Monte Carlo simulations of cosmic-ray induced muons, neutrino induced muons and showers. The effective simulated live time for all simulations was larger than the actual observation period of 335 days in the energy ranges relevant for the two analyses.

The differences between the two strategies are briefly described. The two strategies are denoted as "Analysis A" and "Analysis B".

### 2.1 Analysis A

Analysis A selected events that are fully contained in the detector. Containment was defined by requiring the x-y position (horizontal plane) of the reconstructed event vertex to be within the green polygon displayed in Fig. 1. The z-position of the reconstructed vertex was required to be between $-450$ m $< z <$ 450 m, i.e. at least 50 m inward away from the top and bottom of the detector. Events were also rejected if the DOM that recorded the highest number of photons was located on a string outside the green polygon. Furthermore, only those events that had recorded light patterns on more than three strings were considered.

More complex event properties to remove remaining background events have also been exploited, including the fraction of DOMs that recorded light in a virtual sphere around the reconstructed vertex (Fill Ratio), the difference between earliest expected and actual arrival time of the first photon in the event, and the likelihood value of the vertex, direction and energy reconstruction. At this point, the remaining event sample was dominated by atmospheric neutrinos from $\nu_e$ and $\nu_\mu$ NC interactions that are expected at a rate of $(2.1 \pm 0.1_{\mathrm{stat}})$ $\mu$Hz.

The final step in the event selection was then to choose an energy threshold that optimized the sensitivity of a search for an astrophysical signal based on count statistics. Only events with reconstructed energies of $\log_{10}(E/\mathrm{GeV}) > 4.58$ were used for the likelihood fit. No simulated cosmic ray shower-induced muons with a reconstructed energy above this threshold remained. Based on an extrapolation of the distribution of cosmic ray shower-induced muons below the threshold, their contribution above threshold was estimated to be $0.65 \pm 0.25_{\mathrm{stat}}$ events.

The number of atmospheric neutrinos from $\pi$ and K-meson decay expected in this sample was $2.0 \pm 0.2_{\mathrm{stat}}$ events based on flux calculations of [8] and after applying corrections for slope changes in the CR spectrum described in [9]. An additional atmospheric neutrino contribution from the decay of charmed mesons ("prompt" contribution) has large theoretical uncertainties. Using the model of [10] we can estimate this contribution to $1.3 \pm 0.1_{\mathrm{stat}}$ , however the current experimental upper limits for the atmospheric neutrino flux [11] from charged mesons is several times this flux. Based on MC data an all-flavor sensitivity (see [12]) of the analysis of $2.5 \cdot 10^{-8}\,\mathrm{GeV\,s^{-1}\,sr^{-1}\,cm^{-2}}$ was derived.

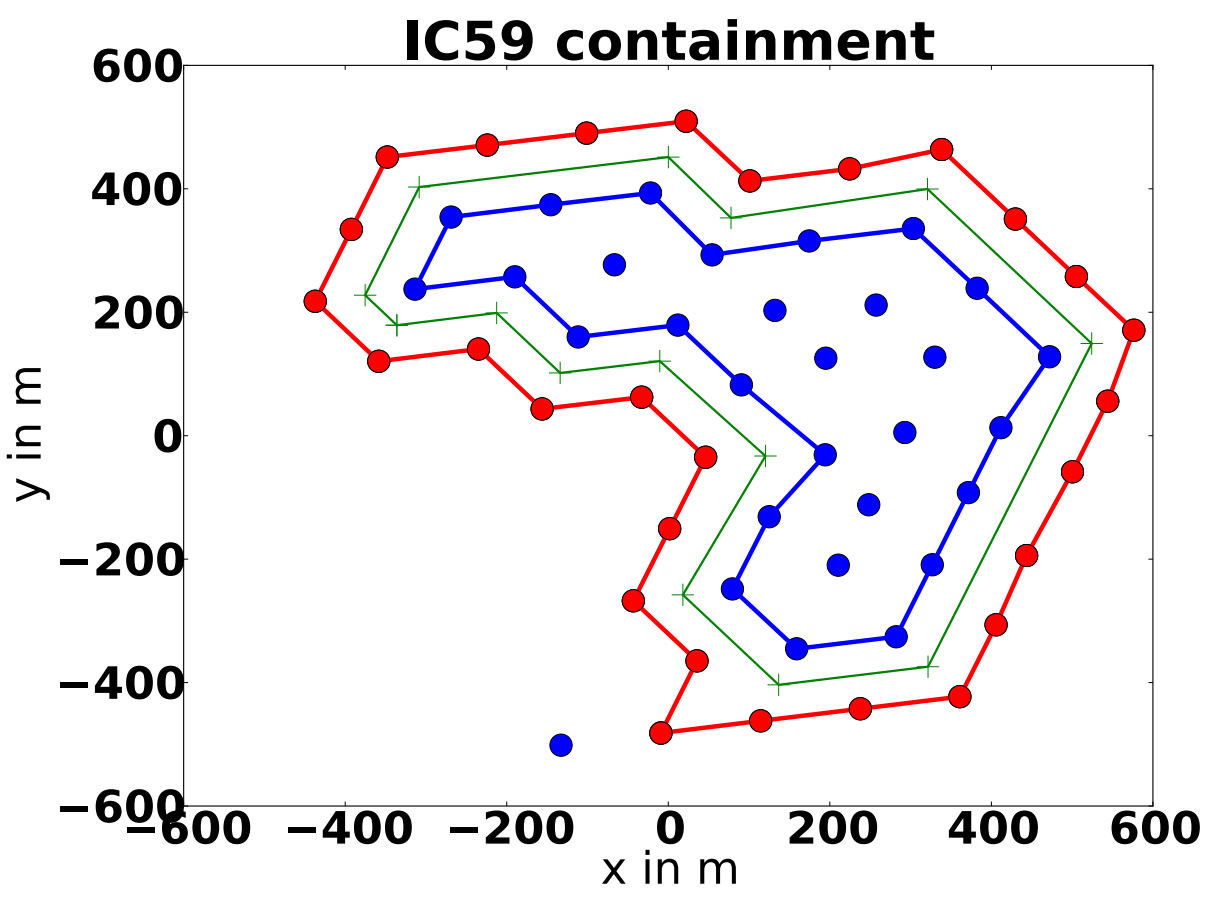

**Fig. 1**: Definition of containment in the x-y (horizontal) plane for the analyses presented here. Each dot represents an IceCube string, the outermost layer is marked by a red line, the second layer by a blue line, the green line is halfway between the two layers of strings. Analysis A requires the reconstructed event vertex to be within the area bounded by the green line, analysis B requires the reconstructed vertex to be within the red line.

### 2.2 Analysis B

The main difference from analysis A is the use of a multivariate analysis (MVA) as part of the final cut level. The machine learning algorithm used for the MVA was a Boosted-Decision-Tree (BDT). Only contained events were used to train the BDT. Containment was defined slightly different in this analysis in comparison to the containment definition described for analysis A. Events were considered contained if they met the following conditions:

- **X-Y-Z Containment:** the reconstructed X-Y position of the event vertex had to lie within the most outer ring of detector strings (red polygon in Fig. 1) and the reconstructed Z position must be $-450\,\mathrm{m} < Z < 450\,\mathrm{m}$, where Z = 0 is the center of the detector along the z axis.

- **Charge Containment:** the DOM with the maximum charge is not allowed to be on the outer ring of detector strings (red polygon in Fig. 1).

- **NString:** number of strings with a hit DOM in an event had to be greater than 3.

The BDT was trained using a sample of electron neutrinos, with an $E^{-2}$ spectrum, for the signal, and a sample of simulated cosmic-ray air showers to describe the atmospheric muon flux, considered to be the dominant background. The BDT was trained using the following variables:

- **NCh:** the number of channels hit during the event.

- **Fill Ratio:** ratio of the DOMs hit within a sphere (with a radius to fully contain the event) around the reconstructed vertex position to the total number of DOM hits.

- **Log-likelihood difference:** Difference in the log-likelihoods for the event, when reconstructed as a cascade or a track.

- **Eigenvalue ratio (of the tensor of inertia):** Topology of the event.
- **Zenith:** Reconstructed zenith angle of the event.
- **Total charge over length ratio:** ratio of the total charge of an event and how elongated an event is.
- **Position:** distance from the centre of the detector.
- **Total charge per string:** amount of charge per string for an event.
- **Difference in linefit velocity:** difference in the z-component of the particle velocity for the two halves of an event.

After BDT training, a BDT response score was calculated for each event. This BDT score influenced the final selection, along with the reconstructed energy of the event. The energy and BDT score range for the final data sample were optimized to deliver the best possible average upper limit on an astrophysical neutrino flux with a power-law spectrum with spectral index $\gamma = -2$. Events with an energy above 50 TeV in energy and a BDT score higher than -0.13 were selected.

## 3 Results

We present results from "Analysis A" which has been applied to the full dataset. Two data events were found that fulfilled all selection criteria in the 10% of the data sample that was used to develop the event selection. (The development of Analysis B is not yet complete, but the same two events were found in the 10% data sample.) Six more events were found in the remaining sample for a total of 8 events (for an expected background of $4.0 \pm 0.3$ ). Table 1 lists the properties of these events. Figure 2 shows the energy distribution of the 8 events which were found between 39 TeV and 67 TeV.

| event | sample | E/TeV | x/m | y/m | z/m | q/pe |
|---|---|---|---|---|---|---|
| 1 | 10% | 67 | 266 | 325 | -397 | 5152 |
| 2 | 10% | 52 | -227 | 213 | 321 | 1404 |
| 3 | 90% | 42 | 452 | 32 | 369 | 1108 |
| 4 | 90% | 39 | 200 | 240 | -259 | 2510 |
| 5 | 90% | 61 | 123 | 8 | 43 | 2552 |
| 6 | 90% | 43 | 442 | 192 | 400 | 567 |
| 7 | 90% | 48 | 422 | 125 | 213 | 1948 |
| 8 | 90% | 39 | 126 | -98 | 61 | 3643 |

**Table 1**: Event parameters in the 10% and 90% data samples: energy in TeV, event coordinates x, y and z, sum of charge over all DOMs in number of photoelectrons (pe).

A binned maximum likelihood fit was applied to the energy distribution of the events in Fig. 2 to determine the uncertain contribution to the atmospheric neutrino flux from charmed meson decay and a possible contribution from an astrophysical neutrino flux. The parameters of the fit are the normalization of the charm contribution, and normalization and spectral index of an astrophysical neutrino flux with a power-law spectrum. The spectral energy distribution of the charm contribution is taken from [10] for this work. Systematic uncertainties on the K/$\pi$ induced atmospheric neutrino flux and spectrum, the absolute scale of the energy reconstruction, and the residual contamination from atmospheric muons have been included in the likelihood fit as nuisance parameters that are allowed to vary within their respective uncertainty ranges. Table 2 shows parameter uncertainty and the pulls of the nuisance parameters of the likelihood fit to the data. The energy scale is slightly shifted upwards by 4%, the normalisation of the atmospheric neutrino flux (from K/$\pi$ decays) is slightly increased by 4% and the normalisation of the muon flux (created in interactions of the cosmic rays within the atmosphere) is slightly increased by 8%.

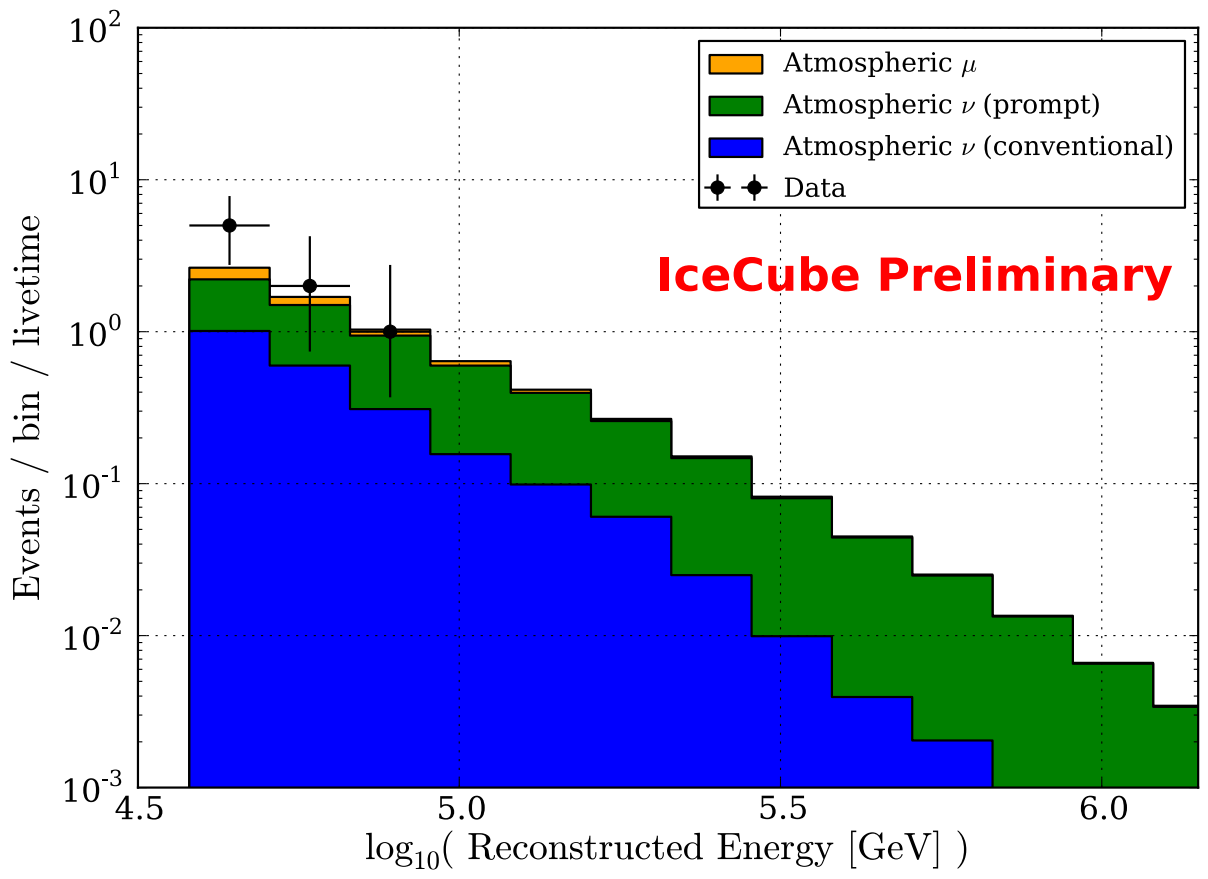

**Fig. 2**: Energy spectrum of the events found in the data sample obtained from IceCube in its 59-string configuration. Data events are marked in black, the yellow area is the predicted contamination with atmospheric muons, the blue area is the expected contribution of atmospheric neutrinos from $\pi$-meson and K-meson decays. The green area represents the best-fit contribution from prompt atmospheric neutrinos corresponding to $2.9^{+3.2}_{-2.6}$ times the flux predicted in [10]. The best likelihood in the fit was obtained without any additional astrophysical neutrino contribution.

| parameter | pull | parameter uncertainty |
|---|---|---|
| $\phi_{conv}$ | 0.16 | $\sigma[\phi_{conv}] = 0.25$ |
| $\phi_{\mu}$ | 0.16 | $\sigma[\phi_{\mu}] = 0.5$ |
| cosmic ray index | 0.1 | $\sigma[\text{index}] = 0.5$ |
| energy scale | 0.27 | $\sigma[\text{energy scale}] = 0.15$ |

**Table 2**: Parameter pulls and parameter uncertainties of the nuissance parameters of the likelihood fit: $\phi_{conv}$ is the change in the normalisation of the atmospheric neutrinos from $\pi$-meson and K-meson decays, $\phi_{\mu}$ is the change in the normalisation of muons produced in the atmosphere, the cosmic ray index is the change of the index of cosmic rays and the energy scale is a possible shift in the energy.

The best fit to the data has no astrophysical neutrino contribution and an atmospheric flux from charmed meson decays corresponding to $2.9^{+3.2}_{-2.6}$ times the flux predicted in [10]. The likelihood profile shown in Figs. 3 and

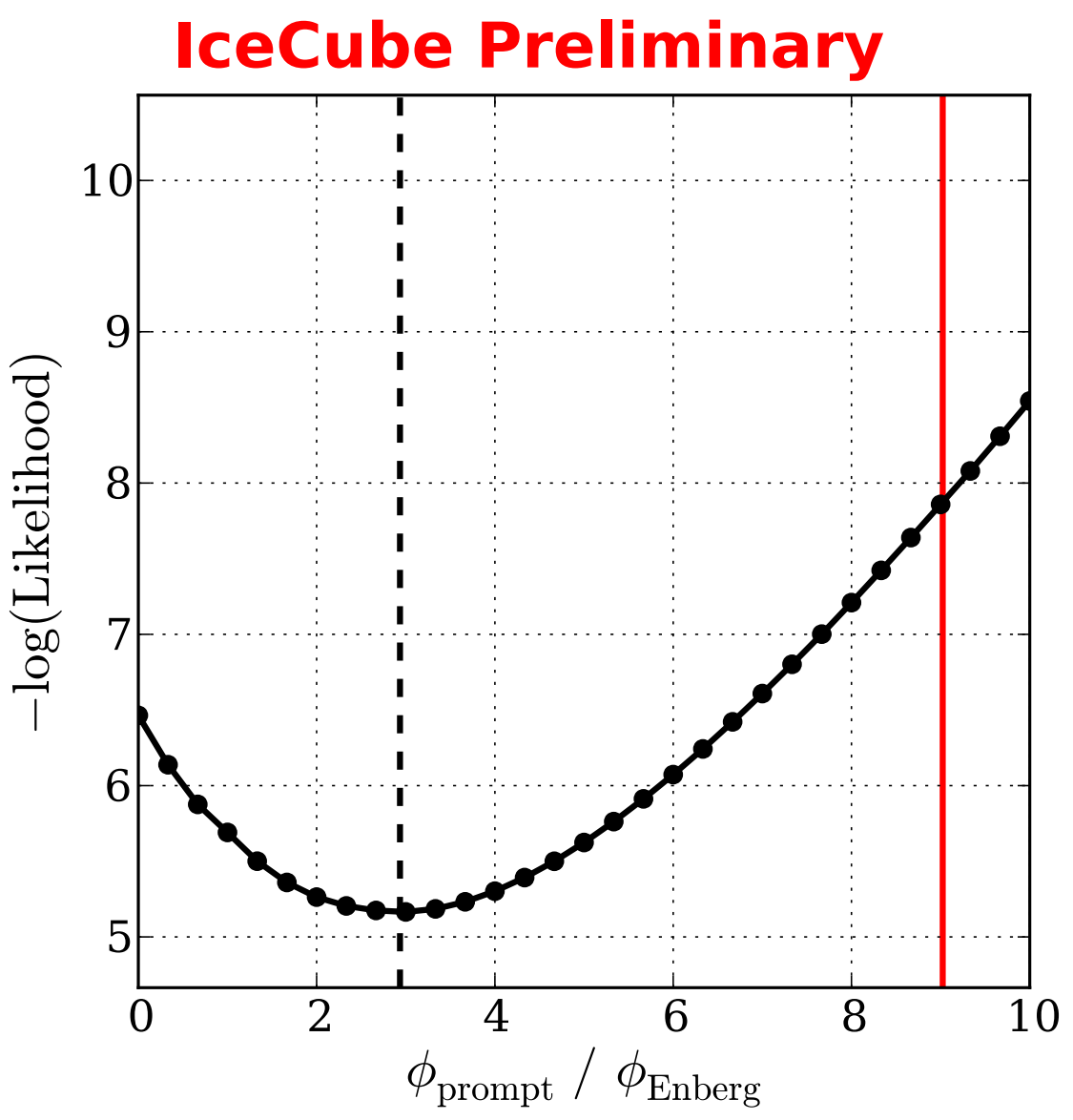

**Fig. 3**: Results of the Likelihood fit. The plot shows the likelihood profile varying the prompt atmospheric neutrino flux in units of the Enberg flux (see [10]). The black dashed line marks the best fit value. The red line shows the upper limit at 90%-confidence level on the prompt atmospheric neutrino flux.

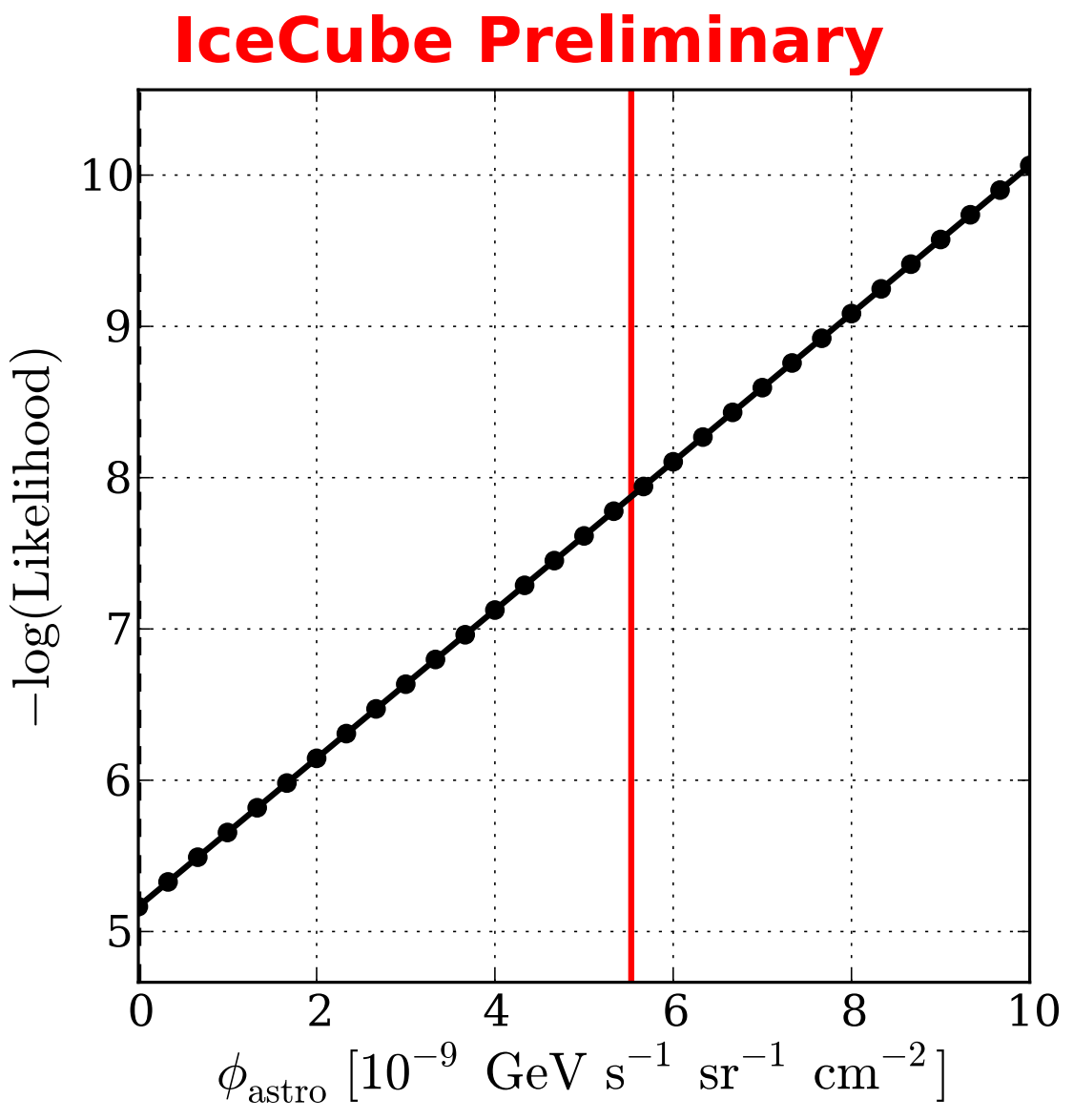

**Fig. 4**: Results of the Likelihood fit. The plot shows the likelihood profile for an astrophysical flux (per flavor) assuming a power-law with an index of $\gamma = -2$. The best likelihood is obtained with no astrophysical flux. An upper limit can be obtained from the likelihood profile. The red line indicates the per-flavor upper limit of the astrophysical flux at a 90%-confidence level.

4 is used to determine upper limits on the prompt atmospheric neutrino and the astrophysical neutrino contributions. To derive the upper limits the spectral index of the astrophysical neutrino contribution in the fit is fixed to $\gamma = -2$, in accordance with the Fermi acceleration prediction. From the likelihood profile we derive a preliminary 90%-confidence upper limit on the astrophysical neutrino flux of $1.7 \cdot 10^{-8}\,\mathrm{GeV\,s^{-1}\,sr^{-1}\,cm^{-2}}$ in the energy range between 43 TeV and 6310 TeV. We note that the limit given above applies to an unbroken power law extending over the entire energy range, and a higher normalization would be allowed in part of the energy range if the spectrum were cut off at some energy. This limit is more than an order of magnitude more constraining than the one from an earlier IceCube analysis (see [7]), which had reported an upper limit on the astrophysical flux of $3.6 \cdot 10^{-7}\,\mathrm{GeV\,cm^{-2}\,s^{-1}\,sr^{-1}}$ in the energy range between 24 TeV and 6600 TeV. A preliminary 90%-confidence upper limit on the contribution of charm decays to the atmospheric neutrino flux was found at 9.0 times the flux predicted by [10] (including a modification to the prompt atmospheric neutrino spectrum of [10] that is described in [9] and caused by a steepening of the primary cosmic-ray spectrum at PeV energies that was not considered in the original calculation).

## 4 Conclusions

We report a search for the isotropic diffuse flux of astrophysical neutrinos in data recorded with the IceCube neutrino observatory between May 2009 and April 2010. No indications for such a flux were found based on the outcome of a likelihood fit to the energy distribution of the observed neutrinos. The observed events are compatible with atmospheric origin and therefore a preliminary 90% upper limit of $1.7 \cdot 10^{-8}\,\mathrm{GeV\,s^{-1}\,sr^{-1}\,cm^{-2}}$ can be set on an astrophysical neutrino flux with an unbroken power-law type spectrum power-law type spectrum of index $\gamma = -2$ in the energy range between 43 TeV and 6310 TeV. This is currently the best upper limit on such an astrophysical flux. Again we stress that this limit is only valid for an unbroken power law extending over the entire energy range.